\documentclass{aa}  
\usepackage{graphicx}
\usepackage{subfigure} %
\usepackage{graphicx}
\usepackage{subfigure}
\usepackage{hyperref}
\hypersetup{
    colorlinks = true,
    citecolor ={blue}
}
\usepackage{txfonts}
\usepackage{verbatim}
\usepackage{soul}
\usepackage{float}
\usepackage{pifont}
\usepackage{tabularx,adjustbox,booktabs}
\begin{document}

\title{Pulsating hydrogen-deficient white dwarfs and pre-white dwarfs 
observed with {\sl TESS}} 
\subtitle{V. Discovery of two new DBV pulsators, WDJ152738.4$-$450207.4 and WD~1708$-$871, 
and asteroseismology of the already known DBV stars PG~1351+489, EC~20058$-$5234, and EC~04207$-$4748}

\author{Alejandro H. C\'orsico\inst{1,2}, 
        Murat Uzundag\inst{3,4},          
        S. O. Kepler\inst{5},
        Leandro G. Althaus\inst{1,2},
        Roberto Silvotti\inst{6},             
        Paul A. Bradley\inst{7},
        Andrzej S. Baran\inst{8},         
        Detlev Koester\inst{9},           
        Keaton J. Bell\inst{10,11},
        Alejandra D. Romero\inst{5},
        J. J. Hermes\inst{12},
        Nicola P. Gentile Fusillo\inst{13}
}
           \institute{Grupo de Evoluci\'on Estelar y Pulsaciones. 
           Facultad de Ciencias Astron\'omicas y Geof\'{\i}sicas, 
           Universidad Nacional de La Plata, 
           Paseo del Bosque s/n, 1900 
           La Plata, 
           Argentina
           \email{acorsico@fcaglp.unlp.edu.ar}
           \and
           IALP - CONICET, La Plata, Argentina
           \and
           Instituto de F\'isica y Astronom\'ia, Universidad de Valpara\'iso, Gran Breta\~na 1111, Playa Ancha, Valpara\'iso 2360102, Chile
           \and
           European Southern Observatory, Alonso de Cordova 3107, Santiago, Chile
           \and
           Instituto de F\'{i}sica, Universidade Federal do Rio Grande do Sul, 91501-900, Porto-Alegre, RS, Brazil
           \and 
           INAF-Osservatorio Astrofisico di Torino, strada dell'Osservatorio 20, 10025 Pino Torinese, Italy
           \and 
           XCP-6, MS F-699 Los Alamos National Laboratory, Los Alamos, NM 87545, USA
           \and
           Uniwersytet Pedagogiczny, Obserwatorium na Suhorze, ul. Podchor\c{a}\.zych 2, 30-084 Krak\'ow, Polska
           \and
           Institut f\"ur Theoretische Physik und Astrophysik, Universit\"at Kiel, 24098 Kiel, Germany
           \and
           DIRAC Institute, Department of Astronomy, University of Washington, Seattle, WA-98195, USA 
           \and
           NSF Astronomy and Astrophysics Postdoctoral Fellow
           \and    
           Department of Astronomy \& Institute for Astrophysical Research, Boston University, 725 Commonwealth Ave., Boston, MA 02215, USA
           \and
           European Southern Observatory, Karl-Schwarzschild-Straße 2, Garching 85748, Germany           
           }
           
\date{Received ; accepted }
\abstract{
The {\sl TESS} space mission has recently demonstrated its great
potential  to discover new pulsating white dwarf and pre-white dwarf
stars, and to  detect periodicities with high precision in already
known white-dwarf pulsators.}  
{We report the discovery of two new pulsating He-rich atmosphere white
  dwarfs (DBVs) and present a detailed asteroseismological analysis of
  three already known DBV stars employing observations collected by
  the {\sl TESS} mission along with ground-based data.}
{We processed and analyzed {\sl TESS} observations of the three
  already  known DBV stars PG~1351+489 (TIC~471015205),
  EC~20058$-$5234 (TIC~101622737),  and EC~04207$-$4748
  (TIC~153708460), and the two new DBV pulsators WDJ152738.4$-$50207.4
  (TIC~150808542) and WD~1708$-$871 (TIC~451533898), whose variability
  is reported for the first time in this paper.  We also carried out a
  detailed asteroseismological analysis using fully evolutionary DB
  white-dwarf  models built considering the  complete evolution of the
  progenitor stars. We constrained  the   stellar  mass  of three of
  these  target stars  by means of the  observed period spacing, and
  derived a representative asteroseismological model using the
  individual periods, when possible.}
{We extracted frequencies from the {\sl TESS} light curves of these
  DBV   stars using a standard  pre-whitening procedure to derive the
  potential pulsation frequencies. All the oscillation frequencies
  that we found  are associated with $g$-mode pulsations with periods
  spanning from $\sim 190$ s to $\sim 936$ s.  We find hints of
  rotation from frequency triplets in some of the targets,  including
  the two new DBVs. For three targets, we find constant period
  spacings,  which allowed us to infer their stellar masses and
  constrain the  harmonic  degree $\ell$ of the modes. We also
  performed period-to-period fit analyses and  found an
  asteroseismological model for three targets, with stellar masses
  generally compatible with the spectroscopic masses. Obtaining
  seismological models allowed us to estimate the seismological
  distances  and compare them with the precise astrometric distances
  measured with {\it Gaia}. We find a good agreement between the
  seismic and the astrometric distances for  three stars (PG~1351+489,
  EC~20058$-$5234, and EC~04207$-$4748), although for the other two stars
  (WD~J152738.4$-$50207 and WD~1708$-$871), the
  discrepancies are substantial.}
{The high-quality data from the {\sl TESS} mission continue to provide
  important clues to  determine the internal structure of pulsating
  pre-white dwarf and white dwarf stars  through the tools of
  asteroseismology.}

\authorrunning{C\'orsico et al.}
\titlerunning{Asteroseismology of DBV stars with {\sl TESS}}
\maketitle

\section{Introduction}

Pulsating white dwarfs (WD) and pre-WDs constitute one of the most
thoroughly studied classes in the zoo of pulsating stars \citep[see,
  e.g,][]{2015pust.book.....C,2021RvMP...93a5001A,2022ARA&A..60...31K}.
Their multiperiodic brightness fluctuations, with periods in the range
$100 - 7\,000$ s, and amplitudes up to $0.4$ mag, are due to
low-degree ($\ell \leq 3$) nonradial gravity ($g$) mode pulsations.
These modes are excited by a driving mechanism related to the partial
ionization of the dominant chemical species in the zone of driving,
which is located at the outer regions of these stars
\citep[][]{2008ARA&A..46..157W,2008PASP..120.1043F,2010A&ARv..18..471A}.
The most populous class of pulsating WDs is that of the ZZ Ceti or DAV
stars, which are DA-type ---hydrogen(H)-rich atmospheres--- at $T_{\rm
  eff} \sim 10\,500 - 13\,000$ K, of which $\sim 500$ objects are
known so far \citep{2019A&ARv..27....7C,2020AJ....160..252V,
  2021ApJ...912..125G,2022MNRAS.511.1574R}.  At higher temperatures
($T_{\rm eff} \sim 21\,000 - 30\,000$ K), and with helium
(He)-dominated atmospheres, we find the pulsating WDs called V777 Her
or DBVs.  Although not many DBVs are known at present \citep[47
  objects; see][]{2019A&ARv..27....7C, 2021ApJ...922....2D,
  2022ApJ...927..158V}, these compact pulsating stars are very
interesting because the origin of DB WDs is not entirely clear. They
are probably descendants of the PG~1159 stars (oxygen (O)-, carbon
(C)-, He-rich atmospheres) after going through the DO WD (very hot
He-rich atmospheres) stage
\citep{1995ApJ...445L.141D,2005A&A...435..631A,2020A&A...638A..30B,
  2022ApJ...930....8B}. A large fraction of PG~1159 stars is believed
to be the result of a born-again episode, i.e., a very late thermal
pulse (VLTP) experienced by a WD during its early cooling phase
\citep{1977PASJ...29..331F,1979A&A....79..108S,1983ApJ...264..605I,2005A&A...435..631A}. During
the VLTP, most of the H content of the remnant is violently burned
\citep{1999A&A...349L...5H,2006A&A...454..845M}.  As a result, the
remnant is forced to evolve rapidly back to the asymptotic giant
branch (AGB) and finally into the central star of a planetary nebula
as a hot H-deficient object.  After that, gravitational settling
acting during the early stages of WD evolution causes He to float and
heavier elements (C, O) to sink, giving rise to an He-dominated
surface, and turning the PG~1159 star into a DO WD
\citep{2000A&A...359.1042U}, and later into a DB WD
\citep{2005A&A...435..631A}.  Alternatively, some DB WDs could be
descendant of DO WDs resulting from evolutionary channels that not
involve only the PG~1159 stars. For instance, they could be the result
of post-merger evolution involving the giant, H-deficient RCrB stars
\citep{2006ASPC..348..194R,2009ApJ...704.1605A}. Also, DB WDs could be
the result of the merger of two WDs \citep{1984ApJ...277..355W,
  2019MNRAS.488..438L}.

DBV stars are found to pulsate with periods between 120 and 1080~s
\citep[][]{2008ARA&A..46..157W}. Their existence 
was anticipated theoretically \citep{1982ApJ...252L..65W} 
before being discovered  shortly after \citep{1982ApJ...262L..11W}. 
$g$ modes in DBVs are thought to be excited by a combination of the $\kappa$ mechanism acting in the He partial ionization zone --- and thus setting the blue edge of the DBV
instability strip \citep{1983ApJ...268L..33W},  and  the ``convective driving''
mechanism \citep{1991MNRAS.251..673B} which is possibly dominant once
the outer convection zone has deepened enough \citep[][]{2017ASPC..509..321V}.

At the beginning of the exploration of pulsating WDs, the discovery of 
variable objects with single-site observations \citep[e.g.][]{1968ApJ...153..151L,1982ApJ...262L..11W} and 
even with multi-site campaigns like those of the Whole Earth Telescope \citep[{\it WET};][]{1990ApJ...361..309N,1994ApJ...430..839W}, 
was challenging and slow. A very interesting account of 
those early days can be found in the article of 
\cite{2021FrASS...8..229B}. Later on, discoveries of pulsating WDs 
increased enormously thanks to the identification of candidates from 
the Sloan Digital Sky Survey \citep[SDSS,][]{2000AJ....120.1579Y,Kleinman13,Kepler15,Kepler16,Kepler19}, 
and in recent years, with space missions. In particular, 
the {\it Kepler/K2} mission \citep{2010Sci...327..977B, 2014PASP..126..398H} 
resulted in the study of 32 ZZ Ceti stars and three DBV stars \citep[][]{2011ApJ...736L..39O,2015ApJ...809...14B,2017ApJS..232...23H,2017ApJ...835..277H,
2017ApJ...851...24B,2020FrASS...7...47C,2021ApJ...922....2D}.
That mission ended in 2018. The Transiting Exoplanet Survey Satellite 
\citep[{\sl TESS} mission,][]{2015JATIS...1a4003R}, 
on the other hand, is currently in its fourth year of operation and has 
completed its initial 2-year mission in 2019 
and 2020 to observe $\sim 85$\% of the whole sky searching for small planets 
orbiting nearby stars. The mission is planned to observe 69 sectors of the sky, covering it completely. Currently (April 2022), the mission is observing sector 51. {\sl TESS} has allowed the discovery of pulsating stars, and in particular, pulsating WDs and pre-WDs  with magnitude G$\lesssim$16--17. 

The first pulsating WD observed extensively with 
{\sl TESS} was the DBV pulsator EC~0158$-$160 (TIC~257459955), 
which was studied and modeled in detail by \cite{2019A&A...632A..42B}.
Recently, a thorough asteroseismological analysis of the DBV star 
GD~358 ---the archetype of the class--- based on the observations gathered by the {\sl TESS} mission 
combined with data taken from the Earth has been carried out by 
\cite{2022A&A...659A..30C}. The authors detected  26  periodicities from 
$\sim 422$ s to $\sim 1087$ s, along with eight combination frequencies with
periods between $\sim 543$ s and $\sim 295$ s. These data, combined with 
a huge amount of ground-based observations \citep{2019ApJ...871...13B}, 
allowed \cite{2022A&A...659A..30C} 
to find a constant period spacing of $39.25$ s, compatible with 
a stellar mass of $M_{\star}= 0.581\pm0.031M_{\odot}$. Also, they found an 
asteroseismological model for GD~358 with a stellar mass $M_{\star}= 0.584^{+0.025}_{-0.019}M_{\odot}$, compatible with the stellar mass 
derived from the period spacing, and in agreement with the spectroscopic mass ($M_{\star}= 0.560\pm 0.028M_{\odot}$).  The seismological distance of GD~358, $d_{\rm seis}= 42.85\pm0.73$ pc, is in good agreement with the precise astrometric distance measured by {\it Gaia} (EDR3),  $d_{\rm Gaia}= 43.02\pm0.04$ 
pc.

In this work, we present new {\sl TESS} observations of the already known DBV stars PG~1351+489, EC~20058$-$5234, and EC~04207$-$4748. In addition, we report for the first
time the variability of the DB WDs, WDJ~152738.4$-$450207.4, and WD~1708$-$871. 
With the discovery of these two new DBVs, the number of known stars of 
this class increases to 49. We perform a detailed asteroseismological analysis of 
these stars on the basis of the 
fully evolutionary models of DB WDs computed by \cite{2009ApJ...704.1605A}. 
The present study is the fifth in our series of papers devoted to the study 
of pulsating H-deficient WDs observed with {\sl TESS}. The first article is focused
on six already known GW Vir stars \citep{2021A&A...645A.117C}, the second  
one is devoted to the discovery of two new GW Vir stars, specifically DOVs \citep{2021A&A...655A..27U}, the third paper is dedicated to 
the DBV star GD~358 \citep{2022A&A...659A..30C}, and the fourth one 
is focused on the discovery of two additional GW Vir stars \citep{2022MNRAS.513.2285U}.

The paper is  organized as  follows.  In Section \ref{targets} 
we provide a brief  account of the main characteristics of
the studied DB stars. In  Sect. \ref{observations}, we
describe the methods we apply to obtain the pulsation periods of each
target star. A brief summary of the stellar
models of DB WD stars employed for the asteroseismological analysis
of these stars is  provided  in Sect.  \ref{models}.
Sect. \ref{asteroseismology} is devoted to the asteroseismological modelling 
of the target stars, including the search for a
constant period spacing in the sets of periods of each star by
applying statistical tests, the assessing of the stellar mass of each object
through the use of the period spacing when possible, and by performing
period-to-period fits with the aim of finding an asteroseismological
model for each DBV star.  Finally, in
Sect. \ref{conclusions}, we summarize  our  main  results  and  make
some  concluding remarks.

\begin{figure}
\includegraphics[clip,width=1.0\columnwidth]{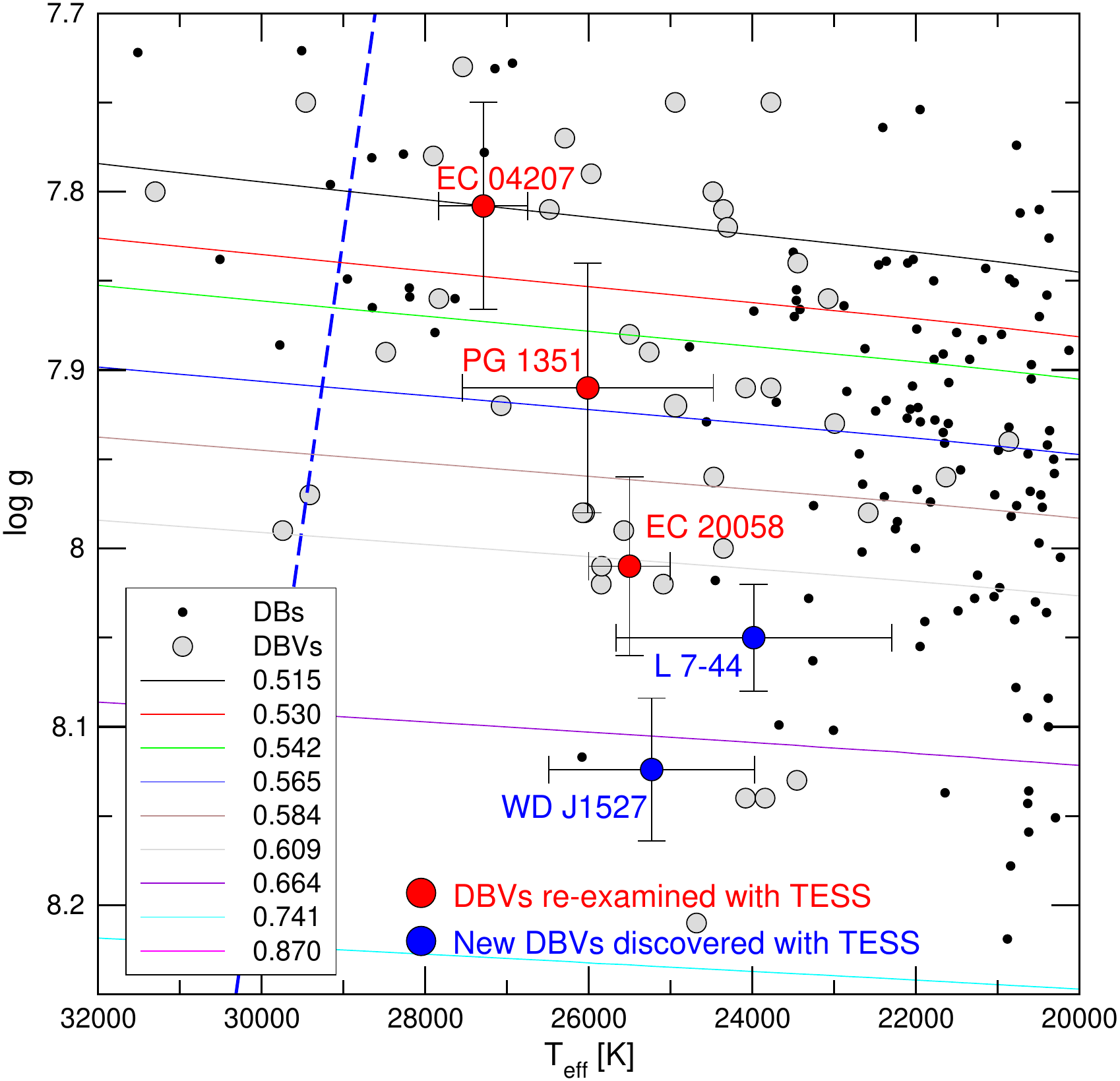}
\caption{Location of DB WDs in the $T_{\rm eff}-\log g$ diagram 
\citep{2019MNRAS.486.2169K}, marked with 
small black circles. Also depicted is the location of the published  DBV  stars
(gray circles), according  to  the  compilation  by  \cite{2019A&ARv..27....7C}
and including also the newly discovered DBV stars by 
\cite{2021ApJ...922....2D} and \cite{2022ApJ...927..158V}. 
The target stars of the present paper are highlighted with 
red circles corresponding to three already known DBVs, 
and with blue circles 
associated to the two new DBVs observed 
with {\sl TESS} and reported here for the first time. 
The DB WD evolutionary tracks of \cite{2009ApJ...704.1605A} 
are displayed with different colors according 
to the stellar-mass values (in solar units).
The blue-dashed line represents the theoretical dipole ($\ell= 1$) blue edge of the DBV instability strip, according to \cite{2009JPhCS.172a2075C}.}
\label{fig:1} 
\end{figure}

\section{The target stars}
\label{targets}

\begin{table*}
\renewcommand{\arraystretch}{1.5}
\centering
  \caption{The list of six DBV (V777 Her)  stars studied in this work. Columns 
  1, 2, 3, 4, 5, and 6 correspond to
  the {\sl TESS} input catalog number, name of the object, effective temperature, surface gravity,  
  parallax, and distance, respectively. For details, see the text.}
  \begin{tabular}{cccccc}
\hline
TIC        & Name             &  $T_{\rm eff}$ & $\log g$  & $\pi$   &  $d$  \\
           &                  &    [K]         & [cgs]     & [mas]   &   [pc]\\
\hline
471015205 & PG~1351+489       & $26\,010\pm1536^{(a)}$ & $7.91\pm0.07^{(a)}$       & $5.69_{-0.05}^{+0.05}$ & $175.73^{+1.53}_{-1.58}$  \\
101622737 & EC~20058$-$5234   & $25\,500\pm500^{(b)}$ & $8.01\pm0.05^{(b)}$        & $8.48_{-0.05}^{+0.05}$ & $117.95^{+0.68}_{-0.75}$  \\
153708460 & EC~04207$-$4748   & $27\,288\pm545^{(c)}$  & $7.808\pm0.058^{(c)}$     & $10.93_{-0.03}^{+0.03}$ & $91.48 ^{+0.22}_{-0.23}$  \\
150808542 & WDJ~152738.4$-$450207.4& $25\,228\pm630^{(d)}$ & $8.124\pm0.010^{(d)}$ & $10.57_{-0.04}^{+0.04}$ & $94.62 ^{+0.34}_{-0.37}$  \\
451533898 & L~7$-$44=WD 1708-891& $23\,980\pm1686^{(e)}$ & $8.05\pm0.03^{(e)}$     & $14.47_{-0.03}^{+0.02}$ & $69.09 ^{+0.14}_{-0.12}$  \\ 
\hline
\end{tabular}
\label{basic-parameters-targets}

{\footnotesize  References: (a) \cite{2011ApJ...737...28B}; (b) \cite{2014A&A...568A.118K}; (c) \cite{2007A&A...470.1079V}; (d) This work; (e) \cite{2018ApJ...857...56R}.}
\end{table*}

In this study, we  announce the discovery of two new DBVs, 
WD~J152738.4$-$450207.4, and WD~1708$-$871 (L~7$-$44), and also report new {\sl TESS} observations of the already known DBV stars PG~1351+489, EC~20058$-$5234,  and  EC~04207$-$4748. The location of the 
five stars in the $\log T_{\rm eff}$ versus 
$\log g$ diagram is displayed in Fig. \ref{fig:1}. 
These stars were classified as white dwarf candidates with a  
a probability of being a white dwarf $P_{\rm WD} \geq 0.99$ by \citet{GentileFusillo19} based on their colors and \textit{Gaia} DR2 parallaxes.
The Gaia EDR3 parallaxes and corresponding distances are given in Table \ref{basic-parameters-targets} from \citet{Bailer-Jones2021}.
We also describe the basic characteristics of these stars below and summarize their stellar properties in Table \ref{basic-parameters-targets}.

PG~1351+489 (hereafter PG~1351; aka TIC~471015205, V*EM UMa) 
is an   known DBV star
discovered by  \cite{1987ApJ...316..305W}. Since the beginning,
it was realized that it could be a candidate for the 
first measurement of a rate of period change in a DBV star, 
because its power spectrum is dominated by a single high-amplitude
pulsation mode with a period at $\sim 489$ s. This star was
reanalyzed by \cite{2011MNRAS.415.1220R}, providing more precise
periods and also an additional low-amplitude mode. More importantly, 
it was possible to obtain, for the first time for a DBV star, an estimate of 
the rate of period change for the period at $\sim 489$ s of 
$\dot{\Pi}= (2.0\pm 0.9) \times 10^{-13}$ s/s \citep{2011MNRAS.415.1220R}. 
Regarding the effective temperature and gravity of PG 1351,
there are several spectroscopic 
determinations  for this star. The first detailed analysis
\cite{1999ApJ...516..887B} provided $T_{\rm eff}= 22\,600 \pm 700$ K, 
$\log g= 7.90 \pm 0.03$ using pure He atmospheres, and 
$T_{\rm eff}= 26\,100 \pm 700$ K, $\log g= 7.89 \pm 0.03$
employing atmospheres with impurities of H. 
The analysis of \cite{2011ApJ...737...28B}
indicates that $T_{\rm eff}= 26\,010 \pm 1536$ K, $\log g= 7.91 \pm 0.07$
(see Fig. \ref{fig:1}), obtained by assuming H contamination 
($\log {\rm H}/{\rm He}= -4.37 \pm 0.82$). 
\citet{2015A&A...583A..86K} estimate $T_\mathrm{eff}=28434\pm 124$, $\log g=7.89\pm 0.02$, $\log {\rm H}/{\rm He}<-3.1$, and $\log {\rm Ca}/{\rm He}<-4.0$.
PG~1351 has been asteroseismologically modeled by 
\cite{2014JCAP...08..054C}, and were able to place constraints 
on the neutrino magnetic dipole moment. The star has also been used 
to infer bounds to the axion mass by \cite{2016JCAP...08..062B}.

EC~20058$-$5234, also known as QU Tel (henceforth EC~20058; aka TIC~101622737) 
is an already known DBV star whose variability was reported by 
\cite{1995MNRAS.277..913K}, who found that the pulsation spectrum appeared to 
be stable.  This star was long scrutinized by \cite{2003ASIB..105..231S,
2005ASPC..334..495S}, \cite{2007ASPC..372..629S,2008MNRAS.387..137S}, and \cite{2017ASPC..509..315S}. In particular, \cite{2008MNRAS.387..137S} 
increased the number of detected frequencies and performed an 
asteroseismic global-model analysis. Interestingly, 
these authors found that EC~20058 had very stable pulsation periods,
which meant that, in principle, it could be possible to measure the rate of period 
change to place constraints on the plasma neutrino emission rate
\citep{2004ApJ...602L.109W,2010AIPC.1273..536D}. However, \cite{2013ApJ...765....5D}
found that the pulsation frequencies of this star undergoes secular changes that are inconsistent with simple neutrino plus photon-cooling models. A detailed 
asteroseismological analysis of EC~20058 was carried out by 
\cite{2011MNRAS.414..404B} to place constraints on helium
diffusion in WD envelopes. The effective temperature and gravity of 
EC~20058 are $T_{\rm eff}= 25\,500\pm500$ K and $\log g= 8.01\pm0.05$
\citep{2014A&A...568A.118K} with H and C impurities. 

EC~04207$-$4748 (hereafter EC~04207; aka TIC~153708460) is an already known DBV WD 
whose variability was discovered by
\cite{2009MNRAS.397..453K}, who found a pulsation spectrum dominated
by a period of  $\sim 447$ s. The star was re-observed by
\cite{2013MNRAS.431..520C} who found at least four independent eigenmodes 
in the star, with the dominant  mode  having  a  period  of  $\sim 447$  s.
The  light-curve  exhibits  distinct  non-sinusoidal shapes, which results 
in significant harmonics of the dominant frequency appearing in the Fourier 
transforms. The effective temperature and gravity of 
EC~04207 are $T_{\rm eff}= 27\,288\pm545$ K and $\log g= 7.808\pm0.058$,
respectively \citep{2007A&A...470.1079V}. 
However, \citep{2014A&A...568A.118K} quote $T_{\rm eff}= 25\,970$ K and $\log g= 7.79$,
with log H/He of $-5.0$. 

\begin{figure} 
\includegraphics[clip,width=1.0\columnwidth]{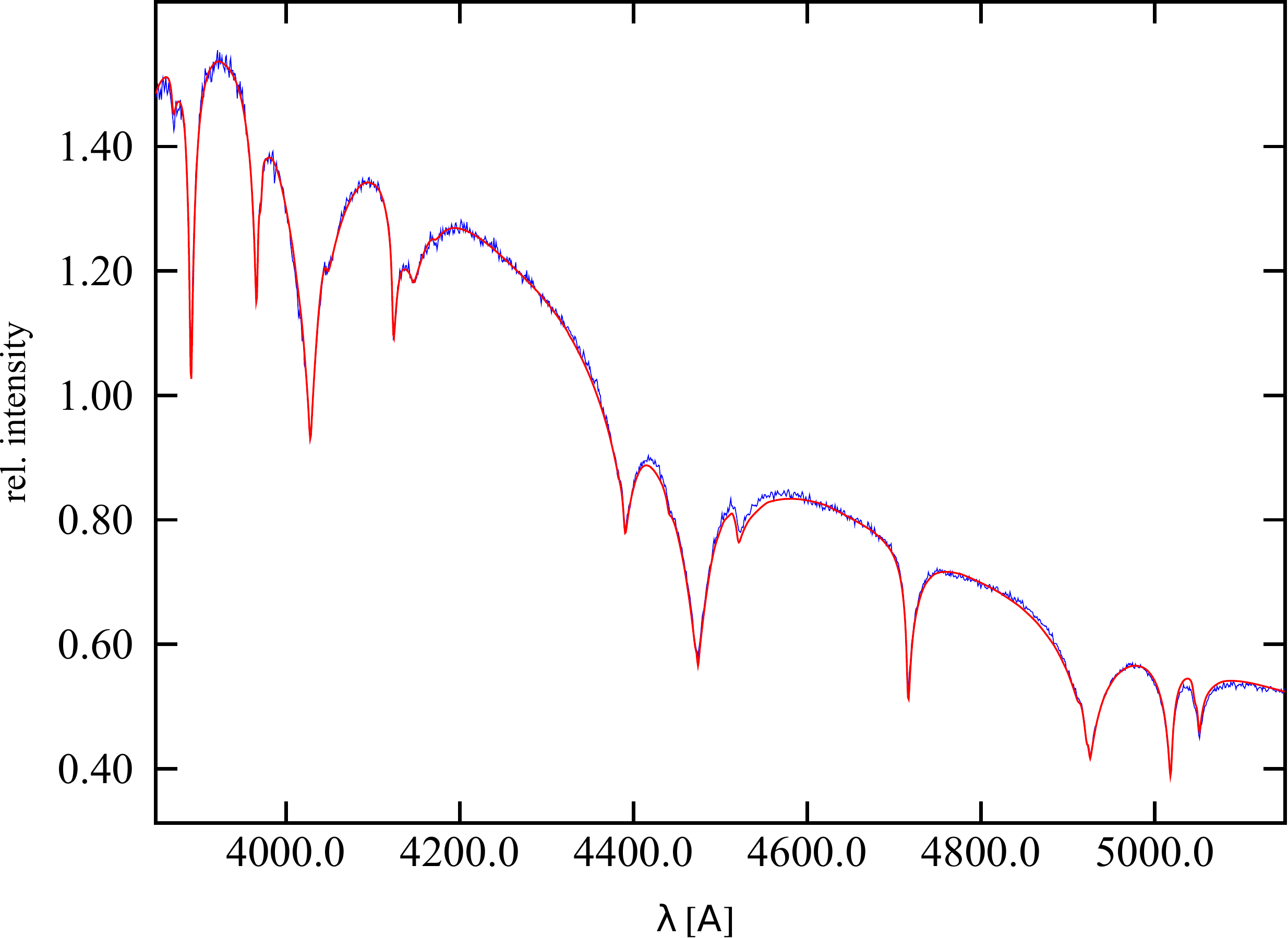}
\caption{Optical spectra from SOAR (blue line) of the new DBV WD\,J1527. Overplotted is the best-fit model atmosphere solution (red line).} 
\label{fig:atmfit} 
\end{figure}

WD~J152738.4$-$450207.4 (hereafter WD~J1527; aka  TIC~150808542) is a new DB WD star.
We observed WD~J1527 with the Southern Astrophysical
Research (SOAR) Telescope, a 4.1-meter aperture optical and near-infrared telescope \citep{clemens2004}, situated at Cerro Pach\'on, Chile.
The Goodman spectrograph with a setup of 400\,l/mm grating with the blaze wavelength 5500 $\AA$  (M1: 3000-7050 \AA) with a slit of 1 arcsec was used. The signal-to-noise ratio of the final spectrum is $\sim$150 at 4250\,$\AA$.
The atmospheric parameters for WD~J1527 are derived by fitting synthetic spectra to the newly obtained low-resolution spectra as described by \cite{2010MmSAI..81..921K}. 
The reduced spectrum is shown in Fig.\,\ref{fig:atmfit} with blue line and the model fits is shown with red line. WD~J1527 has $T_\mathrm{eff} = 25\,228 \pm 630$\,K and $\log g =  8.124 \pm 0.010$, with minor H contamination.
At that effective temperature, the star is well within the DBV instability strip (Fig. \ref{fig:1}). Pulsations with periods between 229 and 1170 s are 
confirmed by observation in {\sl TESS} sectors 12 and 38.

WD~1708$-$871 (hereafter L~7$-$44; aka TIC~451533898) is a DB 
WD star characterized by $T_{\rm eff}= 23\,980 \pm 1686$ K and  
$\log g= 8.05 \pm 0.03$ \citep{2018ApJ...857...56R} and only an upper limit for 
possible H contamination. Observations in Sectors 12, 13, and 39 of {\sl TESS} show 
luminosity variations with periods in the range $466-937$ s.
The location of this star in the $T_{\rm eff}-\log g$ diagram
is displayed in Fig. \ref{fig:1}. 

\begin{table*}
\setlength{\tabcolsep}{3pt}
\renewcommand{\arraystretch}{1.2}
\centering
  \caption{The list of the five DBV (V777 Her) stars reported in this paper from {\sl TESS} observations, including the name of the targets, the {\sl TESS} magnitudes, the observed sectors, and the date and length of the runs (columns 1, 2, 3, 4, 5 and 6). After Fourier
    transform, three different set of parameters (resolution, average
    noise level of amplitude spectra, and detection threshold, which
    is defined as $0.1\%$ false-alarm probability (FAP), are presented
    in columns 7, 8 and 9, respectively.}
  \begin{tabular}{ccccccccc}
\hline
TIC & Name &  $T_{\rm mag}$ & Obs.   & Start Time  &  Run Length & Resolution & Average Noise & $0.1\%$\,FAP \\
    &      &                & Sector & (BJD-2\,457\,000) & [d]     & $\mu$Hz    &  Level [ppt]  &    [ppt]  \\
\hline
471015205 & PG~1351+489     & 17 & 16, 22, 23, 49, 50 & 1738.6492 & 128.89 & 0.1 &  1.32  & 6.3 \\
101622737 & EC~20058$-$5234 & 16 & 13, 27  & 1653.9251 & 52.74  & 0.22 &  0.66  & 2.54 \\
153708460 & EC~04207$-$4748 & 15 & 4, 5, 30, 31, 32  & 1410.9047 & 137.48  & 0.08 &  0.42  & 1.79\\
150808542 & WDJ152738.4-450207.4 & 16  & 12 ,38  & 1627.6494 & 25.24  & 0.45 &  4.90  & 20.7 \\ 
451533898 & L~7$-$44        & 15 & 12, 13, 39   & 1624.9589 & 84.40  & 0.14 &  0.28  & 1.39 \\
\hline
\label{DBVlist}
\end{tabular}
\end{table*}

\section{{\sl TESS} data: observations, data reduction and analysis}  
\label{observations}  

Our analysis is based on 2-min short-cadence (SC) {\sl TESS} data for all targets, or 20-sec ultra-short-cadence when available. 
We used the pre-search data conditioned simple aperture photometry (PDCSAP) data \citep{jenkins2016} downloaded from the  "Barbara A. Mikulski Archive for Space Telescopes" (MAST)\footnote{\url{http://archive.stsci.edu}}. 
We extracted times in Barycentric corrected Julian days ("BJD - 245700"), and fluxes (“PDCSAP FLUX”) by using the Python package $\tt{lightkurve}$ \citep{lightkurve2020}. Afterwards, we removed outliers by applying a running $5\sigma$ clipping mask. 
The fluxes were normalized and transformed to amplitudes in parts-per-thousand (ppt) units. We also detrend the light curves applying a Savitzky–Golay filter with a three-day window length to remove any additional low-frequency systematic.
The target pixel files (TPFs) are examined to estimate the amount of contamination from nearby unresolved targets. We have checked the contamination by looking at the $\tt{CROWDSAP}$ parameter, which is the
ratio between target flux and total flux in the aperture.
If the $\tt{CROWDSAP}$ is significant ($<$ 0.5), then we used {\it Gaia} EDR3 parallaxes to look at the contaminant targets in the field of view. Severe contamination affects the average noise level in Fourier space and thus affects also the false-alarm probability (FAP) thresholds as described in \citet{2020A&A...638A..82B}. 
After creating the final light curves, we calculated their Fourier transforms (FTs) to search for periodic signals.
The detection thresholds were defined as $0.1\%$ FAP significance level so that there is a 0.1\% chance
that the peaks are caused by the noise fluctuations. $0.1\%$ FAP thresholds were calculated by randomization of the data 1000 times, as described in \citet{kepler1993}
\citep[see][]{2021AcA....71..113B}.
We identified frequency, amplitude and phase of each pulsation mode using a nonlinear least square (NLLS) method.
Also, we computed sliding FTs to inspect the temporal evolution of the detected frequencies. 
 To calculate the sliding FTs, we use a 6-day sliding window with a 2-day step size. We then compute the Fourier transform for each subset and trail them in time.
In Table \ref{DBVlist} we list the six DBV stars studied in this paper from {\sl TESS} observations, including observed sectors, {\sl TESS} magnitude
along with name of the targets, the date and length of the runs. 

\subsection{PG~1351+489}
\label{obs-PG1351}

\begin{figure} 
\includegraphics[width=0.47\textwidth]{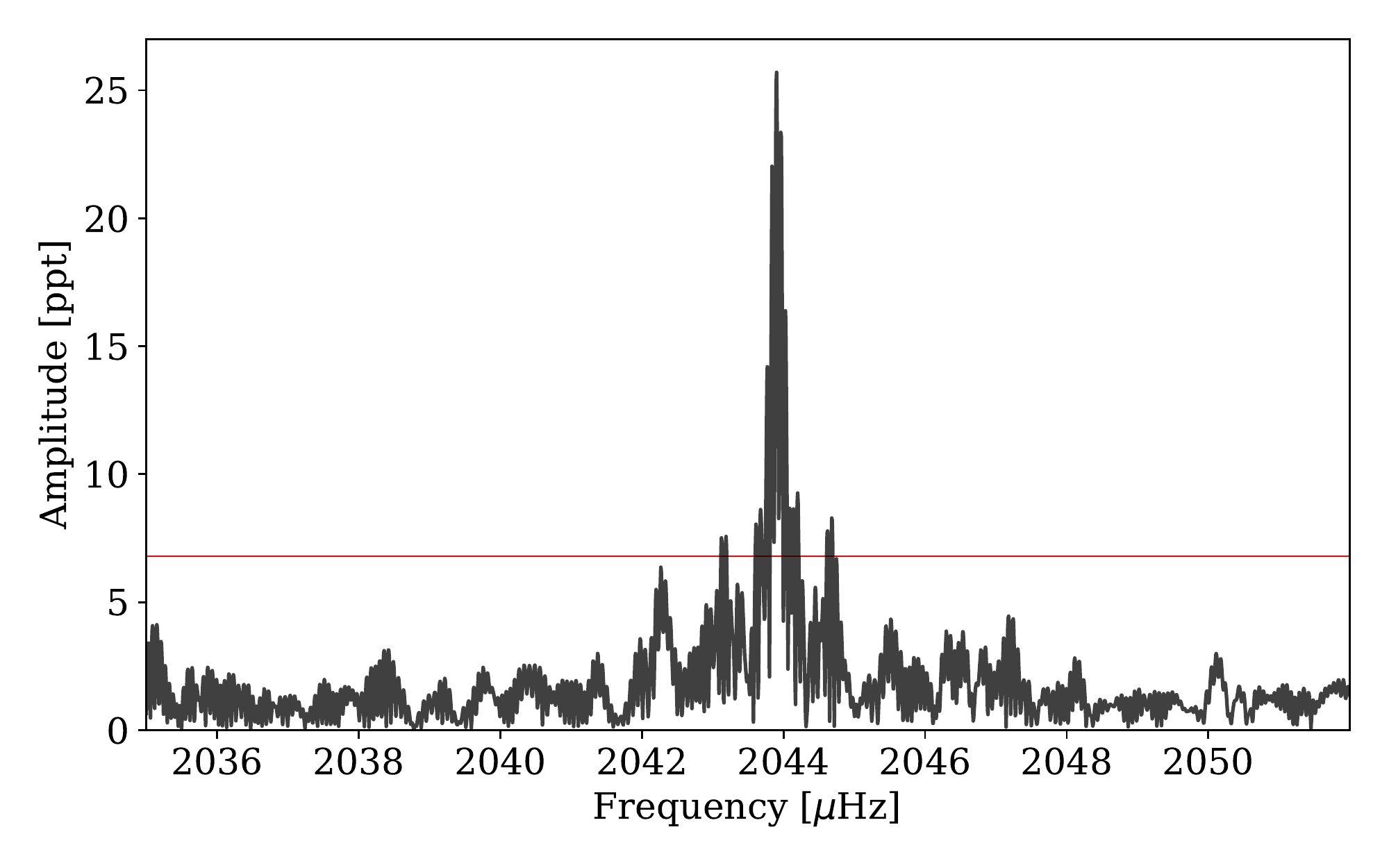}
\includegraphics[width=0.51\textwidth]{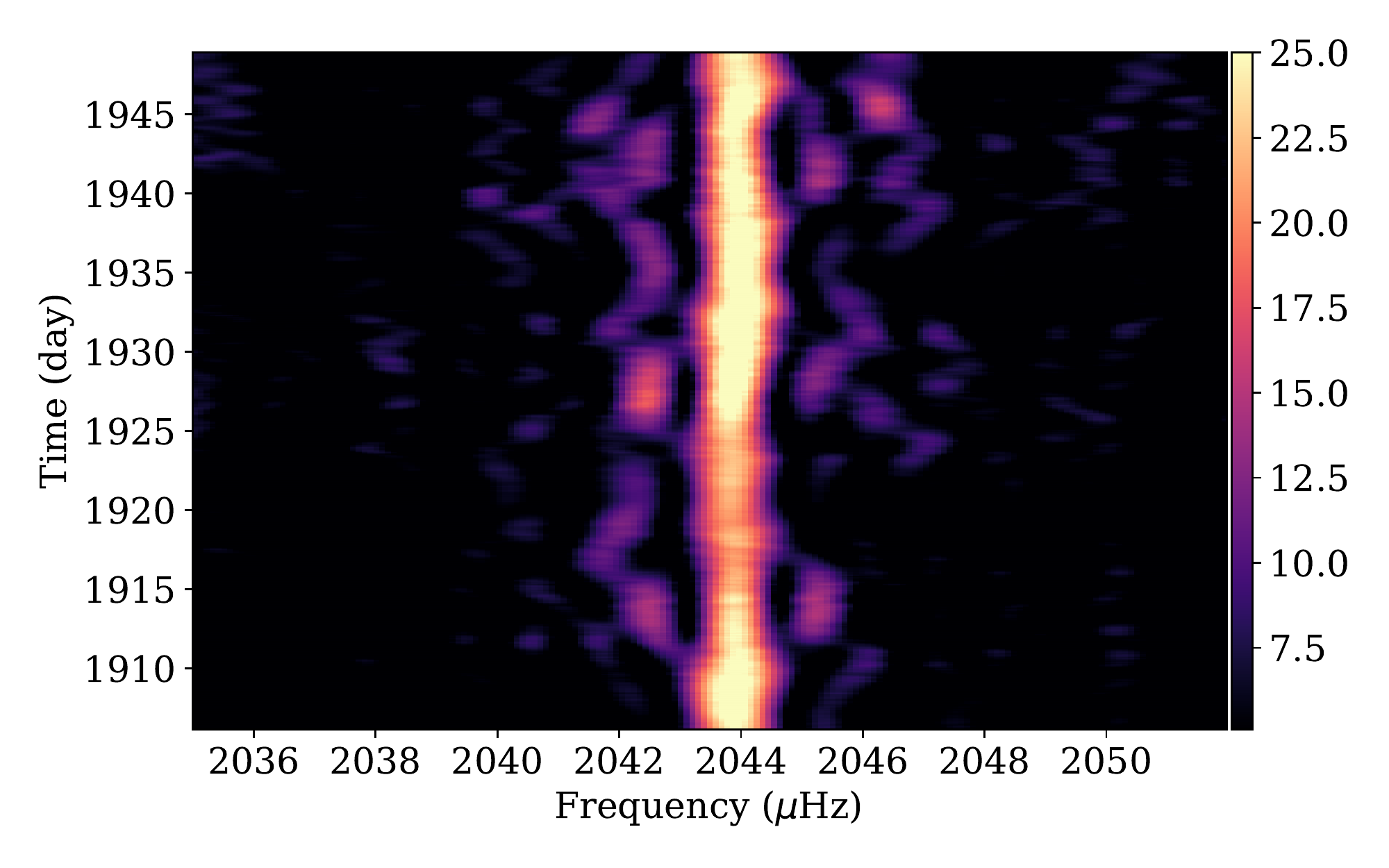}
\caption{{\sc Top:} Main pulsational mode in PG~1351 presented in Table \ref{table:PG1351}. Fourier transform of data taken in sector 16, 22 and 23.
The red dashed line denotes the 0.1\% FAP.
 {\sc Bottom:} Sliding Fourier transform sector 22 and 23 of PG~1351+489 concentrating the main pulsational mode. 
 The color-scale illustrates amplitude in ppt units.}
\label{fig:FTPG1351} 
\end{figure}

PG~1351+489 ($T_{\rm mag}$ = 16.7) was observed in short cadence (SC, 120s exposures) mode during sector 16 between September 11 and October 07, 2019, for 23.04 days.
PG~1351 was also observed during Sector 22 and 23 (2020-Feb-18 to 2020-Apr-16) in SC, providing another 54.74 days of observations. 
It was also observed in fast cadence (20~s exposures) in Sectors 49 and 50 (2022-Feb-26 to 2022-Apr-22), 51.11 days of observations.
We examined the field of view (FOV) of PG~1351 and saw that there is a bright star that is 1.44 arcsec away from PG~1351 in sector 16. 
The problem is that the nearby star is about 2.4 magnitudes brighter than PG~1351, dominating the pixels. The $\tt{CROWDSAP}$ parameter is around 0.3, meaning that only 30 percent of the light is coming from PG~1351, 
affecting the average noise level of the FT, as shown in Table~2. 
Including all sectors, PG~1351 was observed for 128.89 days in total.
From these observations of PG~1351, we extracted only two periodicities from the light curve. The main pulsation frequency at 2043.9 $\mu$Hz is present in all sectors with S/N $\geq$ 19. 
In Fig. \ref{fig:FTPG1351}, we present the temporal FT (upper panel) and sFT (bottom panel) for 2043.9 $\mu$Hz. 
The other peak at 4087.8 $\mu$Hz, which has a lower S/N of about 5.75, is a combination frequency (often called the first harmonic, 2\,$\times$\,2043.9 $\mu$Hz). 
These two identified frequencies are listed in Table \ref{table:PG1351}.  

\begin{table}
\centering
\caption{Identified frequencies and combination frequencies, periods, amplitudes 
(and their uncertainties)  and the signal-to-noise ratio in the  data of PG~1351.}
\begin{tabular}{lcccr}
\hline
\noalign{\smallskip}
Peak & $\nu$    &  $\Pi$  &  $A$   &  S/N \\
 & ($\mu$Hz)      &  (s)   & (ppt)   &   \\
\noalign{\smallskip}
\hline
\noalign{\smallskip}
f$_{\rm 1}$  & 2043.901 (20) & 489.260(22) & 25.78(1.05)  & 19.8 \\
\hline 
2f$_{\rm 1}$ & 4087.802(87) & 244.630(22) & 6.57(1.05) & 5.1 \\
\noalign{\smallskip}
\hline
\end{tabular}
\label{table:PG1351}
\end{table}

\subsection{EC~20058$-$5234}
\label{obs-EC20058}

\begin{figure}
\includegraphics[width=0.47\textwidth]{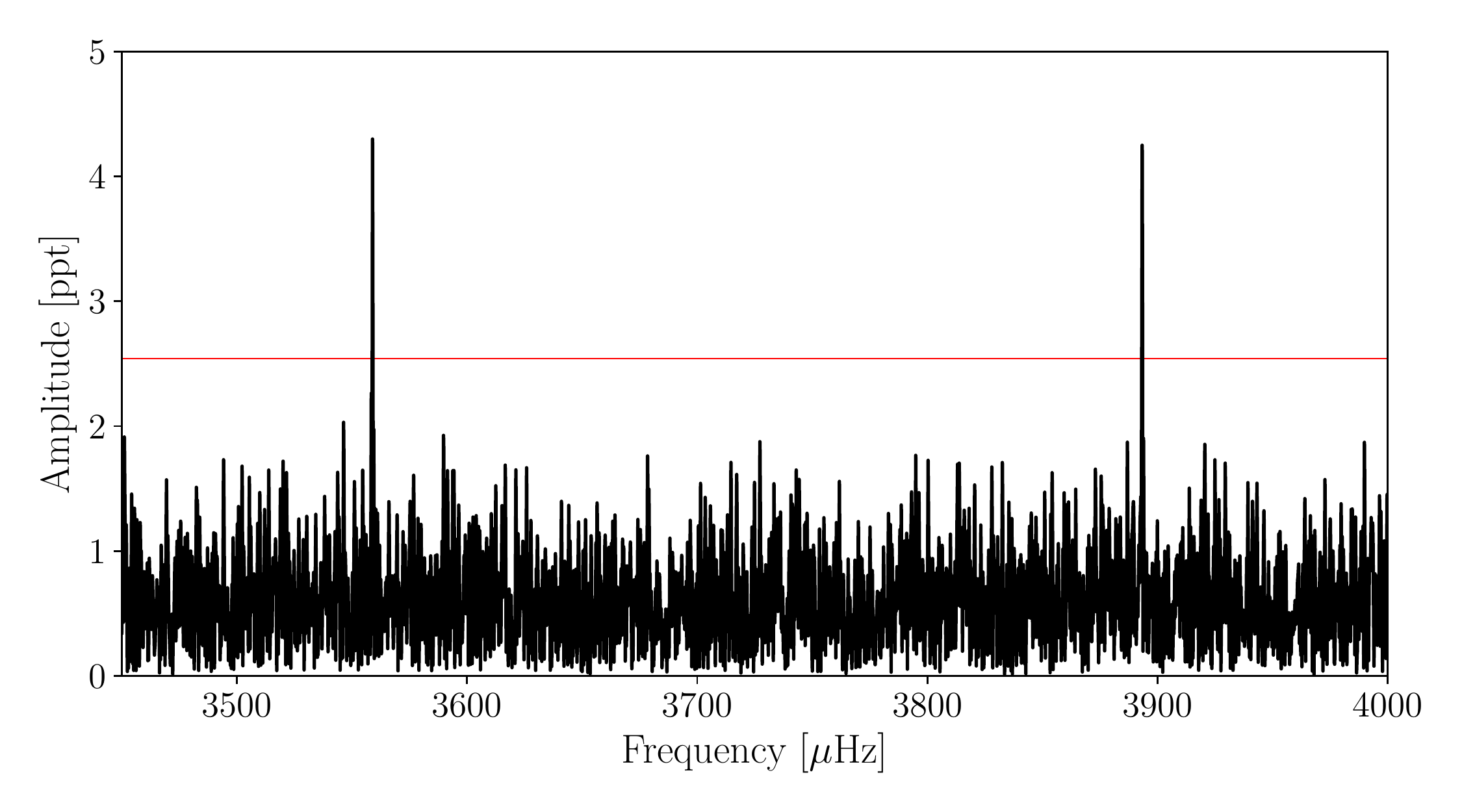}
\caption{Main pulsational modes (f$_{\rm 1}$ and f$_{\rm 2}$) in EC~20058$-$5234 presented in
  Table \ref{table:EC20058} based on sector 13. The red dashed line denotes the 0.1\% FAP.}
\label{fig:EC20058}     
\end{figure}

EC~20058$-$5234 was observed in sector 13 (2019-Jun-19 to 2019-Jul-18) in camera 2. It was also observed in sector 27 (2020-Jul-04 to 2020-Jul-30) in camera 2. Sector 13 contains 19582 measurements that span 28.4 days, while sector 27 includes 16745 measurements that span 24.3 days.
There are two stars within 5 arcsec with brightnesses of 15.7 mag and 16.4~mag close to  EC~20058. These two stars have only a modest effect on the noise level of the FT. The FT average noise level of sector 27 is 0.61 ppt, while it is 0.66 ppt in sector 13. 
We first examined the Fourier transform from each sector. Sector 13 shows two frequencies at 3558 and 3893~$\mu$Hz, which are above the $0.1\%$\,FAP 
of 2.54 ppt. 
These two peaks are also present in sector 27 above the $0.1\%$\,FAP of 2.3 ppt.
We also examined the FT of both sectors together, as shown in Fig. \ref{fig:EC20058}. These two detected frequencies are given in Table \ref{table:EC20058}. 
This target is known to have significant phase and amplitude variations. The sliding Fourier Transform of two reported frequencies was calculated. However, the signal-to-noise ratio is insufficient to show and assess these substantial variations.

\begin{table}
\caption{Independent frequencies, periods, and  amplitudes (and their uncertainties) and the 
signal-to-noise ratio in the  data of EC~20058.}
\begin{tabular}{lcccr}
\hline
\noalign{\smallskip}
Peak & $\nu$    &  $\Pi$  &  $A$   &  S/N \\
 & ($\mu$Hz)      &  (s)   & (ppt)   &   \\
\noalign{\smallskip}
\hline
\noalign{\smallskip}
f$_{\rm 1}$ & 3558.935(27) &	280.983(20) &	4.31(53) & 6.5 \\
f$_{\rm 2}$ & 3893.289(28) &	256.852(20) &	4.25(53) & 6.4 \\
\noalign{\smallskip}
\hline
\end{tabular}
\label{table:EC20058}
\end{table}

\subsection{EC~04207$-$4748}
\label{obs-EC04207}

\begin{figure}
    \includegraphics[width=0.5\textwidth]{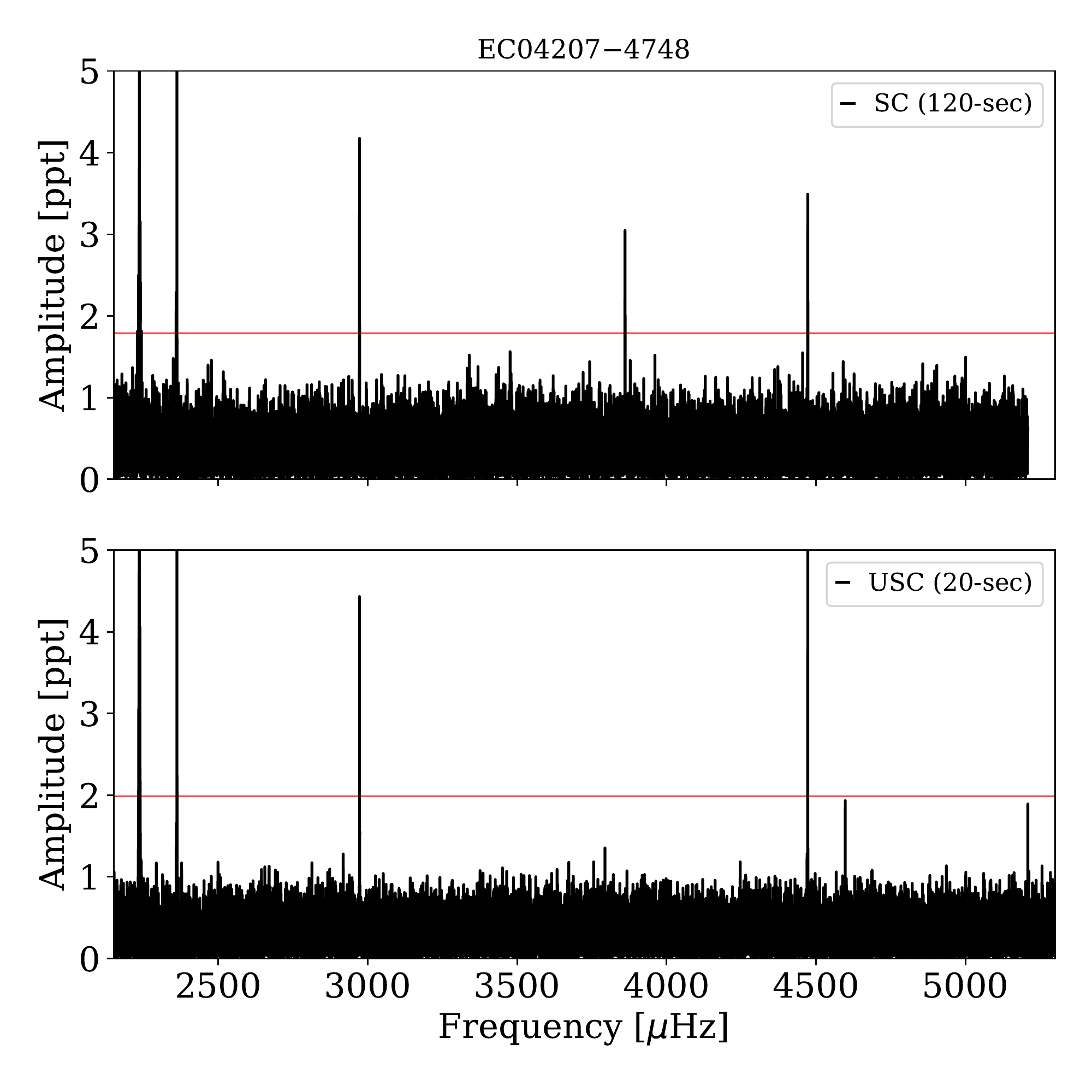} 
 \caption{{\sc Top:} Fourier transform of data taken in sector 4 and 5 of EC~04207. 
 {\sc Bottom:}  Fourier transform of data taken in sector 30, 31 and 32 of EC~04207. The horizontal red line indicates the 0.1\% FAP level.}
    \label{fig:FTEC04207}
\end{figure}

\begin{figure}
    \includegraphics[width=0.5\textwidth]{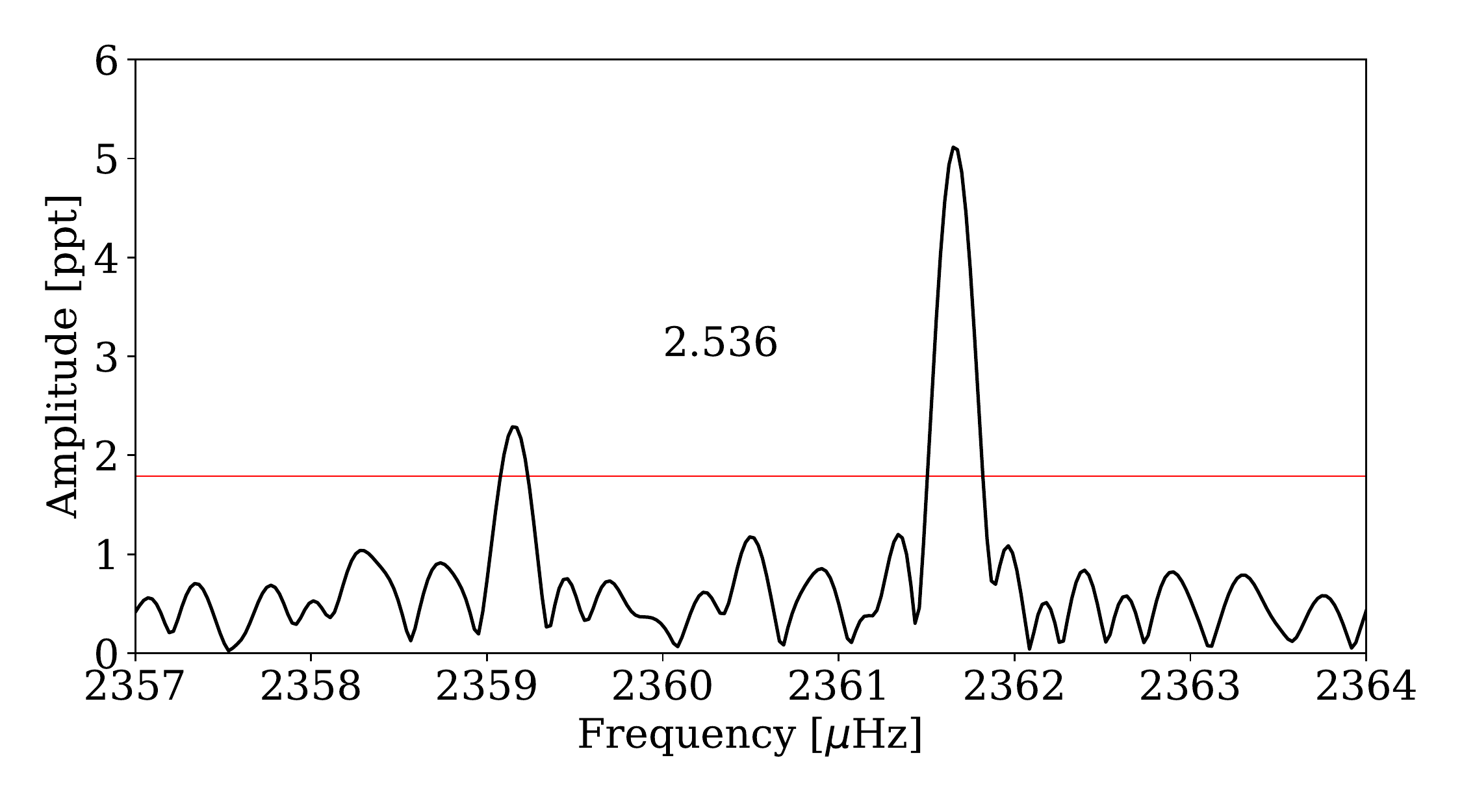} 
 \caption{A close up of the amplitude spectrum of
EC~04207 focusing on the potential rotational multiplet with 
components at 2369 and 2361 $\mu$Hz.}
    \label{fig:FTEC04207_doublet}
\end{figure}

EC~04207 was first observed in SC during cycle 1 in sector 4 and 5 (2018-Oct-18 to 2018-Dec-11). Since in sector 4 the scatter is much higher between 1422.61~d and 1423.51~d, and between 1436.60~d and 1436.82~d, we removed these regions. Then, we added sector 5 and calculated the Fourier transform of 53 days of data
detecting five independent frequencies above $0.1\%$ FAP confidence level, which corresponds to 1.79\,ppt (Fig. \ref{fig:FTEC04207} top). 
The FT median noise level is 0.39\,ppt. The S/N of the detected frequencies ranges from 5 to 69.
EC~04207 was also observed during three consecutive sectors (30, 31 and 32) of the extended mission in ultra-short-cadence (USC) mode.
These observations started on 4 July 2020 and ended on 26 August 2020. 
From this 84-day data set, we calculated the FT of the USC data up to the Nyquist frequency of 25\,000 $\mu$Hz. In this way, we identified 3 additional combination
frequencies that are not detectable in the 120 s cadence data (Fig. \ref{fig:FTEC04207} bottom).  
These combination frequencies are given in Table \ref{table:EC04207}. 
In Fig. \ref{fig:FTEC04207}, we compare the two data sets showing the FT from 120\,s-resolution (upper panel) and 20\,s-resolution data (lower panel). 
From the USC data, we extracted three independent periodicities at 2236 $\mu$Hz (f$_{\rm 1}$), 2361 $\mu$Hz (f$_{\rm 3}$) and 2972 $\mu$Hz (f$_{\rm 4}$). 
2359 $\mu$Hz (f$_{\rm 2}$) and 3861  $\mu$Hz (f$_{\rm 5}$) are not detected in USC data. 
The frequencies at 2359.142 and 2361.678 $\mu$Hz can be considered as two components of a rotationally split dipole mode.
These two peaks are shown in Fig. \ref{fig:FTEC04207_doublet} with frequency splitting of 2.536 $\mu$Hz. 
If we assume that the central azimuthal component ($m = 0$) is missing, then the rotation period of EC~04207 would be 2.28 d. On the other hand, if the missing frequency was one of the side components ($m = +1$ or $-1$), then the rotation period of EC~04207 would be 1.14 d.
This solution is consistent with what has been discovered for other types of pulsating WDs, such as GW Vir pulsating stars, which range from 5 hours to a few days \citep{2019A&ARv..27....7C,2021A&A...645A.117C,2022MNRAS.513.2285U} and DAV pulsating stars, which range from 1 hour to a 4.2 days \citep{2015ASPC..493...65K,2017ApJS..232...23H}. 

Combining the results from sector 4, 5, 30, 31 and 32, we detect five independent frequencies which are concentrated between 2236\,$\mu$Hz and 2972\,$\mu$Hz. 
We also detect three combination frequencies beyond 4400 $\mu$Hz. The extracted frequencies are listed in Table \ref{table:EC04207} with their associated errors and S/N.

\begin{table}
\centering
\caption{Identified frequencies (combination frequencies), periods, and amplitudes (and their uncertainties) and the signal-to-noise ratio in the  data of EC~04207. 
The frequencies that are detected only in sector 4 and 5 are unmarked. 
Frequency and amplitude from sector 30, 31 and 32 are marked with $\dagger$.
Frequency and amplitude that are detected in all sectors marked with $\dagger\dagger$.
}
\begin{tabular}{lcccr}
\hline
\noalign{\smallskip}
Peak & $\nu$    &  $\Pi$  &  $A$   &  S/N  \\
 & ($\mu$Hz)      &  (s)   & (ppt)   &   \\
\noalign{\smallskip}
\hline
\noalign{\smallskip}

f$_{\rm 1}^{\dagger\dagger}$ &  2236.165(00)  & 447.194(02) &  22.87(26) &	69.3   \\	
f$_{\rm 2}$                  &  2359.142(18)  & 423.882(34) &  2.14(33)  &	5.3    \\    	
f$_{\rm 3}^{\dagger\dagger}$ &  2361.678(03)  & 423.427(06) &  5.66(26)  &	17.1  \\	 	
f$_{\rm 4}^{\dagger\dagger}$ &  2972.672(04)  & 336.397(05) &  4.44(26)  &	13.4   \\ 	 	
\hline 
2f$_{\rm 1}^{\dagger}$ &  4472.332(03)  & 223.597(02) &  5.18(26)  &	15.7   \\
(f$_{\rm 1}$+f$_{\rm 3})^{\dagger}$ &  4597.837(10)  & 217.493(05) &  1.93(26)  & 5.8    \\ 
(f$_{\rm 1}$+f$_{\rm 4})^{\dagger}$ &  5208.833(10)  & 191.981(04) &  1.89(26)  & 5.7   \\ 
	
\noalign{\smallskip}
\hline
\end{tabular}
\label{table:EC04207}
\end{table}

\subsection{WDJ152738.4-450207.4} 
\label{obs-WDJ1527}

WDJ152738.4-450207.4 is a new pulsating DBV star that was observed in sector 12 (2019-May-21 to 2019-Jun-19, cycle 1) and in sector 38 (2021-Apr-28 to 2021-May-26) with 20-sec cadence. Within 21 arcsec, the field of view (FOV) includes 21 targets. All these targets are fainter than $G_{mag}= 18$, which is beyond the detection limit of {\sl TESS}. The crowded field causes enough light contamination to significantly raise the FT noise level, as reported in Table \ref{DBVlist}. 
From the short-cadence observations, we detect two significant frequencies
near 1341 and 1539 $\mu$Hz (f$_{\rm 1}$ and f$_{\rm 5}$ in Table \ref{table:WDJ1527}) above the 0.1\%FAP level of 20.7 ppt.

\begin{figure}
    \includegraphics[width=0.5\textwidth]{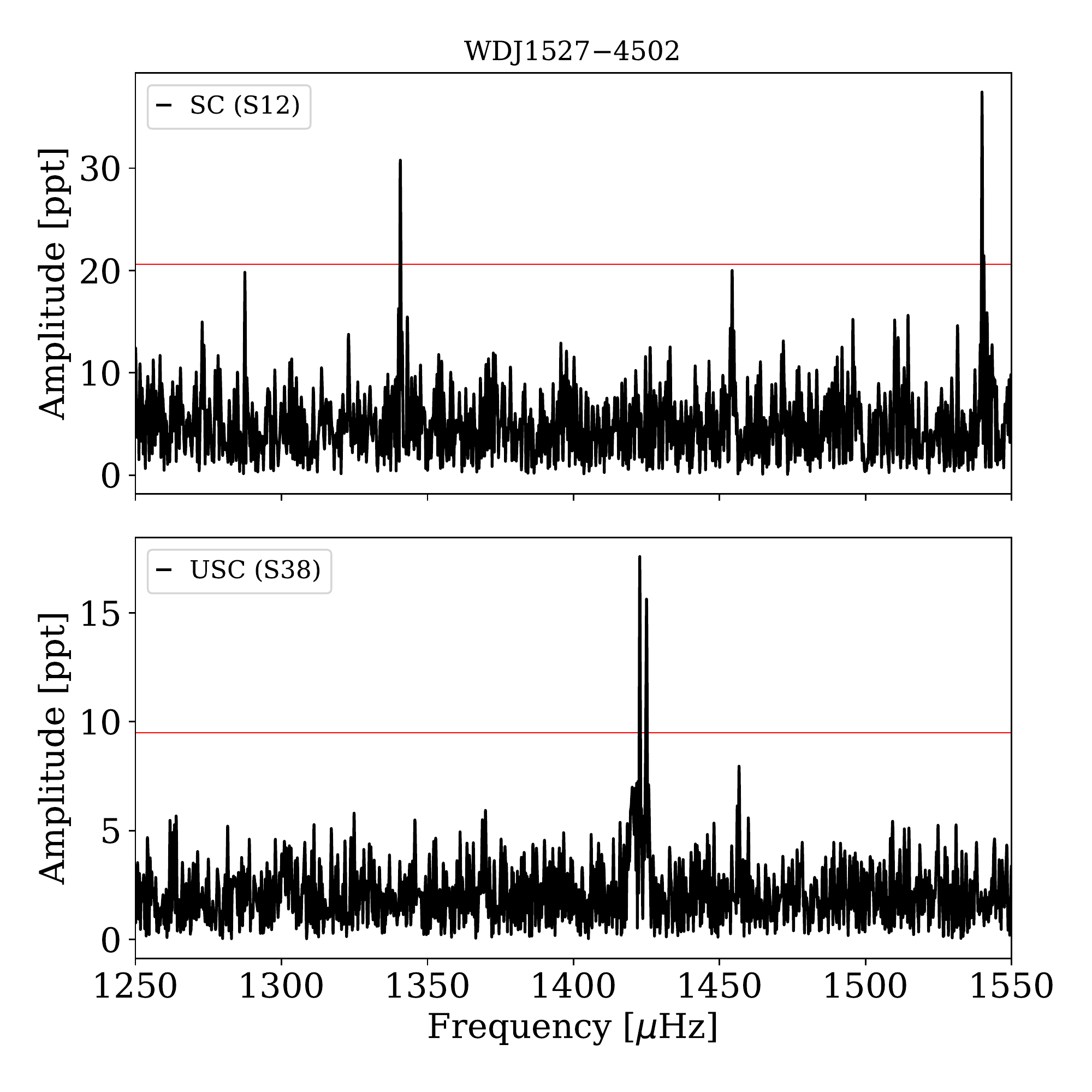} 
 \caption{{\sc Top:} Fourier transform of data taken in sector 12 of WD~J1527. 
 The horizontal red line indicates the 0.1\% FAP level. 
 {\sc Bottom:}  Fourier transform of data taken in sector 38. The horizontal red line indicates the 0.1\% FAP level.}
    \label{fig:FTJ15}
\end{figure}

\begin{figure}
    \includegraphics[width=0.47\textwidth]{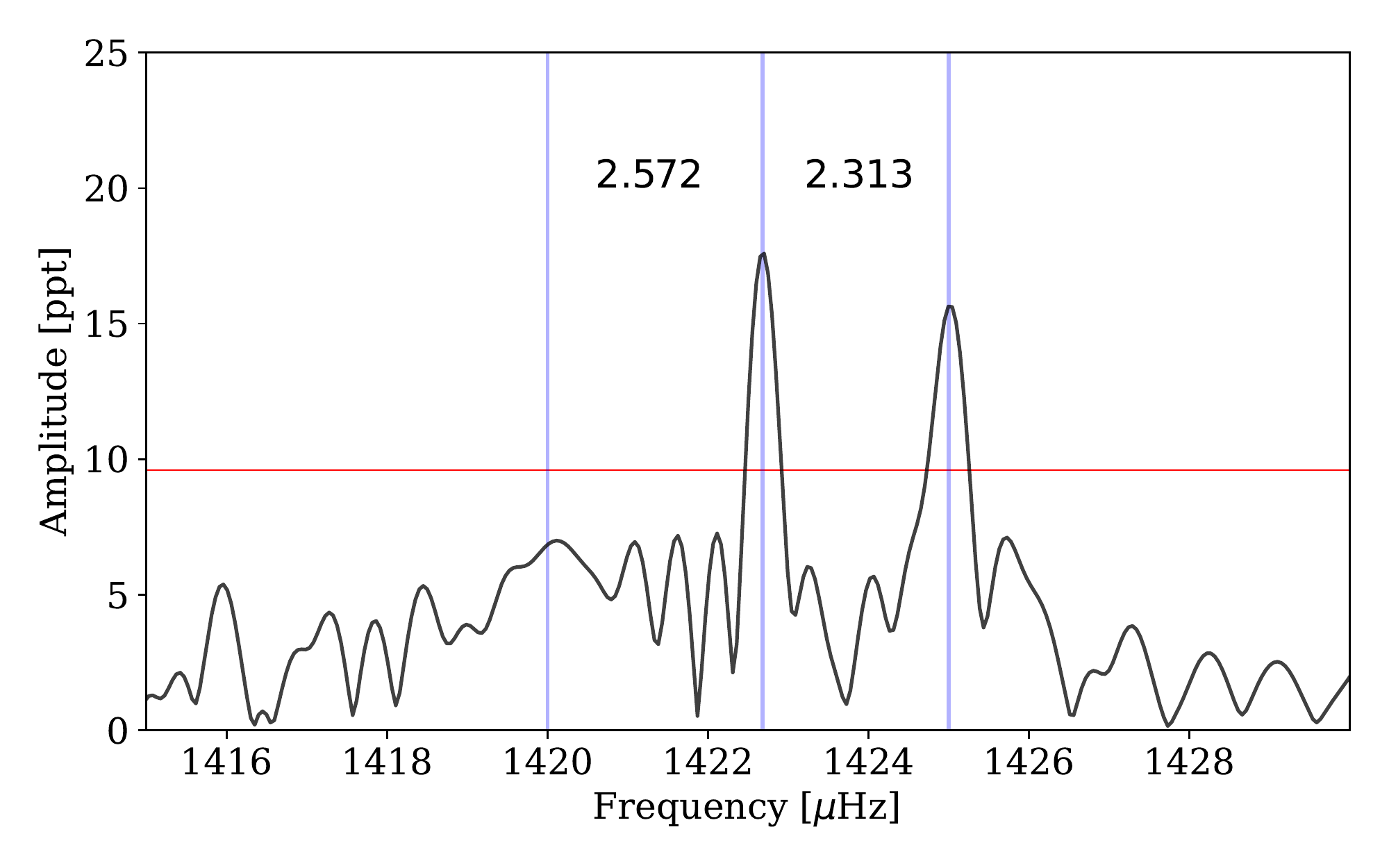} 
    \includegraphics[width=0.51\textwidth]{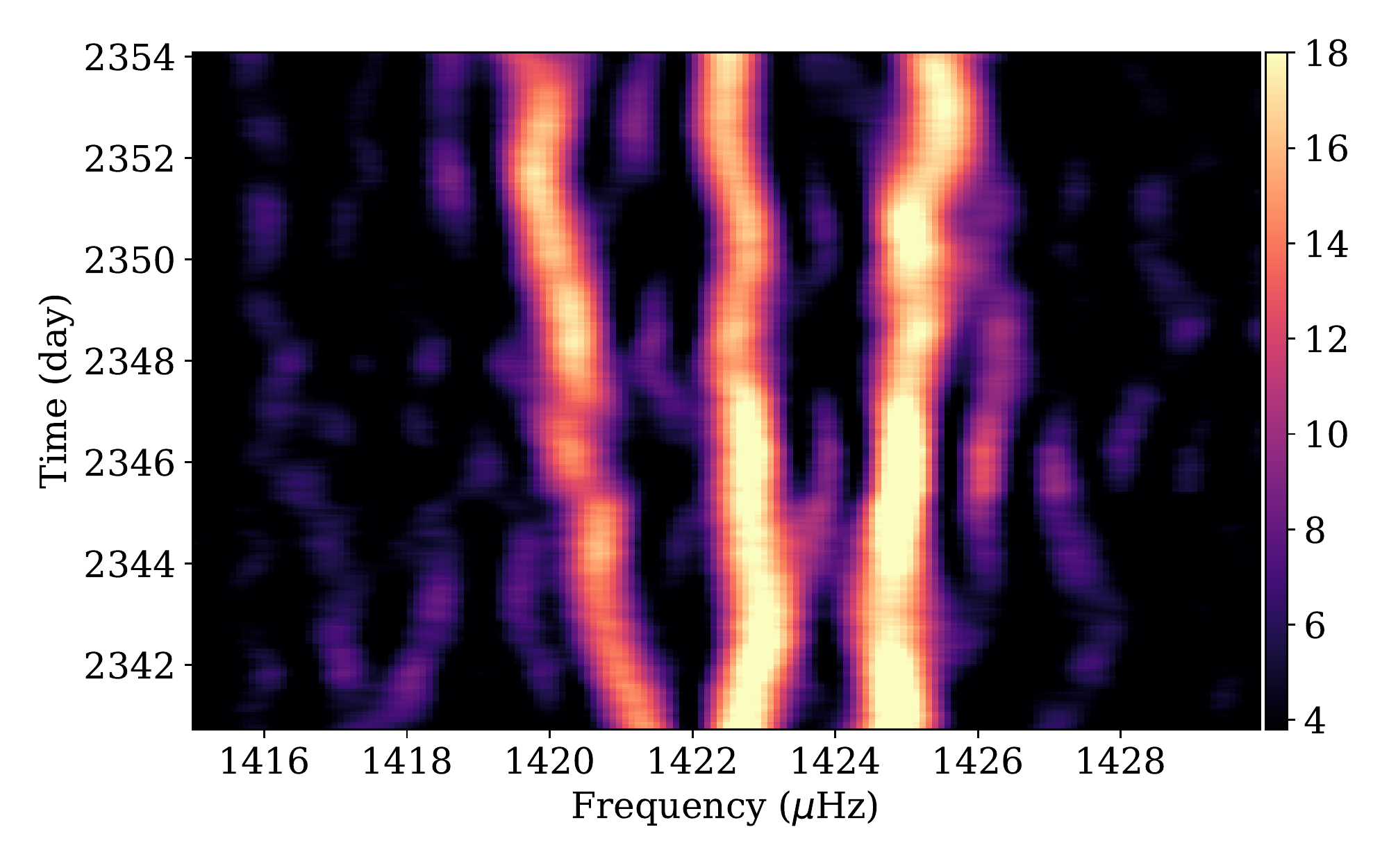} 
 \caption{Detected rotational triplet in WDJ152738.4-450207.4. The top panel shows 
 the Fourier transform of data taken in sector 38. 
 The horizontal red line presents the 0.1\% FAP level. We indicated the azimuthal orders 
 (vertical blue lines) of dipole modes and the rotational splitting. The bottom 
 panel depicts the Sliding Fourier Transform of the same sector covering the same frequency region. The 
 color-scale shows amplitude in ppt units.}
    \label{fig:sFTJ15}
\end{figure}

The data from sector 38 have a time span of 26.33 d (97.5\% duty cycle) including 109476 data points. The frequency resolution is 0.44 $\mu$Hz.  
We detect two independent frequencies near 1422 and 1425 $\mu$Hz. Moreover, we detect a combination frequency at 2845.25 $\mu$Hz ($\simeq$1422\,$\times£$2). 
All the detected frequencies of WD~J1527 are reported in Table \ref{table:WDJ1527}. In Fig. \ref{fig:FTJ15} we display the Fourier transform of 
data taken in sector 12 (upper panel) and sector 38 (lower panel) of WD~J1527.
The horizontal red line indicates the 0.1\% FAP level.
In sector 38, a clear triplet pattern can be seen from the sliding FT of the ultra-short-cadence data of WD~J1527. We detect a candidate for rotational triplet ($\ell= 1$) at 1419.97 (f$_{\rm 2}$), 1422.69 (f$_{\rm 3}$) and 1425.00 $\mu$Hz (f$_{\rm 4}$) with a rotational splitting of $\sim$ 2.51 $\mu$Hz. 
The lower-frequency component at $\sim$1420 $\mu$Hz is not present in overall FT, while it is clearly seen in the sFT as shown in Fig. \ref{fig:sFTJ15}.  
Therefore, even though this side component is below the detection threshold, we include it in our final frequency list with a S/N of 3.5. 
Multiplets in the frequency spectrum of a pulsating WD are useful both to identify the harmonic degree of the pulsation modes and to measure the rotation period of the star. We show in Fig. \ref{fig:sFTJ15} the rotational triplet detected in 
WD~J1527. 
The average of the splitting is $\sim$ 2.51 $\mu$Hz, which translates into a rotation period $P_{\rm rot} \sim 2.3$ days for WD~J1527. 
Since the frequency spectrum of WD J1527 is admittedly scarce,
so that we cannot search for a dipole or quadrupole period
spacing using only these three periods extracted from the 
{\sl TESS} data.

\begin{table}
\centering
\caption{Identified frequencies (combination frequencies), periods, and amplitudes (and their uncertainties)  and the signal-to-noise ratio in the  data of WD~J1527.
The frequencies that are detected only in sector 12 are unmarked.
Frequency and amplitude from sector 38 are marked with $\dagger$.
}

\begin{tabular}{lcccr}
\hline
\noalign{\smallskip}
Peak & $\nu$    &  $\Pi$  &  $A$   &  S/N \\
 & ($\mu$Hz)      &  (s)   & (ppt)   &   \\
\noalign{\smallskip}
\hline
\noalign{\smallskip}
f$_{\rm 1}$ & 1340.703(28)  &  745.877(15) & 30.74(3.89)   & 7.1\\
f$_{\rm 2^{\dagger}}$ &1419.967(55)	& 704.241(27) &	6.91(1.61) & 3.5 \\
f$_{\rm 3^{\dagger}}$ & 1422.687(02)  &  702.895(01) & 17.28(1.61)   & 8.5  \\
f$_{\rm 4^{\dagger}}$ & 1425.000(25)  &  701.754(12) & 15.24(1.61)   & 7.5  \\
f$_{\rm 5}$ & 1539.889(25)  &  649.397(11) & 37.41(3.89)   & 7.9\\
\hline 
2f$_{\rm 3}$ & 2845.254(42)  &	351.462(05) & 	 8.94(0.31) &	4.6  \\
\noalign{\smallskip}
\hline
\end{tabular}
\label{table:WDJ1527}
\end{table}

\begin{figure}
    \includegraphics[width=0.35\textwidth, angle= -90]{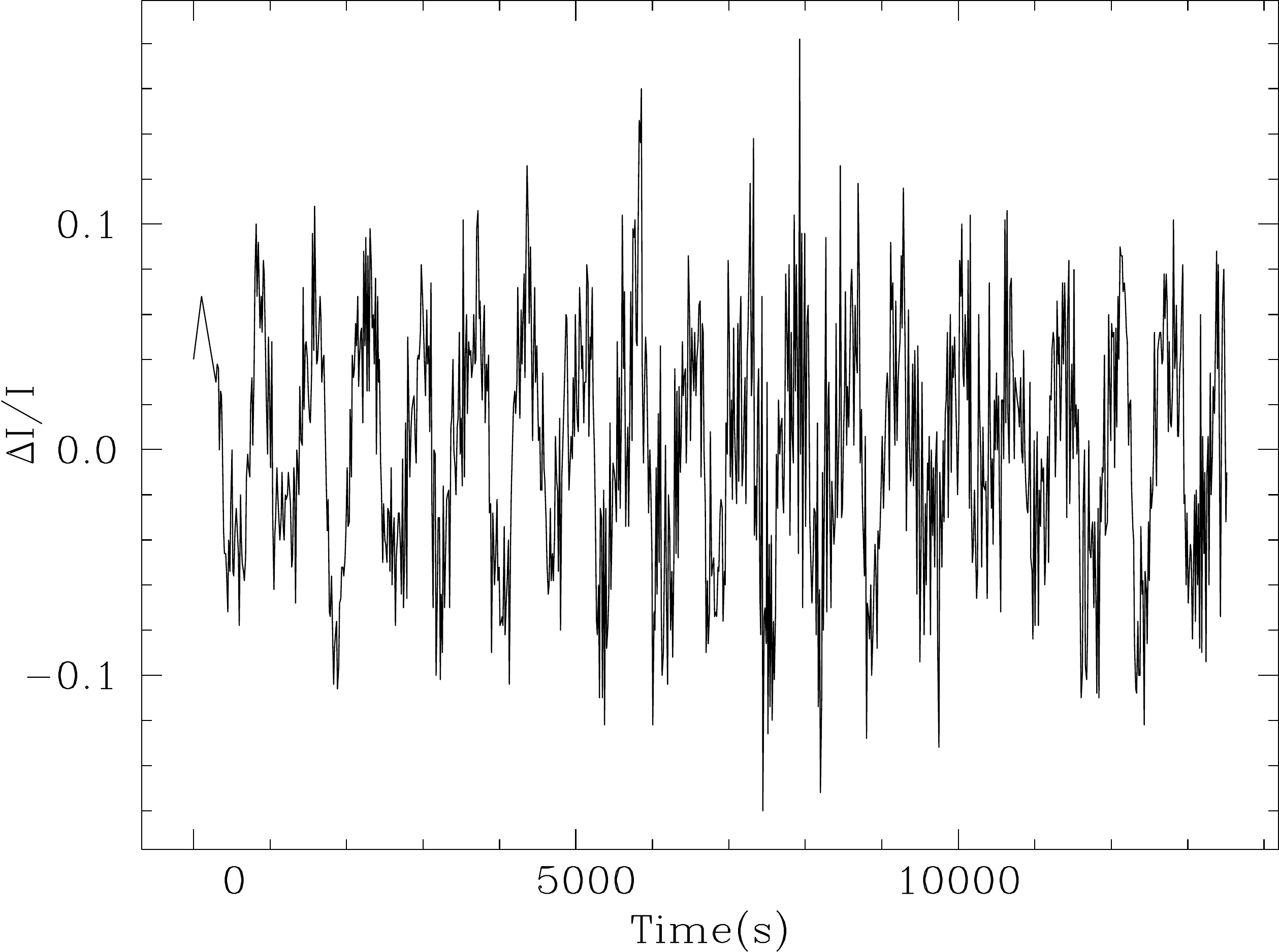} 
    \includegraphics[width=0.35\textwidth, angle= -90]{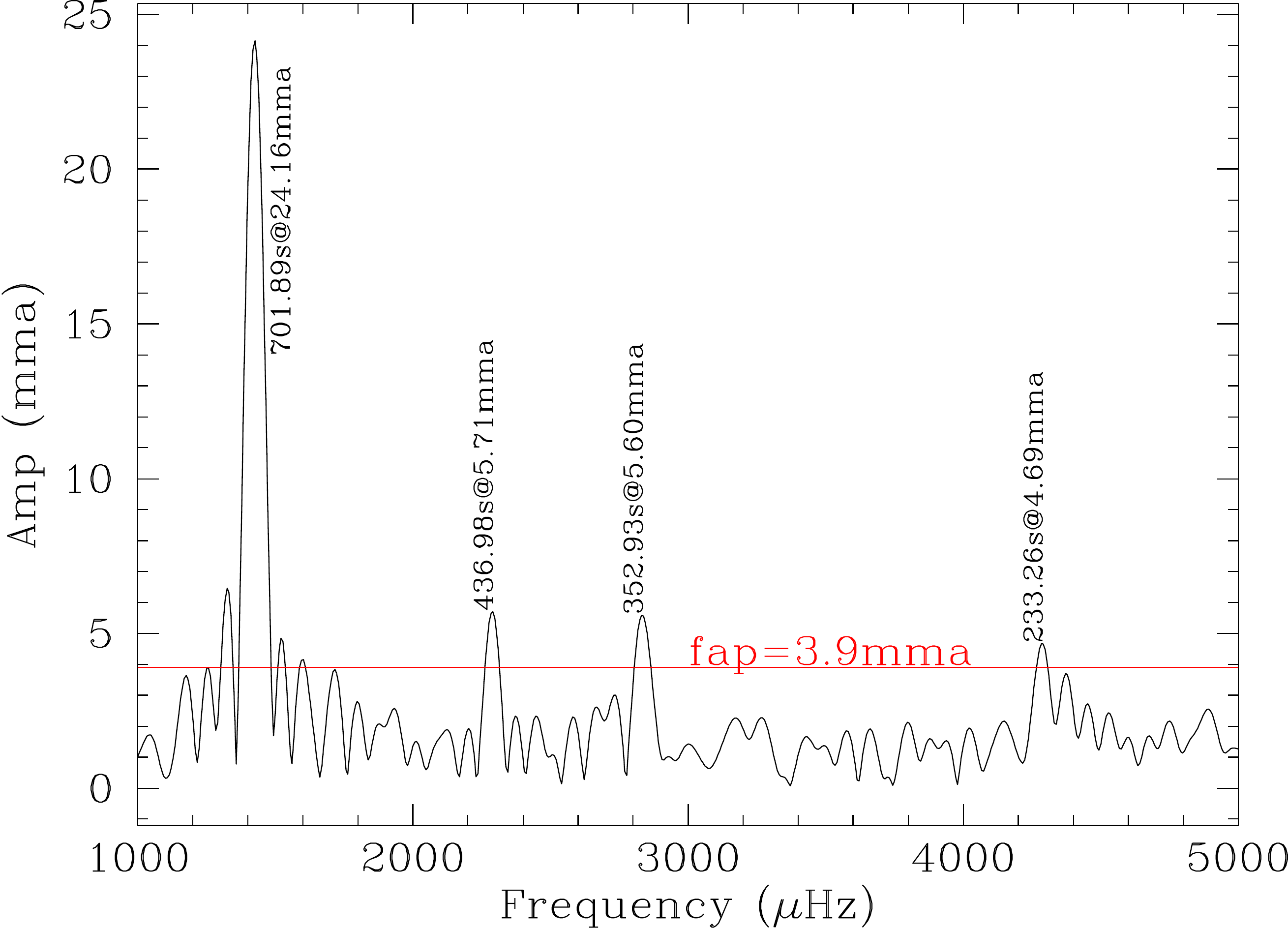} 
 \caption{Upper panel: the light curve of the ground-based 
observations of WD~J1527  
carried out with Pekin Elmer Telescope. The plotted data is one 4h run.
Lower panel: the Fourier transform, with  
a false-alarm pobability FAP(1/1000)= 3.9~mma.}
\label{light_curve_GB}
\end{figure}

Finally, we performed time series photometry of WD~J1527 with the IxON camera on the 1.6-m 
Perkin Elmer Telescope at the Pico Dos Dias Observatory during two nights, one in April and one in June 2021. We observed a total of $\sim$6 h, with integration times from 6 s to 12 s, using a red blocking filter BG40. We reduced the data with the software IRAF, using the task DAOFOT and obtain a light curve using differential photometry. Finally, we obtained the Fourier Transform using the software ${\tt PERIOD04}$ \citep{LB2005}. The list of frequencies, periods and variation amplitudes are listed in table~\ref{table:WDJ1527-OPD}.  In Fig. \ref{light_curve_GB}  we depict the light curve (top) and the Fourier transform (bottom) for WD~J1527 obtained from these ground-based observations. The plotted data is one 4~h run, so there are no gaps to produce aliases. The same peaks are detected when we add the 2~h data from June~21.
We note that the ground-based data can detect periods not seen with {\sl TESS}. This is because the 1.6~m telescope used has an area more than 100 times larger than {\sl TESS} and therefore can reach fainter magnitudes.

\begin{table}
\centering
\caption{Detected periods for WDJ152738.4$-$450207.4 from ground-based observations performed at the Pico do Dias observatory. Column 2 corresponds to the frequencies, 
column 3 shows the periods, and column 4 the amplitudes. Finally, column 5 gives the 
S/N level for each mode.}
\begin{tabular}{ccccc}
\hline
\noalign{\smallskip}
Peak & $\nu$    &  $\Pi$  &  $A$  & S/N  \\
 & ($\mu$Hz)      &  (s)   & (ppt)  &  \\
\noalign{\smallskip}
\hline
\noalign{\smallskip}
f$_{\rm 1}$ & $1425.7(67)$ & $701.89(33)$ & $24.16$ & 18.6 \\
f$_{\rm 2}$ & $2290.6(67)$ & $436.98(13)$ & $ 5.71$ & 4.4  \\
f$_{\rm 3}$ & $2831.2(67)$ & $352.93(8)$ & $ 5.60$ & 4.3  \\
f$_{\rm 4}$ & $4283.9(67)$ & $233.26(4)$ & $ 4.69$ & 3.6  \\
     \noalign{\smallskip}
\hline
\end{tabular}
\label{table:WDJ1527-OPD}
\end{table}

\subsection{L~7$-$44} 
\label{obs-L7-44}

L~7$-$44 is also a new pulsating DBV star, which was observed by {\sl TESS} in short cadence mode during sector 12 and 13 between 21 May 2019 and 18 July 2019, covering about 57 days.
This provides a frequency resolution of 0.14~$\mu$Hz. 
The FT average noise level is 0.28 ppt. The $0.1\%$~FAP confidence level is 1.39~ppt.
We detect four frequencies at 981.22\,(f$_{\rm 1}$), 1067.52\,(f$_{\rm 2}$), 1094.50\,(f$_{\rm 5}$) and 2141.81\,$\mu$Hz (f$_{\rm 6}$). 

\begin{figure} 
\includegraphics[width=0.47\textwidth]{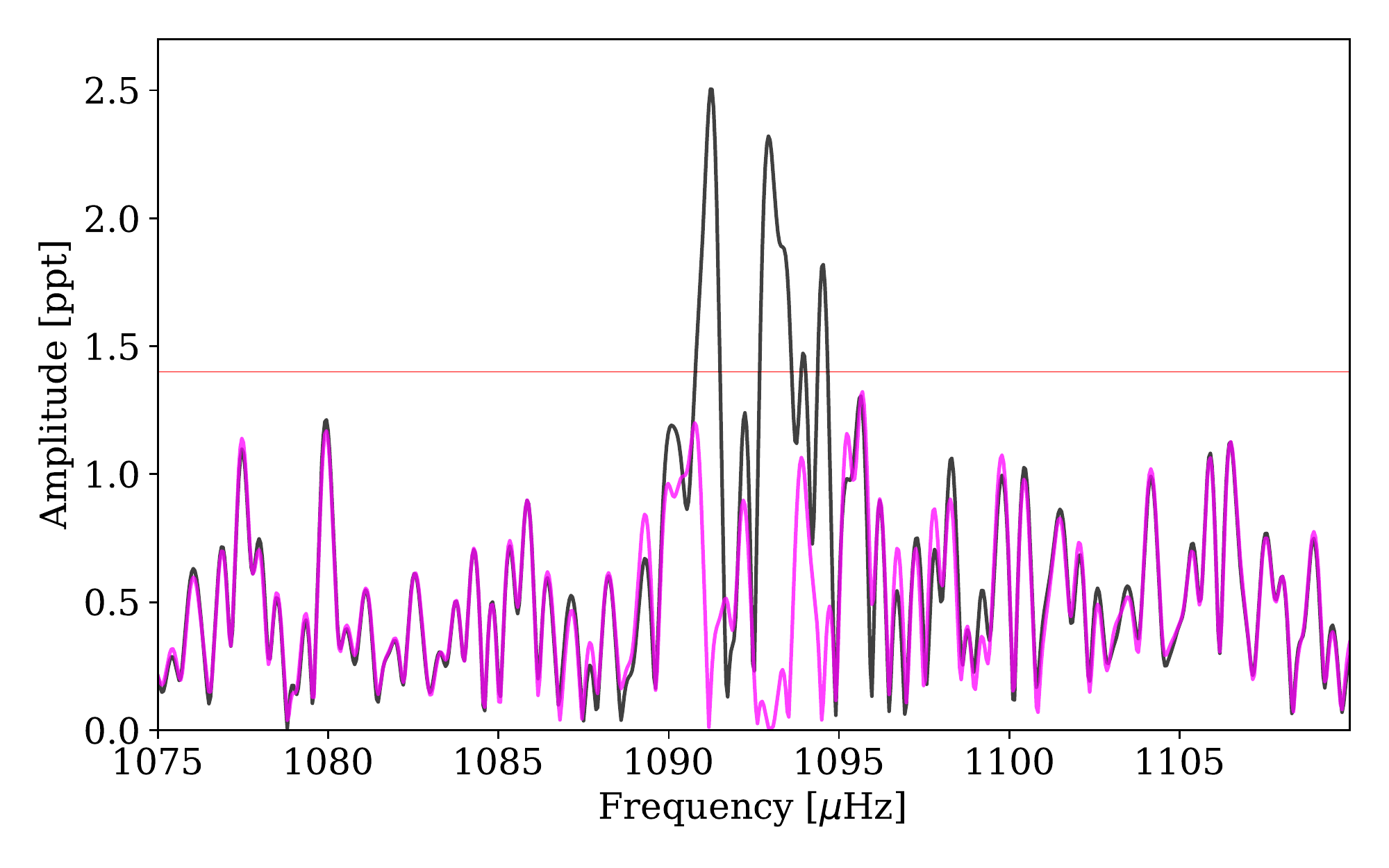}
\includegraphics[width=0.51\textwidth]{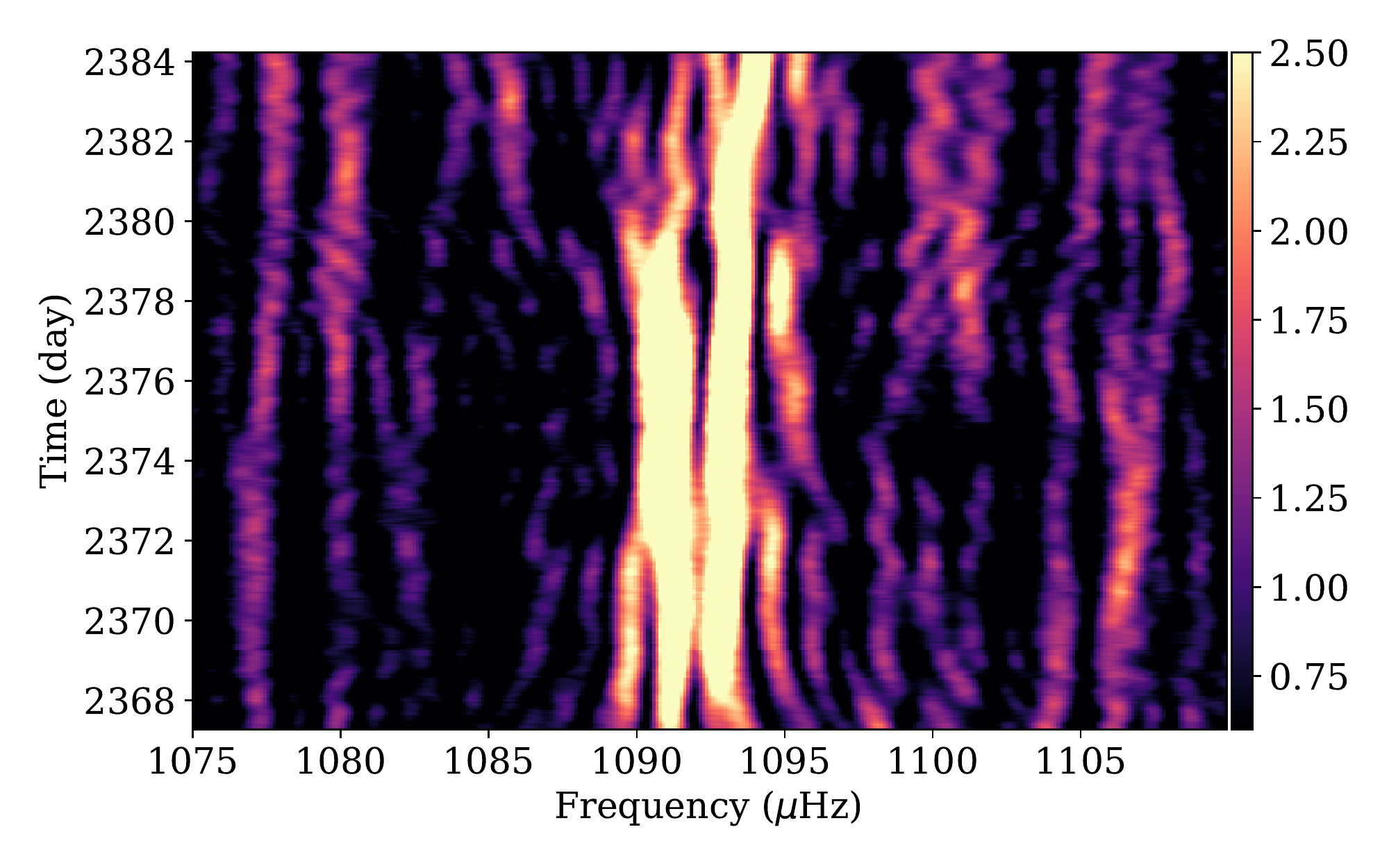}
\caption{{\sc Top:} Fourier transform of sector 39 data of L~7$-$44. The horizontal red line indicates the 0.1\% FAP level. The magenta line is the residual after extraction of the signals. 
 {\sc Bottom:}  Sliding Fourier transform of sector 39 data of L~7$-$44. The color-scale illustrates amplitude in ppt units.}
    \label{fig:FT451} 
\end{figure}

L~7$-$44 was also observed in sector 39 in 20-sec cadence during 27~d. The frequency resolution of this dataset is 0.41 $\mu$Hz. 
We identified four frequencies above $0.1\%$~FAP confidence level of 1.84 ppt, which are listed in Table \ref{table:L7-44}. 
We found a complicated pattern around 1092\,$\mu$Hz, that is
depicted in Fig.~\ref{fig:FT451}.
This unresolved structure is most likely due to frequency and/or phase variations over the length of the data. Also, it could be caused by {\it mode incoherence}, that is, since this is the lowest frequency pulsation mode in this sample, it could be 
incoherent because it is reflecting off the moving base of the outer convection 
zone, as described for DAVs in \cite{2020ApJ...890...11M}. We extracted three frequencies (f$_{\rm 2}$, f$_{\rm 3}$ and f$_{\rm 4}$) from this region. However, only f$_{\rm 3}$ will be used for the asteroseismic modeling of this star since it is the highest FT peak.

\begin{table}
\centering
\caption{Independent frequencies, periods, and 
amplitudes (and their uncertainties) and the 
signal-to-noise ratio in the  data of L~7$-$44.
Frequency and amplitude from both short and ultra-short cadence data are marked with $\dagger\dagger$.
Frequency and amplitude from sector ultra-short cadence data are marked with $\dagger$.
The frequencies that are detected only in sector short-cadence data are unmarked.}
\begin{tabular}{lcccr}
\hline
\noalign{\smallskip}
Peak & $\nu$    &  $\Pi$  &  $A$   &  S/N \\
 & ($\mu$Hz)      &  (s)   & (ppt)   &   \\
\noalign{\smallskip}
\hline
\noalign{\smallskip}

f$_{\rm 1}$ & 981.221(21) &   1019.138(21) &   1.64(31) &   4.6 \\ 
f$_{\rm 2}$ & 1067.521(19) &    936.750(17) &   1.80(31)    &  4.7      \\
f$_{\rm 3^{\dagger}}$ & 1091.211(28) &    916.413(23) &   2.49(30) &   7.1\\
f$_{\rm 4^{\dagger}}$  & 1093.003(27) &    914.910(23) &   2.53(30) &   7.2\\
f$_{\rm 5^{\dagger\dagger}}$  & 1094.528(35) &    913.636(29) &   1.96(30) &   5.6\\
f$_{\rm 6^{\dagger\dagger}}$ & 2141.813(16) &    466.894(4)  &   2.09(31)    &  5.5      \\
\noalign{\smallskip}
\hline
\noalign{\smallskip}
\label{table:L7-44}
\end{tabular}
\end{table}

\section{Evolutionary models and spectroscopic masses}
\label{models}

We employ a  set  of state-of-the-art  DB WD stellar  models  that take  into account the  complete evolution of the progenitor stars, assuming
that DB WDs are descendants of the PG~1159 stars after going through
the DO WD stage. The models were extracted  from the evolutionary  calculations of \cite{2009ApJ...704.1605A} produced with the 
{\tt LPCODE} evolutionary code (see that paper for details). 
These DB WD models have been employed in the asteroseismic analyses of the DBV stars KIC\,8626021 
\citep{2012A&A...541A..42C}, KUV~05134+2605 \citep{2014A&A...570A.116B},  
PG\,1351+489 \citep{2014JCAP...08..054C},  WD 0158$-$160 \citep{2019A&A...632A..42B}, and GD~358 \citep{2022A&A...659A..30C}. 
The sequences of DB WD models have been obtained taking into account a complete 
treatment of  the evolutionary history of progenitors stars, starting from the 
zero-age main sequence (ZAMS), through the thermally pulsing asymptotic giant 
branch (TP-AGB) and born-again (VLTP; very late thermal pulse) phases to the domain of the PG\,1159 stars, and finally the DB WD stage. As such, they are characterized by
evolving chemical profiles consistent with the prior evolution. We  
varied the stellar  mass and the effective temperature in our 
model calculations, while the He content, the chemical structure at the CO core, 
and the thickness of the chemical interfaces were fixed by the 
evolutionary history of progenitor objects. 
The models employ the ML2 prescription of convection with the mixing length parameter, $\alpha$, fixed to 1 \citep{1971A&A....12...21B,1990ApJS...72..335T}.
Specifically, we considered nine DB WD sequences with stellar
masses of:  $0.515, 0.530, 0.542, 0.565, 0.584,  0.609, 0.664, 0.741$,
and $0.870  M_{\odot}$.  These DB WD sequences are characterized  by the
maximum He-rich  envelope that  can be left  by prior evolution  if we
assume that they are the result  of a born-again episode. The value of
envelope  mass ranges  from  $M_{\rm He}/M_*  \sim  2 \times  10^{-2}$
($M_*=  0.515 M_{\odot}$) to  $M_{\rm He}/M_*  \sim 1  \times 10^{-3}$
($M_*=  0.870  M_{\odot}$).   In Figure \ref{fig:1} we 
show the complete set of DB WD evolutionary tracks (with different
colors according to the value  of the stellar  mass) along with the location  of  all  the  
DBVs known  to  date \citep{2019A&ARv..27....7C,2021ApJ...922....2D, 2022ApJ...927..158V} and the two new DBVs reported in this paper for the first time.    

We  computed $\ell=  1, 2$ $g$-mode  pulsation  periods
in the range  $80-1500$ s  with the  adiabatic and
nonadiabatic versions of the pulsation code {\tt LP-PUL}
\citep[][]{2006A&A...454..863C,2006A&A...458..259C} and the same
methods  we employed  in the  previous  works of La Plata Stellar
Evolution and Pulsation Research Group\footnote{\tt
  \url{http://fcaglp.fcaglp.unlp.edu.ar/evolgroup/}}.  
Employing the   evolutionary    tracks   presented   in
Fig.~\ref{fig:1} and the values of the spectroscopic 
surface gravity and temperature  shown in Table \ref{basic-parameters-targets}, 
we derive  a value of the
spectroscopic   mass of each of the five analyzed stars by interpolation. 
It is worth noting that we use the same  set  of  DB  WD  models    
both for estimate the spectroscopic mass and also 
to assess the stellar mass from the period spacing (next Section).  
For PG~1351 we obtain $M_{\star}= 0.558 \pm 0.027 \ M_{\sun}$.
In the case of EC~20058 and EC~04207, we derive
$M_{\star}= 0.614 \pm 0.030 \ M_{\sun}$ and $M_{\star}= 0.515 \pm
0.023 \ M_{\sun}$, respectively. Finally, we obtain  $M_{\star}= 0.675 \pm
0.022 \ M_{\sun}$ for WD~J1527, and $M_{\star}= 0.630 \pm
0.016 \ M_{\sun}$ for L~7$-$44.  The uncertainties in the stellar
mass are estimated from the uncertainties in $T_{\rm eff}$ and
$\log g$ by adopting the extreme values of each parameter when
interpolating between the evolutionary tracks of Fig. \ref{fig:1}.

\section{Asteroseismological analysis}
\label{asteroseismology}

The asteroseismological methods we use in this paper to extract information of the stellar
mass and the internal structure of the DBV stars PG~1351, EC~20058,  EC~04207, 
WD~J1527, and L~7$-$44 are the same 
used in, for instance, \cite{2022A&A...659A..30C} for the DBV star GD~358 observed with {\sl TESS}. 
 
First, we attempt to infer the stellar mass, when possible, by comparing the 
observed mean period spacing, 
$\Delta \Pi$, with the average of the computed period spacings 
($\overline{\Delta \Pi_{k}}$). The mean period spacing is 
assessed through the  Kolmogorov-Smirnov
\citep[K-S;][]{1988IAUS..123..329K}, and the inverse  variance
\citep[I-V;][]{1994MNRAS.270..222O} significance tests.  
On the other hand, the average of the computed period spacings
is calculated  as $\overline{\Delta \Pi_{k}}= (N-1)^{-1} \sum_k \Delta \Pi_{k}$, where the "forward" period spacing ($\Delta \Pi_{k}$) is
defined as $\Delta \Pi_{k}= \Pi_{k+1}-\Pi_{k}$ ($k$ being the radial
order) and $N$ is the number of computed periods laying in the range
of the observed periods.  The
present method relies on the spectroscopic effective temperature, 
so the results are unavoidably affected by its associated uncertainty. This method 
takes full advantage of the fact that the period spacing of 
DBV stars primarily depends on the stellar mass and the effective temperature, 
and very weakly on the thickness of the He envelope 
\citep[see, e.g.,][]{1990ApJS...72..335T}. 

Another asteroseismological tool to disentangle the internal structure
of DBV stars is to seek theoretical models that best match the
individual pulsation  periods  of  the target star. To measure the
goodness of the match between the theoretical pulsation periods
($\Pi_{\ell,k}$) and the observed individual periods ($\Pi_i^{\rm
  o}$), we compute the quality function 
$\chi^2(M_{\star}, T_{\rm eff})= \frac{1}{N} \sum_{i= 1}^{N}
      {\rm min}[(\Pi_{\ell,k}-\Pi_i^{\rm o})^2]$, 
  where $N$ is the number of observed periods. In order to find the
stellar model that best fits the observed periods exhibited by
each target star --- the ``asteroseismological'' model ---, we
evaluate the  function  $\chi^2$ for stellar  masses  
$0.515 \leq M_{\star}/M_{\odot} \leq 0.870$. For
the  effective  temperature, we  employ  a much  finer  grid ($\Delta
T_{\rm eff}= 10-30$ K).  For each target star, the DB WD model that
shows the lowest value of $\chi^2$ is adopted as the best-fit
asteroseismological model. To
quantitatively assess the quality of our period fit, we compute the
average   of   the   absolute   period    differences,
$\overline{\delta \Pi_i}= \left( \sum_{i= 1}^N |\delta \Pi_i|
\right)/N$, where $\delta \Pi_i= (\Pi_{\ell,k}-\Pi_i^{\rm o})$,  
and the root-mean-square residual, $\sigma= \sqrt{(\sum_{i= 1}^N
  |\delta \Pi_i|^2)/N}= \sqrt{\chi^2}$.  To have a global 
  indicator of the goodness of the period fit that takes into 
  account the number of free parameters, the number of fitted periods, 
  and the proximity between
the  theoretical and observed periods, we compute the Bayes
Information Criterion \citep[BIC;][]{2000MNRAS.311..636K}, defined as 
${\rm BIC}= n_{\rm p} \left(\frac{\log N}{N} \right) + \log \sigma^2$,
where $n_{\rm p}$ is the number of free parameters of the stellar
models, and $N$ is the number of observed periods, that is, the number of periods that we want to match. The smaller the
value of BIC, the better the quality of the fit. In our case, $n_{\rm
  p}= 2$ (stellar mass and effective temperature). Note that this 
criterion penalizes for an excess of parameters ($n_{\rm p}$). Below, we employ the tools described above to extract information from the DBV stars considered in this work.

\subsection{PG~1351+489}
\label{pg1351}

\begin{table*}
\centering
\caption{List of periods of PG~1351 available for the asteroseismological analysis. 
Column 1 corresponds to four $\ell= 1$ periods measured by \cite{2011MNRAS.415.1220R}
  (REA11), and column 2 corresponds to the single period detected by {\sl TESS}   
  (Table \ref{table:PG1351}). The third, fourth and fifth columns give the 
  theoretical periods of the asteroseismological model of PG~1351, the harmonic degree and the radial order, respectively. Columns 6 and 7 correspond to the period difference ($\delta \Pi_k= \Pi_i^{\rm O}- \Pi_k$) and the rate of period change, respectively. Finally, the last column provides information about the pulsational stability or instability nature of the modes.}
\begin{tabular}{cc|cccccc}
\hline
\noalign{\smallskip}
$\Pi_i^{\rm O}$  & $\Pi_{i}^{\rm O}$  & $\Pi_{\ell,k}$ & $\ell$ & $k$ & $\delta \Pi_{\ell,k}$ & $\dot{\Pi}_{\ell, k}$ & Unstable\\
  (s) & (s)  & (s) & & & (s) & $(10^{-13})$ s/s & \\
  REA11 & {\sl TESS}  & & & & & & \\
\noalign{\smallskip}
\hline
\noalign{\smallskip}     
335.26 &        & 336.81 & 2 & 13 & $-1.55$ & 0.60 & yes \\
489.33 & 489.26 & 489.47 & 1 & 11 & $-0.21$ & 0.81 & yes \\
584.68 &        & 586.99 & 2 & 25 & $-2.31$ & 1.02 & yes \\
639.63 &        & 639.37 & 1 & 15 & $0.26$  & 1.19 & yes \\
\noalign{\smallskip}
\hline
\end{tabular}
\label{table:PG1351-extended}
\end{table*}

\begin{figure} 
\includegraphics[clip,width=1.0\columnwidth]{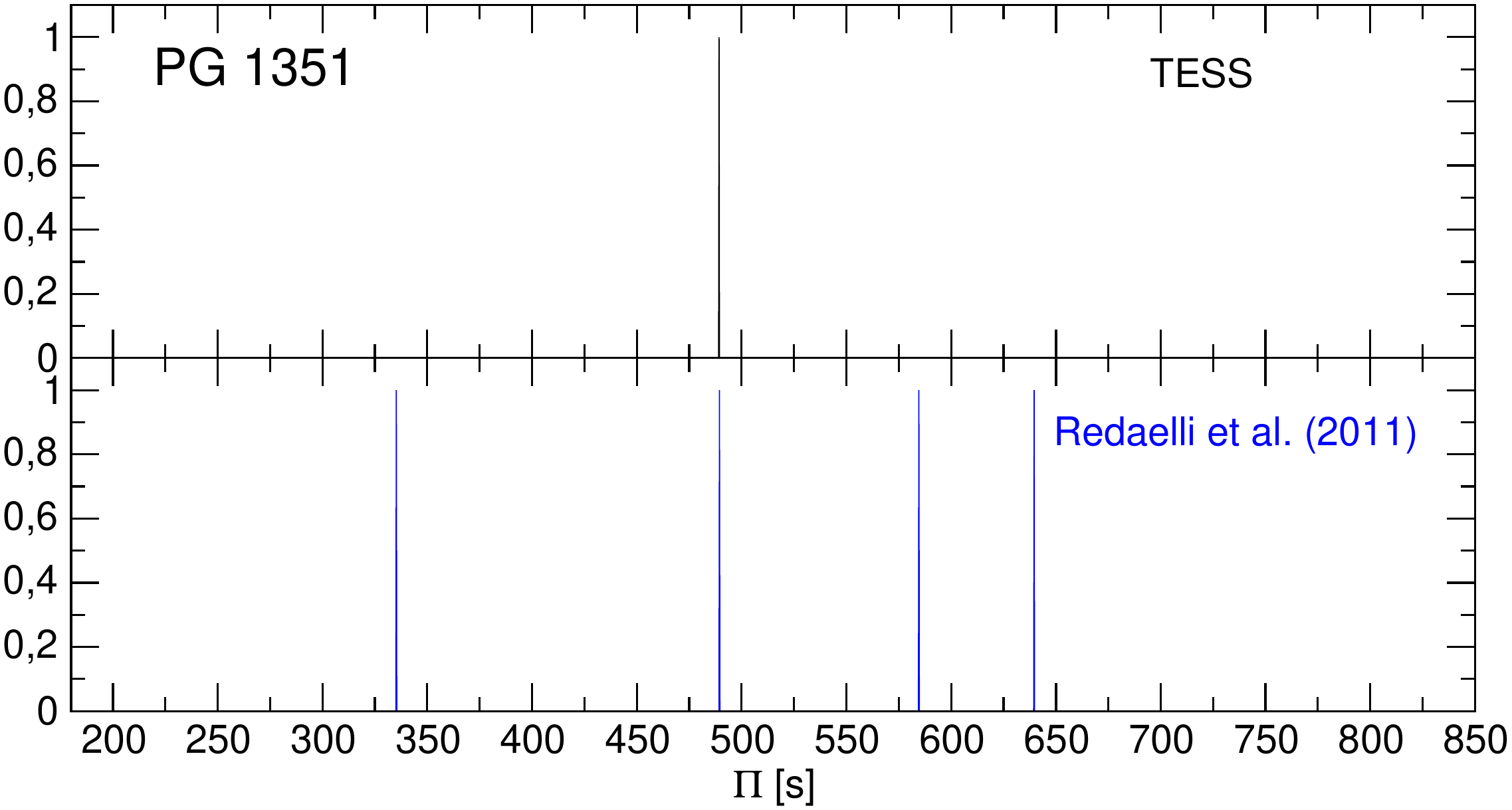}
\caption{Schematic distribution of the pulsation periods of PG~1351
  according  to {\sl TESS} (6 periods, black lines, upper panel), and
  according to \cite{2011MNRAS.415.1220R}  (4 periods, blue lines,
  lower panel). The amplitudes have been arbitrarily  set to one for
  clarity.}
\label{fig:compara-tess-rea11} 
\end{figure}

For this star, {\sl TESS} detects a single period of $489.26$ s which is likely 
associated to an eigenmode of the star (see Table \ref{table:PG1351}).  The other 
period present ($\sim 244.630$ s) corresponds to twice the frequency of that single period. The period of $\sim 489$ s is present also in the data acquired through the ground-based monitoring by \cite{2011MNRAS.415.1220R}, who identified four independent periods (see their Table 2).  In Fig. \ref{fig:compara-tess-rea11} we 
schematically show the period spectrum detected with {\sl TESS} (upper
panel), and the periods detected by \cite{2011MNRAS.415.1220R} (lower
panel), with amplitudes set to unity to facilitate visualization. Unfortunately, the {\sl TESS} data does not broaden the period spectrum of PG~1351 in relation to the periods already known with observations from the ground. The list of periods of the star  is shown in Table \ref{table:PG1351-extended}. With 
this very limited list of periods, we have tried to find a possible mean period spacing 
to infer some hint of the stellar mass of PG~1351.  

\begin{figure} 
\includegraphics[clip,width=1.0\columnwidth]{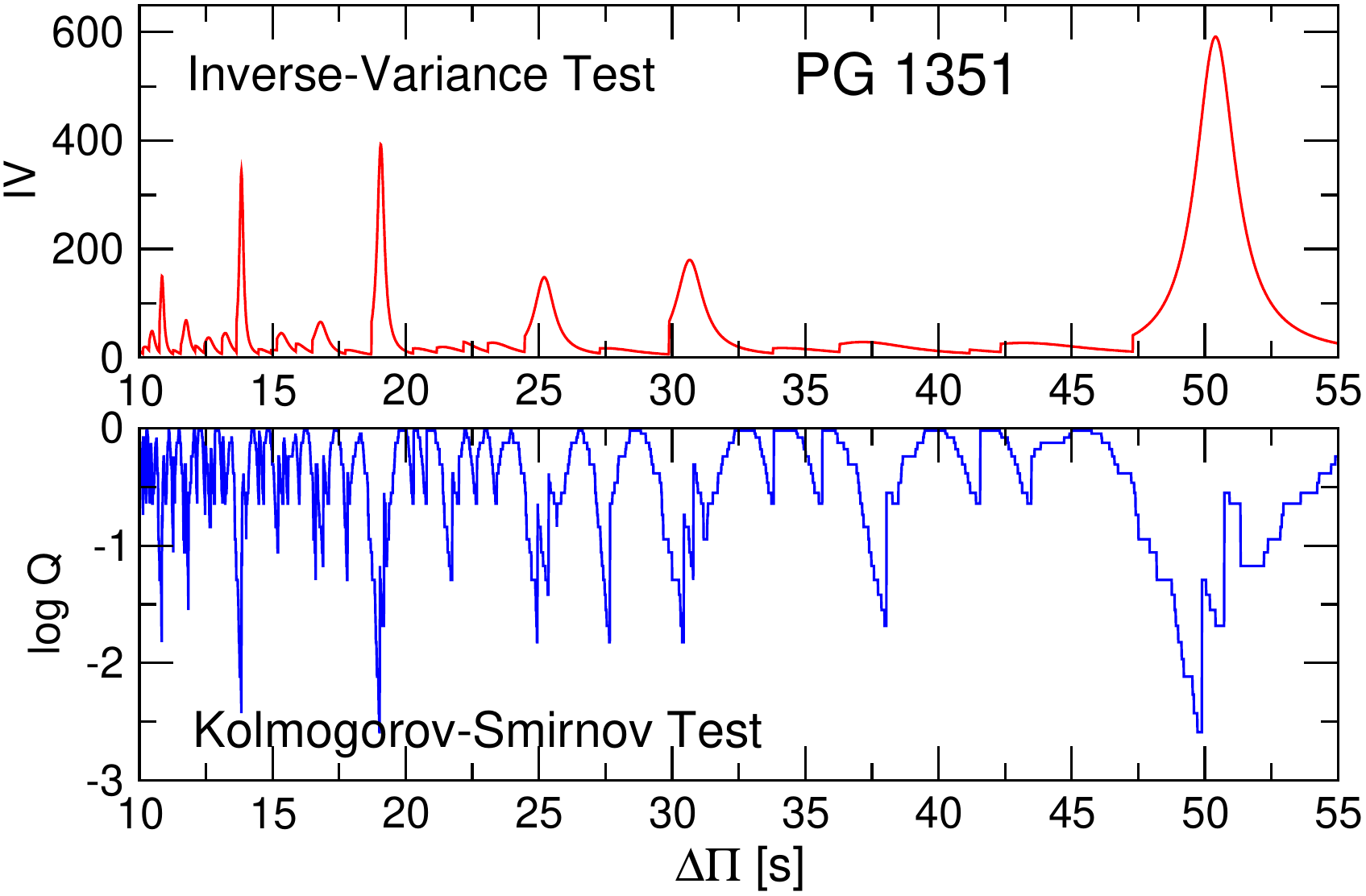}
\caption{I-V  (upper panel) and   K-S  (lower panel)  significance  tests  
to  search  for  a constant  period
  spacing  in PG~1351. The tests are applied to the set of 4
  pulsation periods of Table \ref{table:PG1351-extended}, that includes 
  the {\sl TESS} and   \cite{2011MNRAS.415.1220R}'s periods. 
  Several possible constant period spacings are evident, the most notorious
  being that of $\sim 50$ s. See text for details.}
\label{fig:tests-PG1351} 
\end{figure} 

\begin{figure} 
\includegraphics[clip,width=1.0\columnwidth]{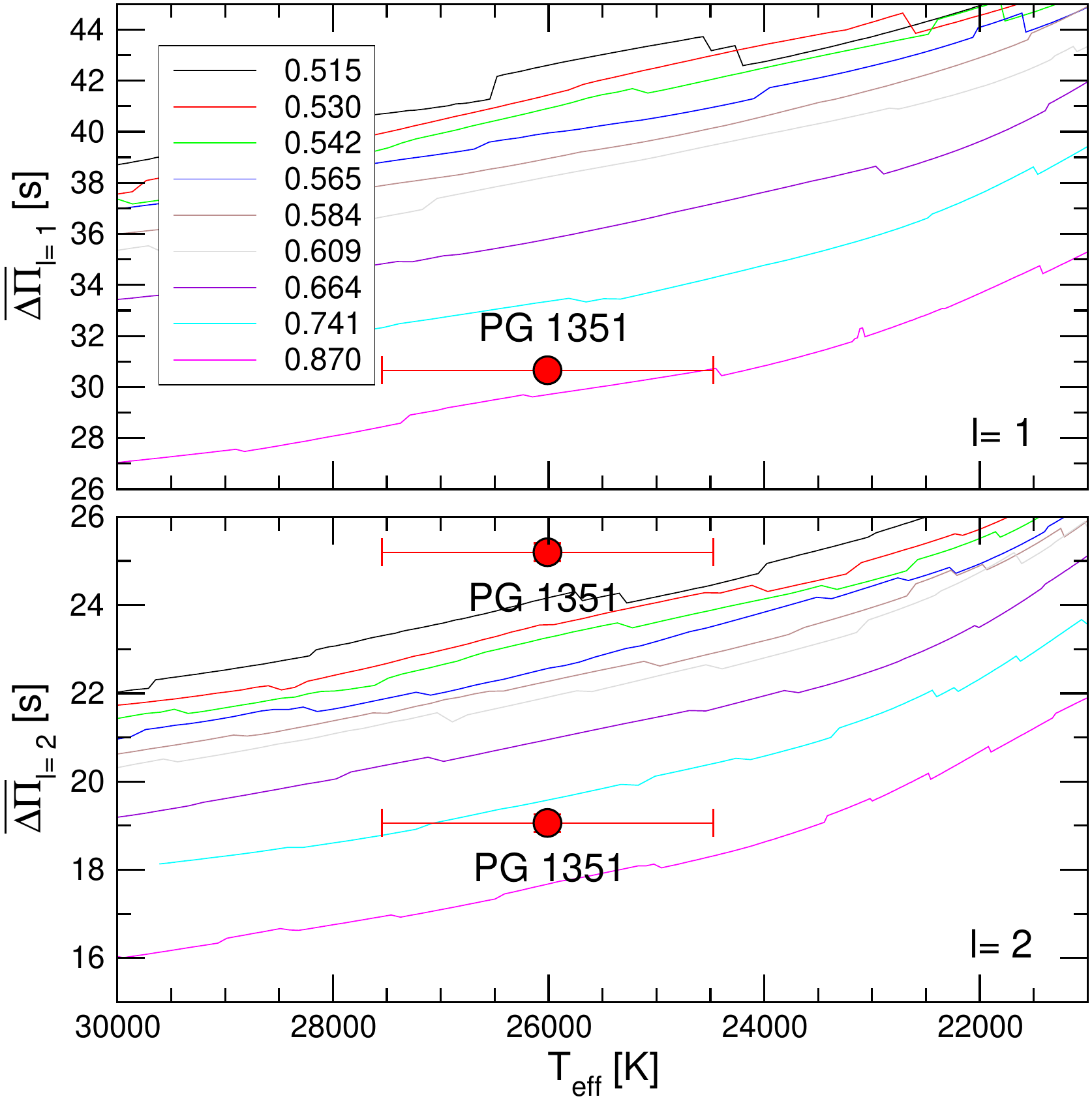}
\caption{Dipole (upper panel) and quadrupole (lower panel) average 
of the computed period spacings,
  $\overline{\Delta \Pi_k}$, assessed  in  a  range  of  periods  that
  includes  the  periods  observed  in PG~1351, shown as curves 
  of different colors according to the different stellar masses. 
  We consider the effective temperature $T_{\rm eff}= 26\,010\pm 1536$~K 
  \citep{2011ApJ...737...28B}, and the possible period spacing values derived
  from the four periods of PG~1351 according to Table \ref{table:PG1351-extended}. 
  We include the error bars associated to the uncertainties in
  $\overline{\Delta \Pi_k}$ and $T_{\rm eff}$.}
\label{fig:psp-teff-PG1351} 
\end{figure}

In Fig.~\ref{fig:tests-PG1351} we show the results of applying the
I-V and K-S statistical tests to the set of four periods of Table
\ref{table:PG1351-extended}. We are aware that with only four periods, 
it is feasible to find different period spacing values that fit the 
distribution of those periods. Indeed, the two tests show hints of several possible constant period spacings of $\sim 11$ s, $\sim 14$ s, $\sim 19$ s, $\sim 25$ s,  $\sim 31$ s, and another at $\sim 50$ s. Since the period spacing in DB WDs depends mainly on the stellar mass and the effective temperature, we can explore which of these possible period spacings are compatible with the range of stellar masses of DB WDs. We have computed the average of the theoretical period spacings for $\ell= 1$ and $\ell= 2$, $\overline{\Delta \Pi_{k}}$,  in terms of the effective temperature for all the masses considered and a period interval of $300-650$ s. In  Fig.~\ref{fig:psp-teff-PG1351} we show  $\overline{\Delta  \Pi_{k}}$ with curves of different colors to distinguish the different stellar masses, corresponding to $\ell= 1$ (upper panel) and $\ell= 2$  
  (lower panel) modes. We note that the only period spacings compatible with 
reasonable stellar masses are $\sim 31$ s for $\ell= 1$ and $\sim 19$ s for $\ell= 2$. 
Both possible period spacings point to a stellar mass somewhat high for DB WD standards,
but still possible, in the range $0.74-0.87 M_{\odot}$\footnote{Note that there is also a possible period spacing of $\sim 25$ s. If this were a $\ell= 1$ period spacing, the mass of PG~1351 should be very high ($\sim 1 M_{\odot}$). Alternatively, if this were a $\ell= 2$ period spacing, then the mass of PG~1351 should be too low ($\sim 0.40 M_{\odot}$). Both cases 
have to be discarded because they are in serious conflict with the spectroscopic determination of $\log g$, that indicates a stellar mass in the range $0.53\lesssim  
M_{\star}/M_{\odot} \lesssim 0.59$ (see Fig. \ref{fig:1}).}.  However, we emphasize that, since the number of available periods is very low, the derivation of the period spacing is not robust, and the stellar mass by this mean is not significantly constrained. 

\begin{figure} 
\includegraphics[clip,width=1.0\columnwidth]{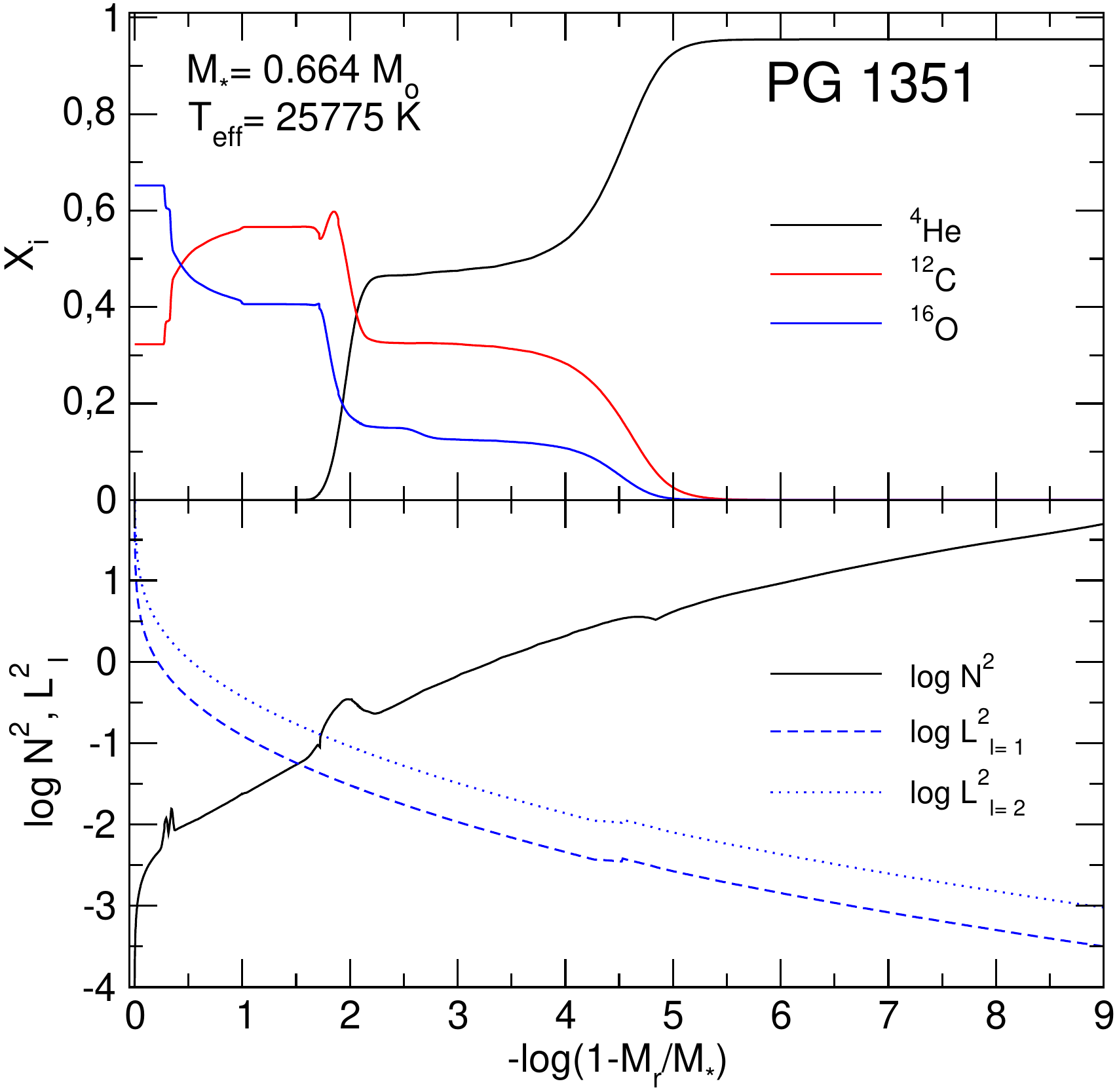}
\caption{Chemical  profiles  (upper panel)  and  the  squared 
Brunt-V\"aïs\"al\"a  and  Lamb  frequencies  for $\ell= 1$ and $\ell= 2$ (lower panel)  corresponding  to  our  asteroseismological  DB  WD  model  of PG~1351 with  a  stellar  mass $M_{\star}= 0.664 M_{\odot}$ and an effective temperature $T_{\rm eff}= 25\,775 $ K.}
\label{PG1351-modelo} 
\end{figure}

\begin{table}
\centering
\caption{The main characteristics of the DBV star PG~1351.} 
\begin{tabular}{l|cc}
\hline
\hline 
Quantity & Spectroscopy &  Asteroseismology \\
         & Astrometry   &         \\ 
\hline
$T_{\rm eff}$ [K]                           & $26\,010 \pm 1536$ & $25\,775\pm150$       \\
$M_{\star}$ [$M_{\odot}$]                   & $0.558\pm0.027$    & $0.664\pm0.013$     \\ 
$\log g$ [cm/s$^2$]                         & $7.91\pm0.07$      & $8.103\pm0.020$     \\ 
$\log (L_{\star}/L_{\odot})$                & $\hdots$           & $-1.244\pm0.03$     \\  
$\log(R_{\star}/R_{\odot})$                 & $\hdots$           & $1.912\pm0.015$     \\  
$(X_{\rm C}, X_{\rm O})_{\rm c}$            & $\hdots$           & $0.32, 0.65$        \\
$M_{\rm He}/M_{\star}$                      & $\hdots$           & $5.42\times10^{-3}$ \\
$d$  [pc]                                   & $175.73^{+1.53 (a)}_{-1.58}$ & $167.05^{+2.31}_{-2.26}$    \\ 
$\pi$ [mas]                                 & $5.69\pm0.05^{(a)}$   & $5.99^{+0.08}_{-0.09}$\\ 
\hline
\hline
\end{tabular}
\label{astero-PG1351}
{\footnotesize  References: (a) Gaia EDR3.}
\end{table}

A detailed asteroseismological analysis of PG~1351   
employing the same DB WD models considered in this paper was already carried out by \cite{2014JCAP...08..054C}. The only difference is that in that work, the authors
used a period of 489.33 s, instead of 489.26 s as in the present paper (see Table \ref{table:PG1351}). Since this difference is completely negligible to the limit of our investigation, the asteroseismological model that we obtain in this work is the same as 
in \cite{2014JCAP...08..054C} and we will not repeat that analysis here. 
We show in Table \ref{table:PG1351-extended} the theoretical periods and the harmonic degree and radial order of this model, along with the period differences and the rates of period change. We obtain $\overline{\delta \Pi_i}= 1.08$ s, $\sigma= 1.40$ s, and BIC= 0.59 for the asteroseismological model. We note that \cite{2022A&A...659A..30C} obtain ${\rm BIC}= 1.13$ for the asteroseismological model of 
GD~358 and  \cite{2019ApJ...871...13B} obtain ${\rm BIC}= 1.2$ for their best period fit to the same star. So, the low value of BIC obtained in this work for PG~1351 indicates that  our period fit is very good. We show the characteristics of this model ($M_{\star}= 0.664 M_{\sun}, T_{\rm eff}= 25\,775$ K) in Table  \ref{astero-PG1351}\footnote{Here, $M_{\rm He}/M_{\star}$ is the total mass content of He in units of the stellar mass.}, and we plot its internal chemical profiles (upper panel) and the logarithm of the squared Brunt-V\"ais\"al\"a and Lamb frequencies (lower panel) in Fig. \ref{PG1351-modelo}. 

\cite{2014JCAP...08..054C} show that there are other possible asteroseismological solutions, as can be seen in their Figure 9 and Table 1.  We note (as those authors do) that 
the possible solution that assumes only $\ell= 1$ modes has a very low quality of the period fit, and can be discarded. The two solutions with $M_{\star}= 0.87 M_{\odot}$ and assuming a 
mix of $\ell=1$ and $\ell= 2$ modes can be discarded too, in one case due to poor quality of the period fit, and in the other one because the $T_{\rm eff}$ is too low. As an additional argument to discard these three possible solutions, we emphasize the fact that they 
correspond to a stellar mass ($M_{\star}= 0.87 M_{\odot}$) that is too large as compared with the spectroscopic one ($M_{\star}= 0.56 M_{\odot}$),  and in severe contradiction with the surface gravity derived for this star. Finally, none of these three possible solutions fits the Gaia's distance better than the asteroseismological model of Table \ref{astero-PG1351} does 
(see below).

We have examined the pulsational stability/instability nature of the modes
associated with the periods fitted to the observed ones.
We adopted the frozen-in convection approximation \citep{1989nos..book.....U}. 
In particular,  we examined the sign and magnitude of the linear nonadiabatic 
growth rates, $\eta_k= -\Im(\sigma_k)/\Re(\sigma_k)$,  where $\Re(\sigma_k)$ and $\Im(\sigma_k)$ are  the  real  and  the imaginary parts, respectively, of the complex eigenfrequency $\sigma_k$. We find that the periods of the asteroseismological model
of PG~1351 are associated with unstable modes, in line with the observational evidence. 
A common fact in all nonadiabatic calculations of pulsations in WDs, but also in any other class of pulsating stars, is that in general the models predict many more unstable modes than are observed. PG~1351 is no exception. There must be some (unknown) filter mechanism that selects some unstable modes and gives them large, observable amplitudes, while the other unstable modes never reach those amplitudes and are not observed. 

Finally, using the asteroseismological model of PG~1351, we can assess the seismic 
distance of the star. We employed the effective temperature and gravity of our best-fit 
model to infer the absolute $G$ magnitude ($M_{G}$) in the {\it Gaia} photometry.
This was done by computing a DB WD atmosphere model using $T_{\rm eff}$ and $\log g$ of the 
asteroseismological model. We find $M_G= 10.564$ mag. With the observed {\it Gaia} magnitude,
$m_{G}= 16.678$ mag, a distance modulus was calculated as $m_G-M_G$, and this was 
converted to a seismological distance $d$ using $\log(d_{\rm s})= (m_G-M_G+5)/5$.  We  obtained  
$d_{\rm s}= 167.05^{+2.31}_{-2.26}$ pc, where the errors were calculated from the 
errors of $T_{\rm eff}$, $\log g$, and $m_G$. The atmosphere models used $\log({\rm H}
/{\rm He})= -5$. A test calculation showed that the difference to pure He models is negligible. The seismological distance is in agreement with the value derived by \cite{2011ApJ...737...28B} ($\sim 166$ pc), and is just $\sim 5 \%$ lower than the astrometric distance measured by {\it Gaia} EDR3, 
of $175.73^{+1.53}_{-1.58}$ pc. 

Because {\sl TESS} has detected a single eigenfrequency in PG~1351 ($\nu \sim 2044~\mu$ Hz), we have not been able to confirm (nor discard) the rotation 
period of 8.9 h derived by  \cite{2011MNRAS.415.1220R}.

\subsection{EC~20058$-$5234}
\label{EC20058}

\begin{figure} 
\includegraphics[clip,width=1.0\columnwidth]{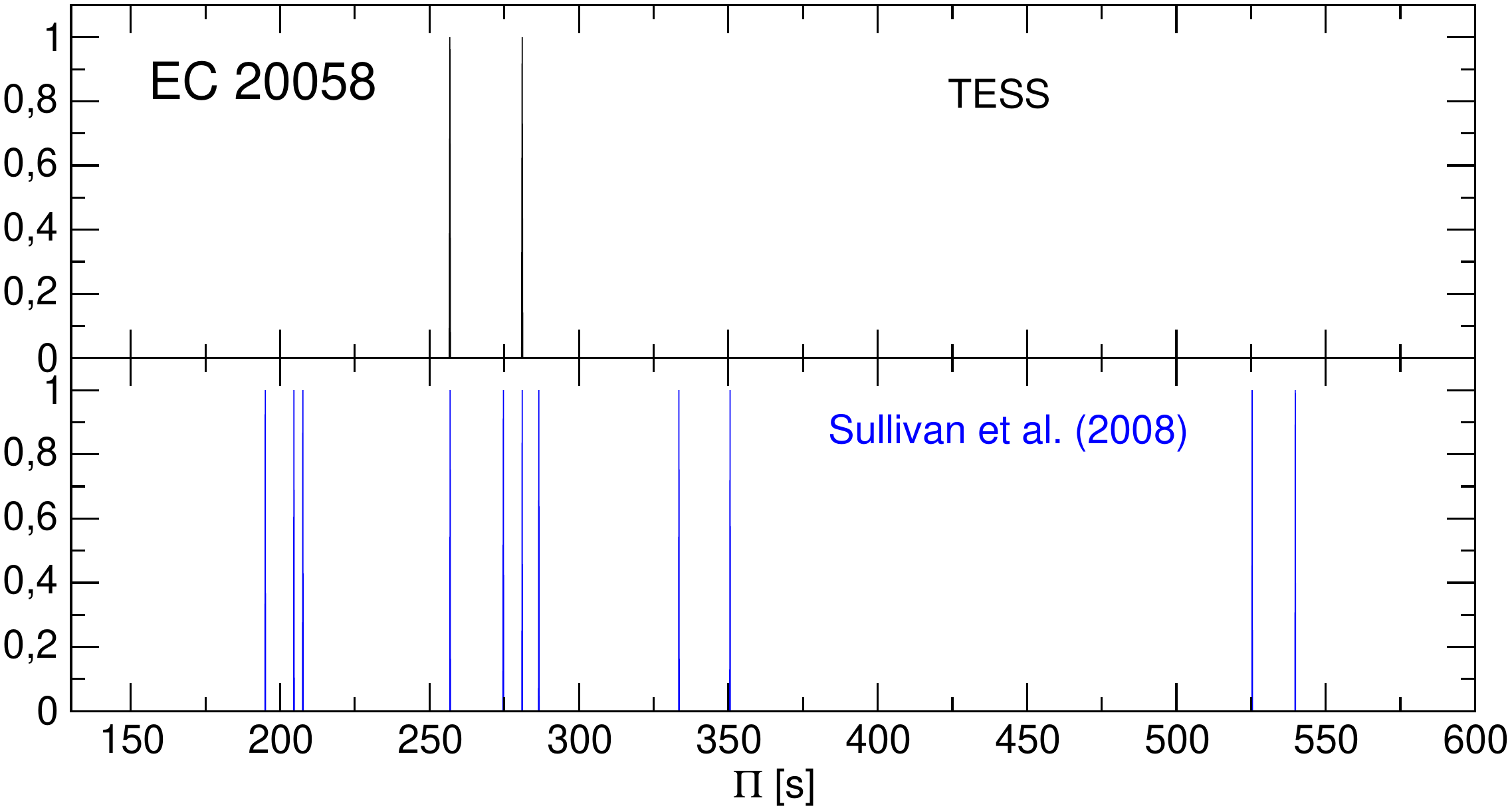}
\caption{Schematic distribution of the pulsation periods of EC~20058
  according  to {\sl TESS} (2 periods, black lines, upper panel), and
  according to \cite{2008MNRAS.387..137S}  (11 periods, blue lines,
  lower panel). The amplitudes have been arbitrarily  set to one for
  clarity.}
\label{fig:compara-tess-sea08} 
\end{figure}

Similar to the case of PG~1351, the pulsation spectrum of EC~20058  
detected by {\sl TESS} is not rich. Indeed, {\sl TESS}  
  detected only two large-amplitude periods, compared to a total of 11 periods   
  measured with observations from the ground by  \cite{2008MNRAS.387..137S} (see Table 
  \ref{table:EC20058-extended}). In Fig.  \ref{fig:compara-tess-sea08} we 
schematically show the two periods detected with {\sl TESS} (upper
panel), and the periods detected by \cite{2008MNRAS.387..137S} (lower
panel). For our analysis, in the case of the two periods that 
are common to both sets, we adopt those detected by {\sl TESS}, as they 
are more accurate. 

\begin{table*}
\centering
\caption{Enlarged list of periods of EC~20058. Column 1 corresponds to 11 $\ell= 1$
periods derived by \cite{2008MNRAS.387..137S} \citep[SEA08; see also Table 1 of][]{2011MNRAS.414..404B}, 
and column 2 corresponds to  two periods detected by {\sl TESS}   
  (Table \ref{table:EC20058}).  The periods with one (two) asterisk(s) are  
   used in a linear least square fit for modes with $\ell= 1$ ($\ell= 2$).}
\begin{tabular}{lc|ccccc}
\hline
\noalign{\smallskip}
$\Pi_i^{\rm O}$ (s) & $\Pi_{i}^{\rm O}$ (s) & $\Pi_{\rm fit}^{\ell= 1}$ (s) & $\delta\Pi^{\ell= 1}$ (s) & $\Pi_{\rm fit}^{\ell= 2}$ (s) & $\delta\Pi^{\ell= 2}$ (s) & $\ell^{\rm O}$ \\
  SEA08 & {\sl TESS} & & & & \\
\noalign{\smallskip}
\hline
\noalign{\smallskip}       
195.0(*)  &             & 193.662 & 1.338     &          &          & 1  \\ 
204.6     &             &         &           &          &          & ?  \\
207.6(**) &             &         &           & 208.572  & $-0.972$ & 2  \\
256.9     & 256.852(**) &         &           & 256.162  & 0.69     & 2  \\
274.7(*)  &             & 276.485 & $-1.785$  &          &          & 1  \\
281.0     & 280.983(**) &         &           & 279.957  & 1.026    & 2  \\
286.6     &             &         &           &          &          & ?  \\
333.5     &             &         &           &          &          & ?  \\
350.6(**) &             &         &           & 351.342  & $-0.742$ & 2  \\
525.4(*)  &             & 524.954 & 0.446     &          &          & 1  \\
539.8     &             &         &           &          &          & ?  \\
\noalign{\smallskip}
\hline
\end{tabular}
\label{table:EC20058-extended}
\end{table*}

\begin{figure} 
\includegraphics[clip,width=1.0\columnwidth]{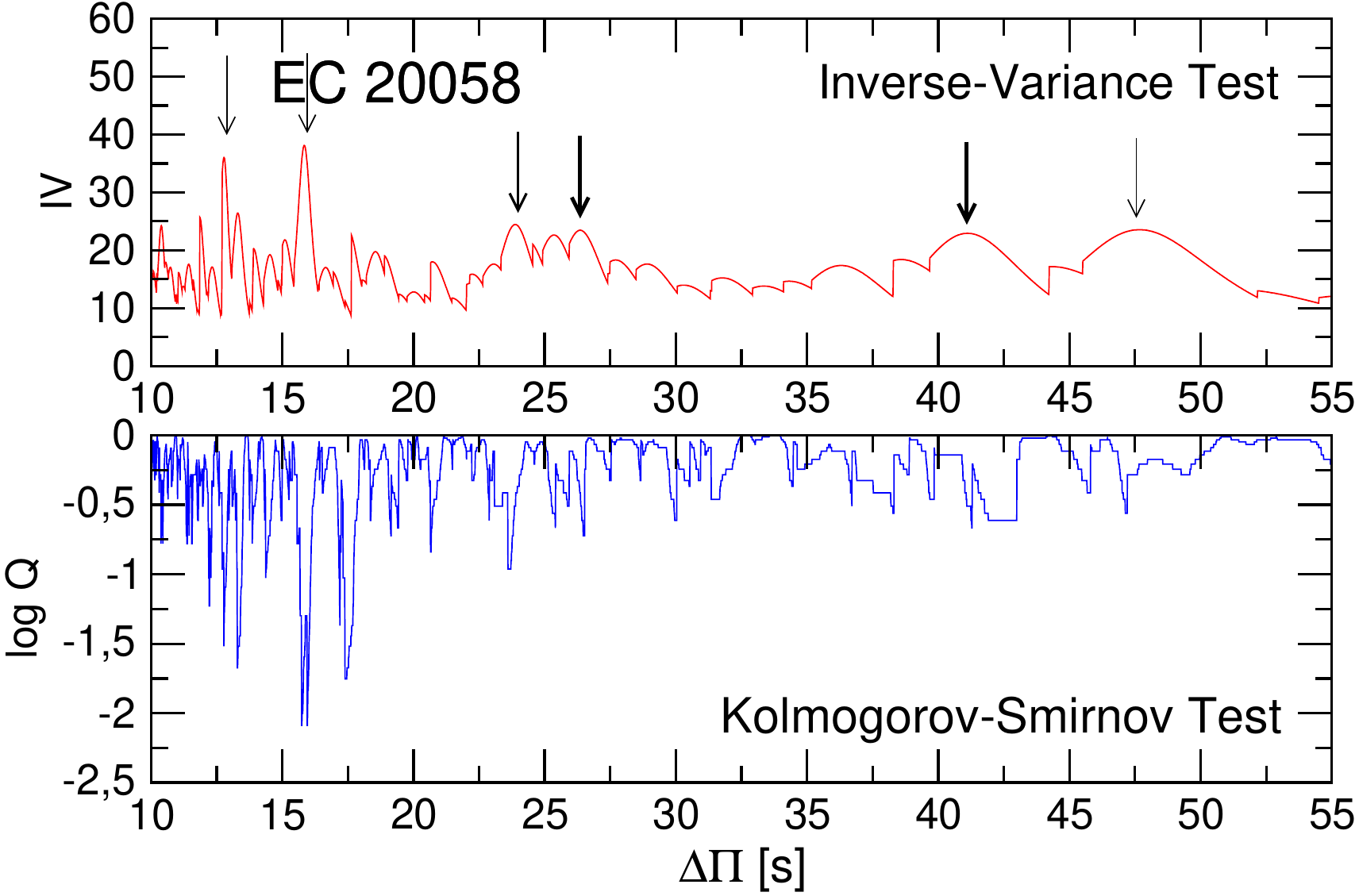}
\caption{I-V  (upper panel)  and  K-S  (lower panel)  
significance  tests  to  search  for  a constant  period
  spacing  in EC~20058. The tests are applied to the set of 11
  pulsation periods of Table \ref{table:EC20058-extended}, that includes 
  the {\sl TESS} plus   \cite{2008MNRAS.387..137S}'s periods. Possible
 values of constant period spacing are marked  with arrows in the upper panel.}
\label{fig:tests-EC20058} 
\end{figure}

\begin{figure} 
\includegraphics[clip,width=1.0\columnwidth]{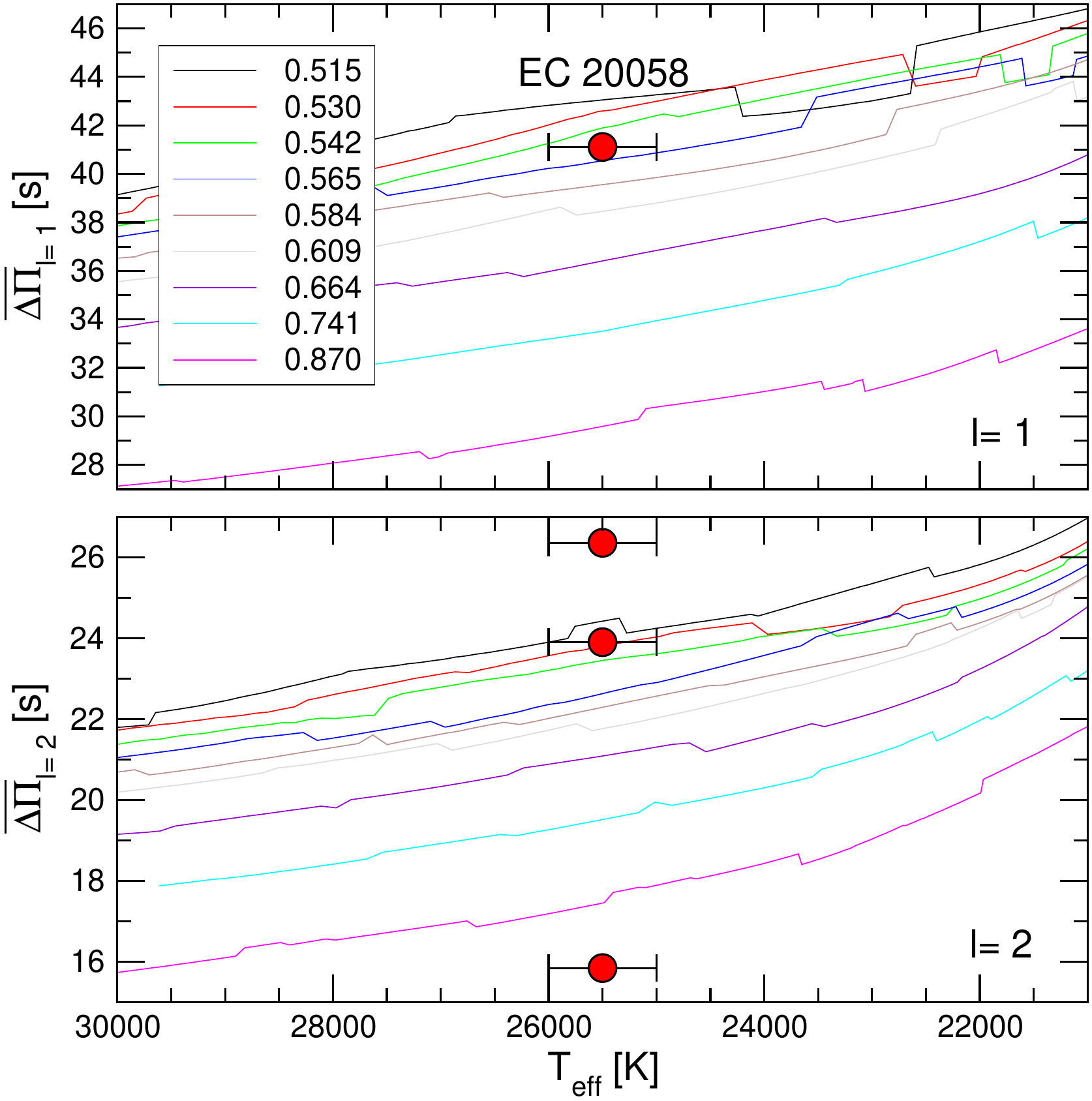}
\caption{Dipole ($\ell= 1$, upper panel) and quadrupole ($\ell= 2$, lower panel) average 
of the computed period spacings,
  $\overline{\Delta \Pi_k}$, assessed  in  a  range  of  periods  that
  includes  the  periods  observed  in EC~20058, shown as curves 
  of different colors according to the different stellar masses. 
  We consider the effective temperature $T_{\rm eff}= 25\,500\pm 500$~K 
  \citep{2014A&A...568A.118K}, and the possible period spacing
  values $\Delta \Pi= 12.78, 15.84, 23.90, 26.36, 41.11, 47.67$ s 
  (see Fig. \ref{fig:tests-EC20058}). For a discussion, see the text.}
\label{fig:psp-teff-EC20058} 
\end{figure}

In Fig.~\ref{fig:tests-EC20058} we display the results of applying the
statistical tests to the set of 11 periods of the first column of Table
\ref{table:EC20058-extended}. 
The tests show hints of several possible constant 
period spacings, $\Delta \Pi= 12.78, 15.84, 23.90, 26.36, 41.11, 47.67$~s 
(emphasized with vertical arrows in the figure), 
suggesting that no simple pattern of period spacing is apparent in the EC~20058
spectrum, in line with previous works \citep{2008MNRAS.387..137S}. 
Fig.~\ref{fig:psp-teff-EC20058} depicts  $\overline{\Delta
  \Pi_{k}}$ (computed considering a period interval of $200-600$ s) 
  with curves of different colors to distinguish the different stellar 
  masses, corresponding to $\ell= 1$ (upper panel) and $\ell= 2$  
  (lower panel) modes. Based on this figure, the possible extreme values for 
  constant period spacings, that is 12.78~s and 47.67~s, can be discarded from the analysis
because they imply a mass too high or extremely low, respectively. The remaining four
values could actually be constant period spacings, and they are plotted in  Fig.~\ref{fig:psp-teff-EC20058} with red  circles. Concentrating on 
the upper panel (dipole modes), we see that a 
possible spacing of periods of $41.11$ s s would indicate a stellar mass of $\sim 0.55 M_{\odot}$ if it corresponds to modes with $\ell= 1$. 
Regarding the lower panel ($\ell= 2$), we can discard the 
possible period spacing of $\sim 26$ s because it would indicate a very low mass for EC~20058 ($\lesssim 0.4 M_{\odot}$). The possible solutions that remain indicate a stellar mass of $\sim 0.530 M_{\odot}$ if $\Delta \Pi= 23.90$ s, or a stellar mass $\sim 0.95 M_{\odot}$ if $\Delta \Pi= 15.84$ s. We note that a stellar mass as high as $\sim 0.95 M_{\odot}$ is not compatible with the spectroscopic mass of this star, $\sim 0.61 M_{\odot}$. On the other hand, such massive DB WDs are very unusual. For these reasons, we discard the possible solution of $\sim 16$ s for the period spacing. Then, we are left with the possible period spacing of $23.90$ s and $\ell= 2$. 

\begin{figure} 
\includegraphics[clip,width=1.0\columnwidth]{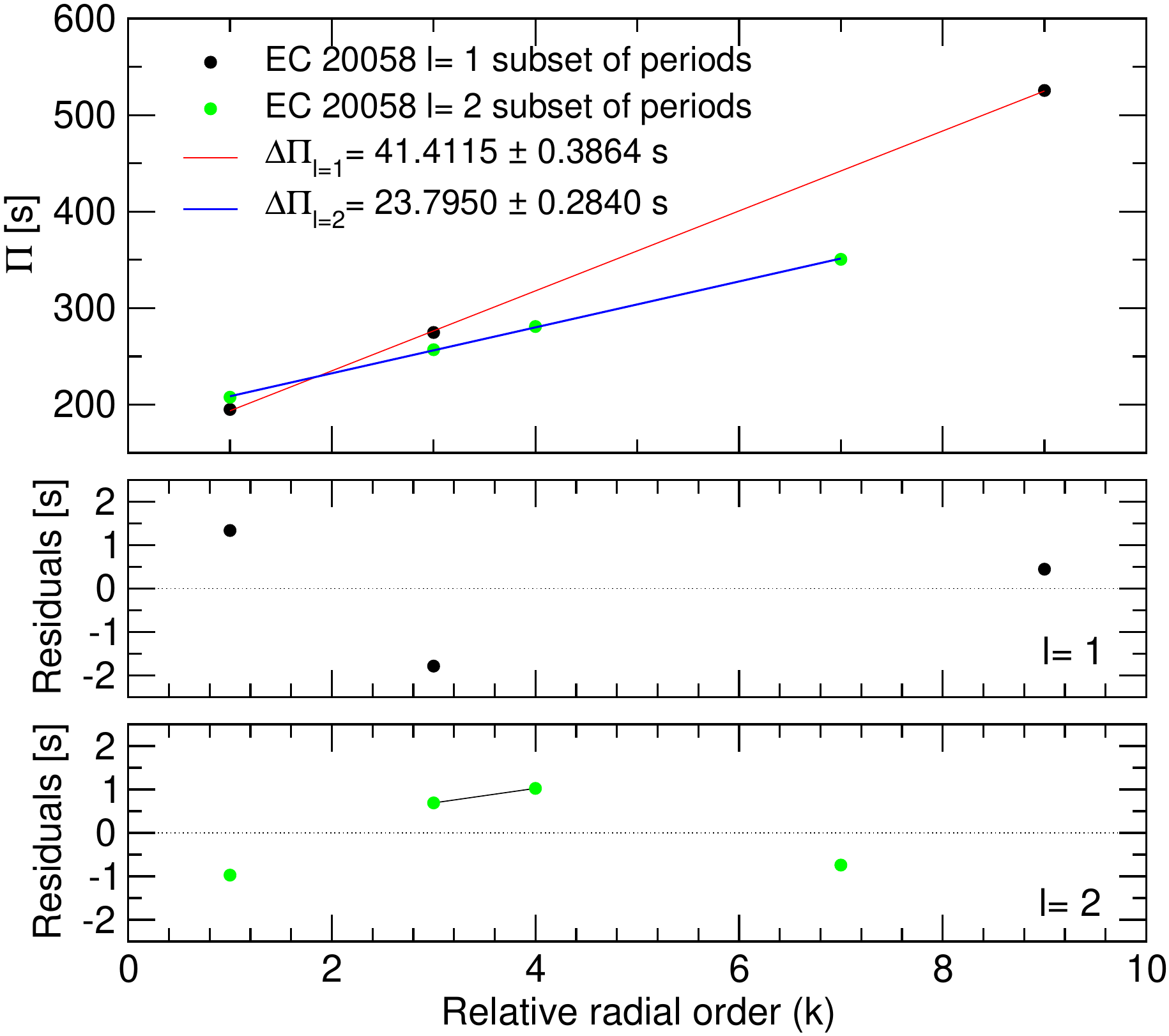}
\caption{Upper panel: linear least-squares fits to the three dipole periods of  EC~20058 marked  
with  an  asterisk  and  the  four  quadrupole  periods marked with two asterisks in Table \ref{table:EC20058-extended}. The derived period spacings from these fits are
$\Delta \Pi_{\ell=1}= 41.4115 \pm 0.3864 $~s and $\Delta \Pi_{\ell=1}= 23.795 \pm 0.284 $~s, 
respectively. Middle panel: residuals of the $\ell=1$ period distribution relative 
to the dipole mean period spacing. Lower panel: residuals of the $\ell=2$ period 
distribution relative to the quadrupole mean period spacing. Modes with consecutive 
radial order are connected with a thin black line.}
\label{fig:fit-EC20058} 
\end{figure}

In summary, the two most probable period spacings present in the star 
are $\Delta \Pi_{\ell= 1}= 41.11$ s and $\Delta \Pi_{\ell= 2}= 23.90$s.  
We immediately note that $41.11/23.90= 1.72$, which is very close to the asymptotic 
prediction, $\Delta \Pi_{\ell= 1}/\Delta \Pi_{\ell= 2}= \sqrt{3} \sim 1.732$ 
\citep{1990ApJS...72..335T}. While it is appealing to conclude that the pulsational spectrum of EC~20058 is composed of two patterns of periods, one with $\ell = 1$ 
and the other with $\ell = 2$, it is not trivial to explain why both patterns 
suggest different stellar masses ($\sim 0.53$ and $\sim 0.55 M_{\odot}$).  
However,  this small discrepancy  could be alleviated if  more realistic  
uncertainties  in the  effective temperature  were adopted. Beyond this, the fact that the ratio of the period spacings is so close to that predicted by the asymptotic theory is quite encouraging, 
and prompts us to  try to identify the harmonic degree of each period of 
EC~20058. To determine the precise period spacings, we performed linear least-squares 
fits (plotted in the upper panel of Fig. \ref{fig:fit-EC20058}) 
using three periods 
for the case of $\ell= 1$, and four periods for the case of $\ell= 2$,  marked
with a single asterisk and two asterisks, respectively, 
in Table~\ref{table:EC20058-extended}.  We obtain a dipole period spacing 
$\Delta \Pi_{\ell=1}= 41.41 \pm 0.39 $~s and  a quadrupole period spacing 
$\Delta \Pi_{\ell=2}= 23.80 \pm 0.28 $~s. We have four remaining periods 
(204.6 s, 286.6 s, 333.5 s, and 539.8 s) that do not fit in any of the 
two period spacing patterns found. They  could have harmonic degree 
$\ell= 1$ and/or $\ell= 2$ and the reason for their deviation from the spacings derived for both $\ell = 1$ and $\ell = 2$ modes could be that they are affected by mode trapping \citep[see, e.g.,][]{1993ApJ...406..661B}.

We investigated how the results of the K-S test change
with variations in the input frequency list, that is, excluding different periods, one at a time, 
and ran the test again.  We have found that the $\sim 24$ sec period  spacing 
is  very robust, since, in general,  the corresponding minimum in the  
K-S test becomes steeper  or stays  the same  when one  ignores each  period.  
On  the  other  hand,  the  $\sim 41$ sec  period spacing  is  much  more
sensitive to the specific list of periods on which the test is computed.

A completely different assumption than the one described above 
is that the period spectrum of EC~20058 has a combination of 
$\ell= 1$ and $\ell= 2$ modes, but with the possible presence of 
rotational frequency multiplets. It has been the hypothesis 
favored by \cite{2008MNRAS.387..137S} (see their Fig. 8). In this scenario, 
the frequency $4816.8~\mu$Hz (period  207.6 s) could be one component of a rotational 
triplet centered at $4887.8~\mu$Hz (204.6 s), the other component being inhibited by an unknown mechanism, and the frequencies $3640.1~\mu$Hz
(274.7 s) and $3489.0~\mu$Hz (286.6 s) could be the two non-zero $m$ components of a rotational 
triplet centered at  $3558.935~\mu$Hz (280.983 s). Implicitly, the periods 204.6 s and 280.983
are assumed here to be $\ell= 1$. Using a mean frequency splitting of $\sim 70~\mu$Hz and the 
asymmetry observed in the frequency splitting of the triplet centered at the frequency 
$3558.935~\mu$Hz, \cite{2008MNRAS.387..137S} derived a rotation period of $\sim 2$ hours and
a magnetic field of $\sim 3$ kG for EC~20058. Under the present hypothesis, it is 
interesting to look for a constant period spacing with the statistical tests, this time 
neglecting the periods 207.6 s, 274.7 s, and 286.6 s. Our exercise does not indicate any 
clear constant period spacing in this reduced set of periods. The non-existence of a 
constant period spacing could be explained by the fact that some of the periods exhibited 
by the star are not in the asymptotic regime. 

Putting it all together, we face two possible scenarios for EC~20058. According to one of them, 
we are seeing a combination of ($m= 0$) $\ell= 1$ and $\ell= 2$ excited modes, that conform 
two patterns of constant period spacing. In this scenario, no frequency 
multiplets due to rotation are present. In this case, the assignment of the $\ell$ 
value for some of the periods is that of Table \ref{table:EC20058-extended} (column 7). 
In addition, following \cite{2008MNRAS.387..137S}, we can assume the existence of 
rotational triplets, in this way identifying the periods 204.6 s and 280.983 s 
with $m= 0$ $\ell= 1$ modes. We cannot rule out or accept conclusively 
either of these two scenarios. Thus, when performing the period-to-period 
fits (see below), we  have to consider both possibilities for the identification of 
the harmonic degree of the modes. 

\begin{table*}
\centering
\caption{Observed and theoretical periods of the asteroseismological
  solution for EC~20058 in the case 1, using 11 periods [$M_{\star}= 0.664 M_{\odot}$, 
  $T_{\rm eff}=26\,380$ K]. Periods are in
  seconds  and rates of period change  (theoretical) are  in units of
  $10^{-13}$ s/s. $\delta \Pi_i= \Pi^{\rm O}_i-\Pi_k$ represents  the
  period differences, $\ell$ the harmonic degree, $k$ the radial
  order, $m$ the azimuthal index.  The last column gives information
  about the pulsational stability/instability   nature  of  the
  modes.}
\begin{tabular}{lc|ccccccc}
\hline
\noalign{\smallskip}
$\Pi_i^{\rm O}$ & $\ell^{\rm O}$ & $\Pi_k$ & $\ell$ & $k$ & $m$ & $\delta \Pi_k$ & $\dot{\Pi}_k$ & 
Unstable\\ 
(s) & & (s) & & & & (s) & ($10^{-13}$ s/s) &  \\
\noalign{\smallskip}
\hline
\noalign{\smallskip}       
195.0   & 1  & 193.39  & 2 &  6  &  0   &  1.61   & 0.40   & yes \\ 
204.6   & ?  & 204.19  & 1 &  3  &  0   &  0.41   & 0.44   & yes \\
207.6   & 2  & 206.45  & 2 &  7  &  0   &  1.15   & 0.47   & yes \\
256.852 & 2  & 248.42  & 2 &  9  &  0   &  8.43   & 0.60   & yes \\
274.7   & 1  & 275.87  & 1 &  5  &  0   & $-1.17$ & 0.53 & yes \\
280.983 & 2  & 275.87  & 1 &  5  &  0   &  5.11   & 0.53   & yes \\
286.6   & ?  & 293.03  & 2 &  11 &  0   & $-6.43$ & 0.69 & yes \\
333.5   & ?  & 332.28  & 2 &  13 &  0   &  1.22   & 0.750   & yes \\
350.6   & 2  & 352.48  & 2 &  14 &  0   & $-1.88$ & 0.62 & yes \\
525.4   & 1  & 524.05  & 1 &  12 &  0   &  1.35   & 1.11  & yes \\
539.8   & ?  & 538.01  & 2 &  23 &  0   &  1.79   & 1.24  & yes \\
\noalign{\smallskip}
\hline
\end{tabular}
\label{table:EC20058-asteroseismic-model-1}
\end{table*}

\begin{table*}
\centering
\caption{Same as Table \ref{table:EC20058-asteroseismic-model-1}, but for case 2, in which eight periods are employed in the period fit. The possible asteroseismological solution for EC~20058 corresponds in this case to a DB WD model with  $M_{\star}= 0.664 M_{\odot}$, $T_{\rm eff}= 25\,467$ K.}
\begin{tabular}{lc|ccccccc}
\hline
\noalign{\smallskip}
$\Pi_i^{\rm O}$ & $\ell^{\rm O}$ & $\Pi_k$ & $\ell$ & $k$ & $m$ & $\delta \Pi_k$ & $\dot{\Pi}_k$ & 
Unstable\\ 
(s) & & (s) & & & & (s) & ($10^{-13}$ s/s) &  \\
\noalign{\smallskip}
\hline
\noalign{\smallskip}       
195.0   & ? & 196.99 & 2 &  6  & 0  &  $-1.99$ &  0.29 & yes \\ 
204.6   & 1 & 208.24 & 1 &  3  & 0  &  $-3.64$ &  0.33 & yes \\ 
256.852 & ? & 254.22 & 2 &  9  & 0  &     2.63 &  0.48 & yes \\
280.983 & 1 & 280.94 & 1 &  5  & 0  &     0.04 &  0.42 & yes \\
333.5   & ? & 339.06 & 2 & 13  & 0  &  $-5.56$ &  0.54 & yes \\
350.6   & ? & 349.46 & 1 &  7  & 0  &     1.14 &  0.43 & yes \\
525.4   & ? & 526.18 & 2 & 22  & 0  &  $-0.78$ &  0.92 & yes \\
539.8   & ? & 534.71 & 1 & 12  & 0  &     5.09 &  0.88 & yes \\
\noalign{\smallskip}
\hline
\end{tabular}
\label{table:EC20058-asteroseismic-model-2}
\end{table*}

\begin{figure} 
\includegraphics[clip,width= 1.0\columnwidth]{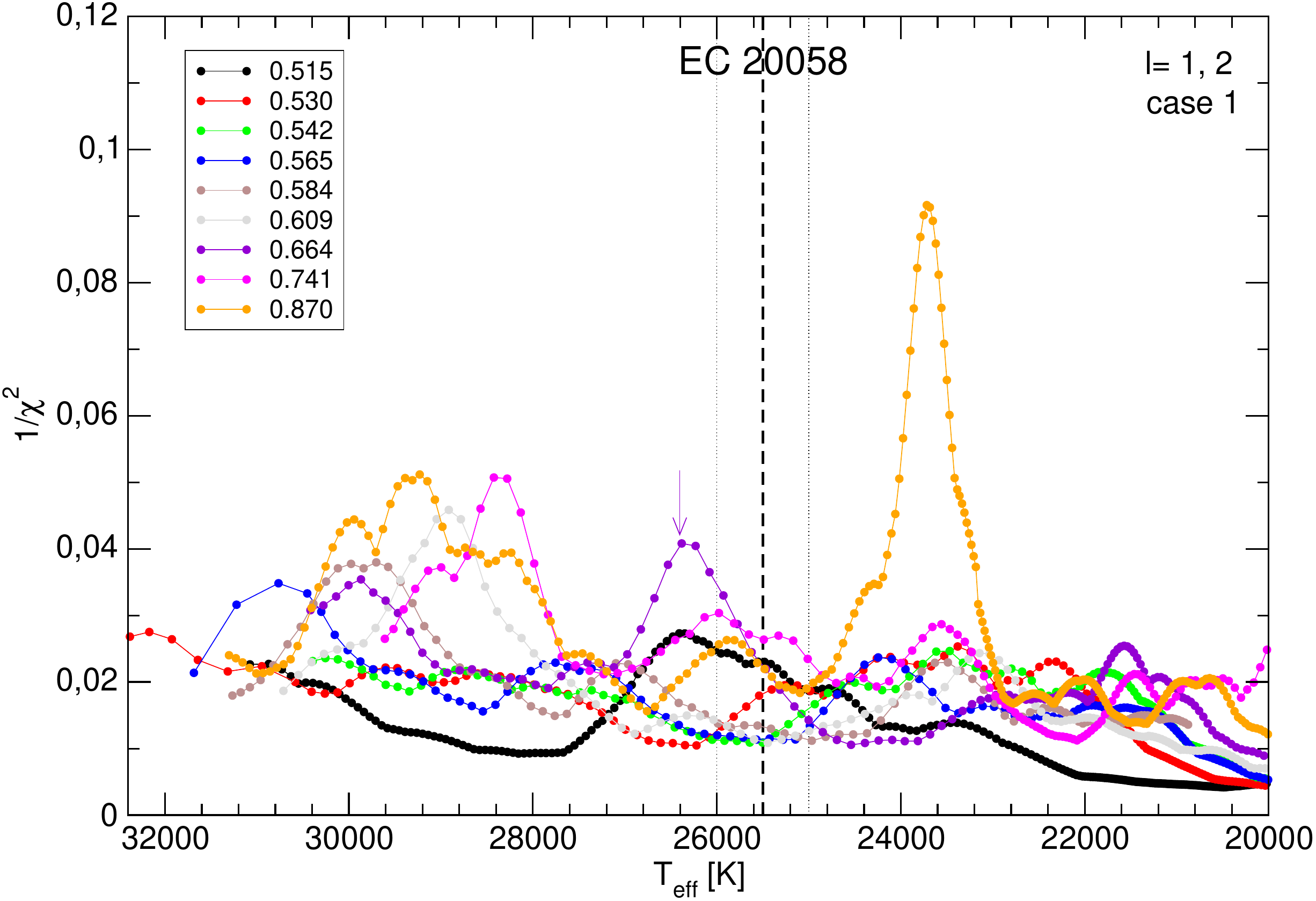}
\caption{The inverse of the quality function of the period fit in terms of 
the effective temperature, shown with different colors for the different 
stellar masses, corresponding to case 1 (11 periods). The vertical black dashed line corresponds to 
the spectroscopic $T_{\rm eff}$ of EC~20058 and the vertical dotted lines 
its uncertainties \citep[$T_{\rm eff}= 25\,500\pm 500$ K;][]{2014A&A...568A.118K}. A local maxima (marked with an arrow), corresponds to a possible asteroseismological solution compatible with spectroscopy 
(see the text).}
\label{chi2-EC20058-caso1} 
\end{figure}

\begin{figure} 
\includegraphics[clip,width= 1.0\columnwidth]{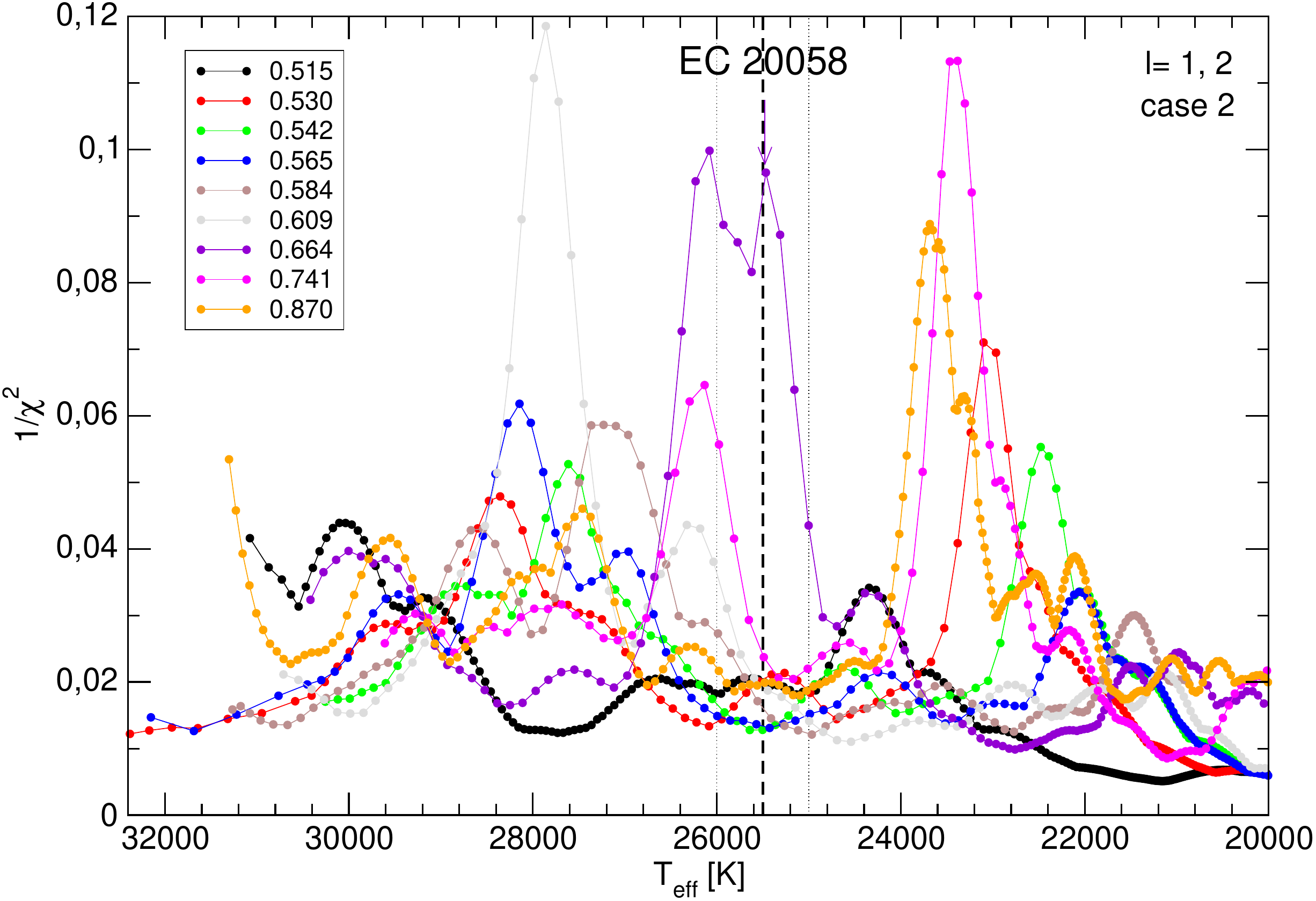}
\caption{Same as in Fig. \ref{chi2-EC20058-caso1}, but for case 2, which involves eight periods. A clear maxima for a model with $T_{\rm eff}= 26\,080$ K and $M_{\star}= 0.664 M_{\sun}$, compatible with 
the spectroscopic effective temperature of EC~20058, 
is marked with an arrow (see the text).}
\label{chi2-EC20058-caso2} 
\end{figure}

In the second part of our analysis of EC~20058, we have performed period fits considering two sets of periods and 
identifications. One of them corresponds to that indicated in Table 
\ref{table:EC20058-asteroseismic-model-1} (case 1), in which we follow the 
$\ell$ identification derived from the period spacing analysis (see Table \ref{table:EC20058-extended}), and the other one 
to that shown  in Table \ref{table:EC20058-asteroseismic-model-2} (case 2), 
that corresponds to the identification assumed by 
\cite{2008MNRAS.387..137S}. We show our results in 
Figs. \ref{chi2-EC20058-caso1}  and \ref{chi2-EC20058-caso2}, 
where we display the inverse of the quality function for case 1 and case 2, 
respectively. Focusing on case 1, there is no good possible seismological 
solution in the range of effective 
temperatures allowed by spectroscopy, so we can adopt a relatively good fit 
model with $T_{\rm eff}= 26\,380$ K (slightly outside the allowed range of 
$T_{\rm eff}$) and $ M_{\star}= 0.664 M_{\odot}$, marked with a down arrow in Fig. \ref{chi2-EC20058-caso1}. 
The non-existence of a defined and clear seismological model in this case 
would indicate that the identification of the modes established in Table 
\ref{table:EC20058-extended} based on possible period spacings  with 
$\ell= 1$ and $\ell= 2$ could not be robust. In case 2, however, it 
is possible to find a small family of possible seismological solutions consisting of models with $M_{\star}= 0.664 M_{\odot}$ and $T_{\rm eff}$ in the range $25\,000-26\,000$ K. 
In particular, there is a good asteroseismological solution for a model with 
$T_{\rm eff}= 25\,467$ K (marked with a down arrow in Fig. \ref{chi2-EC20058-caso2}), well within the allowed range. Curiously, the mass of this seismological model is the same as that of the model with good period fit in case 1, although with a much lower effective temperature and in excellent agreement with the spectroscopic $T_{\rm eff}$ of EC~20058 ($T_{\rm eff}= 25\,500\pm500$ K). The theoretical periods of the models  selected for case 1 and case 2 are
included in Tables \ref{table:EC20058-asteroseismic-model-1} and \ref{table:EC20058-asteroseismic-model-2}, along with the harmonic degree, the radial order, the difference with the observed periods, the rates of period change, and the pulsational stability nature of the modes. In particular, our identification of the harmonic degree and radial order
of the modes for the asteroseismological solution of case 2 is  remarkably similar to the \cite{2008MNRAS.387..137S}'s identification (see their Table 5), even though those authors used DB WD models completely different than ours. Similar considerations hold when we compare the results of our case 2 with the results of \cite{2011MNRAS.414..404B}, even when they consider higher effective temperatures for EC~20058. A similar overall agreement between seismological results derived using the La Plata group evolution/pulsation models (using the {\tt LPCODE} and {\tt LP-PUL} numerical codes) and those from the Texas group (using the {\tt WDEC} code) has been found in the case of the DBV pulsator WD~0158$-$160  (TIC~257459955), the first DBV studied with {\sl TESS}
\citep{2019A&A...632A..42B}. This agreement is encouraging, since the global asteroseismological results are not severely dependent on the WD modeling and period-fit methods used, which makes this type of analysis more robust.

We have also carried out period fits assuming that all the modes exhibited by EC~20058 are $\ell= 1$, but we have not found any possible seismological solution.
Indeed, in general it is not possible with our models to reproduce the periods satisfactorily if they are considered as $\ell= 1$ only. This confirms that the pulsation spectrum of EC~20058 must be a mixture of dipole and quadrupole modes. Finally, we have tried leaving the identification free of all the modes (allowing them to be $\ell= 1$ or $\ell= 2$ from the outset). In this case, we find the same asteroseismological model as in case 2, although 
with a slightly poorer agreement between observed and theoretical periods.  
 
In view of this analysis, we adopt the DB WD model with $M_{\star}= 0.664 M_{\odot}$ and $T_{\rm eff}= 25\,467$ K, that corresponds to the best fit of periods of the case 2 (see Table \ref{table:EC20058-asteroseismic-model-2}), as the asteroseismological model 
for EC~20058.  We obtain $\overline{\delta \Pi_i}= 2.61$ s, $\sigma= 3.22$ s, and BIC= 1.24. We show the characteristics of this model in Table  \ref{astero-EC20058}. The 
internal chemical profiles and the Brunt-V\"ais\"al\"a and Lamb frequencies of this model are quite similar to those corresponding 
to the asteroseismological model of PG~1351 (Fig. \ref{PG1351-modelo}). Indeed, the only difference 
between both models is the effective temperature ($\Delta T_{\rm eff} \sim 300$ K), which has a negligible impact on the chemical profiles and critical frequencies. Thus, Fig. \ref{PG1351-modelo} is also representative of the seismological model of EC 20058.  We mention that the 
possible asteroseismological solution for case 1 is characterized 
by $\overline{\delta \Pi_i}= 2.77$ s, $\sigma= 3.74$ s, and BIC= 1.34, which reflects that the period fit is worse than that of the adopted asteroseismological model (case 2). In addition, the model associated to the possible solution for 
case 1 is outside the range of the effective temperature allowed by spectroscopy. 

Regarding the stability/instability of the modes of the asteroseismological model, 
we find that all the periods of EC~20058 are predicted to be excited  according to our nonadiabatic computations, in line with the very existence of these oscillation periods in the spectrum of EC~20058 (see last column of Table \ref{table:EC20058-asteroseismic-model-2}). The same happens  
with regard to the possible seismological solution in case 1 (see last column of Table \ref{table:EC20058-asteroseismic-model-1}). 

Based on the asteroseismological model for EC~20058, we derive a seismological distance, as we did for PG~1351.  We find a magnitude absolute in the {\it Gaia} photometry of $M_G= 10.518$ mag 
using a DB WD model atmosphere. Using the observed 
{\it Gaia} magnitude, $m_{G}= 15.756$ mag, we obtained  
$d_{\rm s}= 111.60^{+1.53}_{-1.50}$ pc. The seismological distance is $\sim 5 \%$ lower than the astrometric distance measured by {\it Gaia} EDR3, of $117.95^{+0.68}_{-0.75}$ pc, similar to we 
found for PG~1351.

\begin{table}
\centering
\caption{The main characteristics of the DBV star EC~20058 according to the adopted asteroseismological model (case 2).} 
\begin{tabular}{l|cc}
\hline
\hline 
Quantity & Spectroscopy &  Asteroseismology \\
         & Astrometry   &         \\ 
\hline
$T_{\rm eff}$ [K]                           & $25\,500 \pm 500$ & $25\,467\pm130$       \\
$M_{\star}$ [$M_{\odot}$]                   & $0.614\pm0.030$    & $0.664\pm0.013$     \\ 
$\log g$ [cm/s$^2$]                         & $8.01\pm0.05$      & $8.062\pm 0.020$     \\ 
$\log (L_{\star}/L_{\odot})$                & $\hdots$           & $-1.223\pm0.03$     \\  
$\log(R_{\star}/R_{\odot})$                 & $\hdots$           & $-1.901\pm 0.015$     \\  
$(X_{\rm C}, X_{\rm O})_{\rm c}$            & $\hdots$           & $0.32, 0.65$        \\
$M_{\rm He}/M_{\star}$                      & $\hdots$           & $5.42\times10^{-3}$ \\
$d$  [pc]                                   & $117.95^{+0.68 (a)}_{-0.75}$ & $111.60^{+1.53}_{-1.50}$ \\ 
$\pi$ [mas]                                 & $8.48\pm0.05^{(a)}$       & $8.96\pm 0.12$ \\ 
\hline
\hline
\end{tabular}
\label{astero-EC20058}
{\footnotesize  References: (a) Gaia DR3.}
\end{table}

\subsection{EC~04207$-$4748}
\label{EC04207}

\begin{table}
\centering
\caption{Enlarged list of periods of EC~04207. Column 1 corresponds to four 
periods derived by \cite{2013MNRAS.431..520C}, and column 2 corresponds to  
the periods detected by {\sl TESS} (Table \ref{table:EC04207}). In the case 
of the couple of periods of {\sl TESS} of 423.882 s and 423.427 s, we adopt a 
single value of 423.655 s (see the text).}
\begin{tabular}{lc|ccc}
\hline
\noalign{\smallskip}
$\Pi_i^{\rm O}$ (s) & $\Pi_{i}^{\rm O}$ (s) & $\Pi_{\rm fit}$ (s) & $\delta\Pi$ (s) & $\ell^{\rm O}$ \\
  CEA13 & {\sl TESS} & & & \\
\noalign{\smallskip}
\hline
\noalign{\smallskip}       
336.4     & 336.397     & 335.035 & $1.362$  &  1 \\ 
423.5     & 423.655     &         &          &  ? \\ 
447.2     & 447.194     & 448.116 & $-0.922$ &  1 \\ 
599.1     &             & 598.891 & $0.209$  &  1 \\ 
\noalign{\smallskip}
\hline
\end{tabular}
\label{table:EC04207-extended}
\end{table}

\begin{figure} 
\includegraphics[clip,width=1.0\columnwidth]{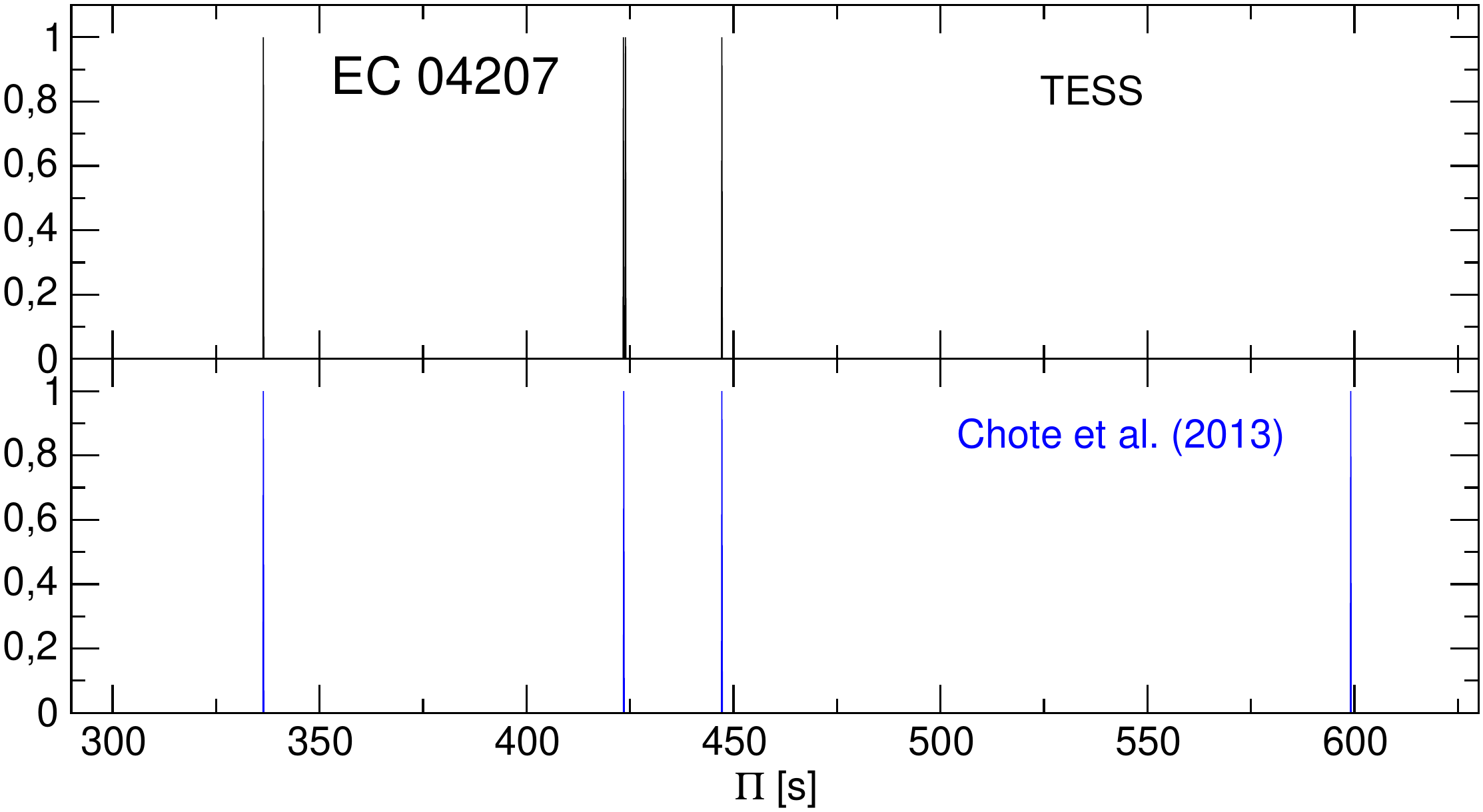}
\caption{Schematic distribution of the pulsation periods of EC~04207
  according  to {\sl TESS} (3 periods, black lines, upper panel), and
  according to \cite{2013MNRAS.431..520C}  (4 periods, blue lines,
  lower panel). Actually, the period of $\sim 423$ s from {\sl TESS} is 
a couple of very similar periods, of 423.882 s and 423.427 s.  
The amplitudes have been arbitrarily  set to one for clarity.}
\label{fig:compara-tess-chea11} 
\end{figure}

In Fig.  \ref{fig:compara-tess-chea11}, we 
schematically show for the DBV star EC~04207 the periods detected 
with {\sl TESS} (upper
panel), and the periods detected by \cite{2013MNRAS.431..520C} (lower
panel). There are three common periods which are detected in both data sets 
($\sim 336$ s, $\sim 423$ s, and $\sim 447$ s), and there is a period 
detected by \cite{2013MNRAS.431..520C} ($\sim 599$ s) 
but not by {\sl TESS}. In the case of the three periods that are common to 
both sets of data, we adopt for the asteroseismological analysis those 
detected by {\sl TESS} as 
they are more precise. In summary, we have four periods in total to 
carry out the asteroseismological analysis. In the {\sl TESS} data, 
the period of $\sim 423$ s is 
actually a pair of periods very close to each other, of 423.882 s and 423.427 s.
Assuming that they are the $m= -1$ and $m= +1$ component of a rotational triplet, 
the period corresponding to the central component ($m= 0$) is 423.655 s. 
We adopt this value of the period for this mode in the following 
analysis\footnote{We have also considered the possibility that the 
periods of 423.882 s and 423.427 s are the $m= 0$ and $m= \pm 1$ 
components of a rotational triplet. The 
results we obtain in these cases are exactly the same as if we assume 
that the periods correspond to the components $m= +1$ and $m= -1$.}.

\begin{figure} 
\includegraphics[clip,width=1.0\columnwidth]{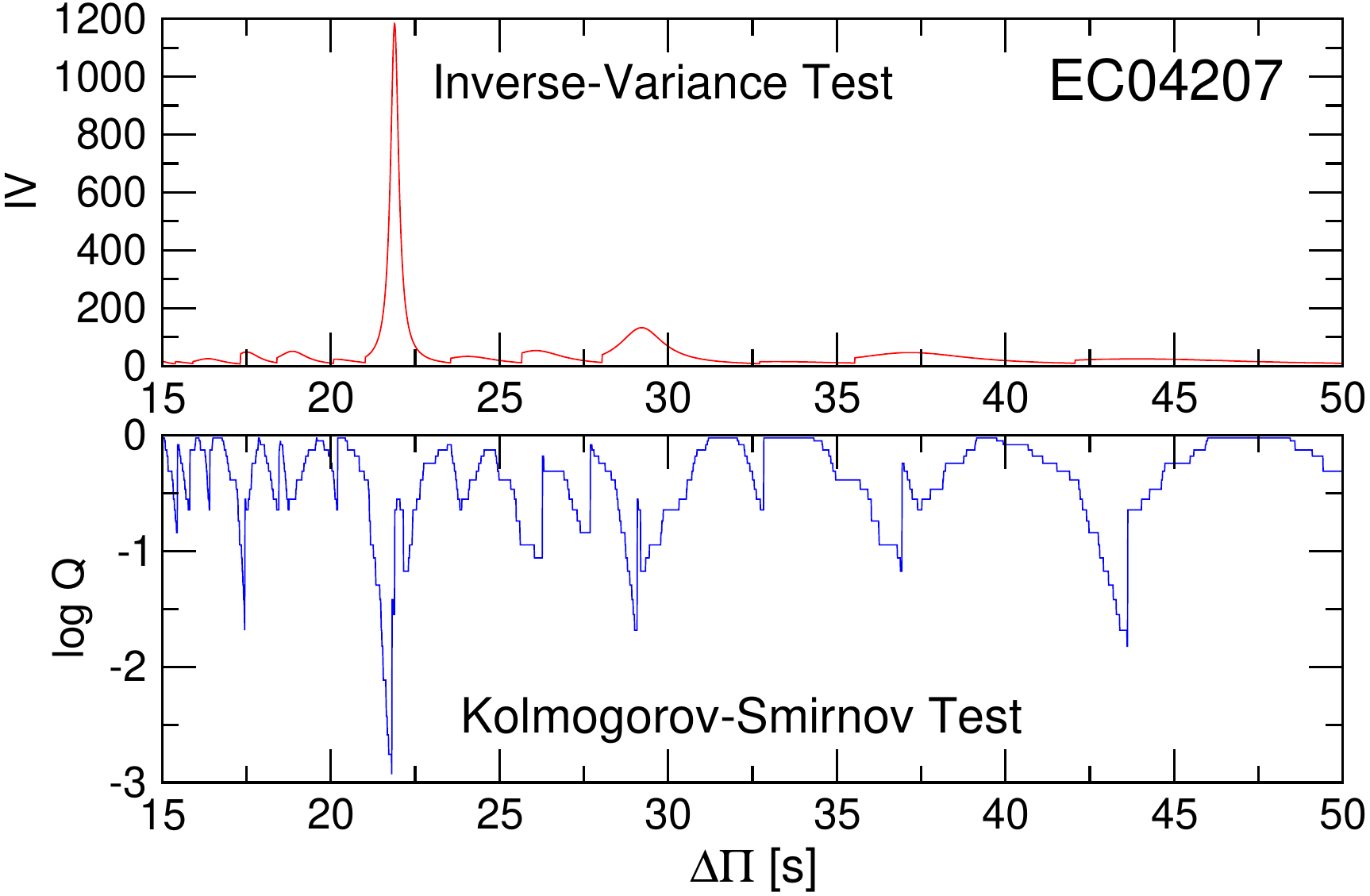}
\caption{I-V  (upper panel)  and  K-S  (lower panel)  
significance  tests  to  search  for  a constant  period
  spacing  in EC~04207. The tests are applied to the set of 4
  pulsation periods of Table \ref{table:EC04207-extended}, that includes 
  the {\sl TESS} plus   \cite{2013MNRAS.431..520C}'s periods. A period spacing 
  of $\sim 38$ s and its sub-harmonic of $\sim 19$ s are evident.}
\label{fig:tests-EC04207} 
\end{figure} 

  \begin{figure} 
\includegraphics[clip,width=1.0\columnwidth]{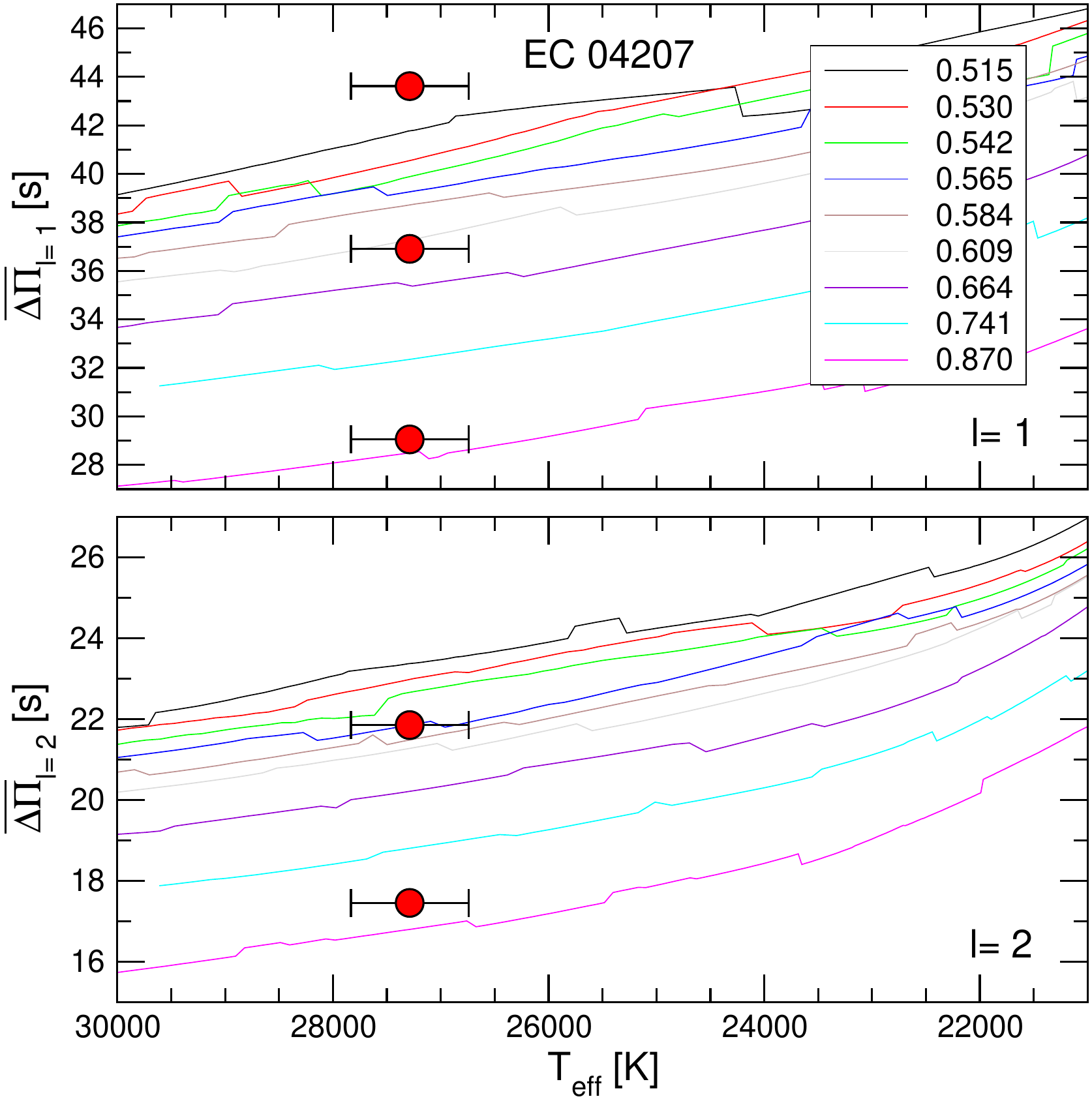}
\caption{Upper panel: dipole ($\ell= 1$) average of the computed period spacings,
  $\overline{\Delta \Pi_k}$, assessed  in  a  range  of  periods  that
  includes  the  periods  observed  in EC~04207, shown as curves 
  of different colors according to the different stellar masses. 
  We consider the effective temperature $T_{\rm eff}= 27\,288\pm 545$~K 
  \citep{2007A&A...470.1079V}. Lower panel: same as in upper panel, but for 
  quadrupole ($\ell= 2$) modes. The red circles correspond to different possible
  period spacings as predicted by the statistical tests (see Fig. \ref{fig:tests-EC04207}).}
\label{fig:psp-teff-EC04207} 
\end{figure}
  
\begin{figure} 
\includegraphics[clip,width= 1.0\columnwidth]{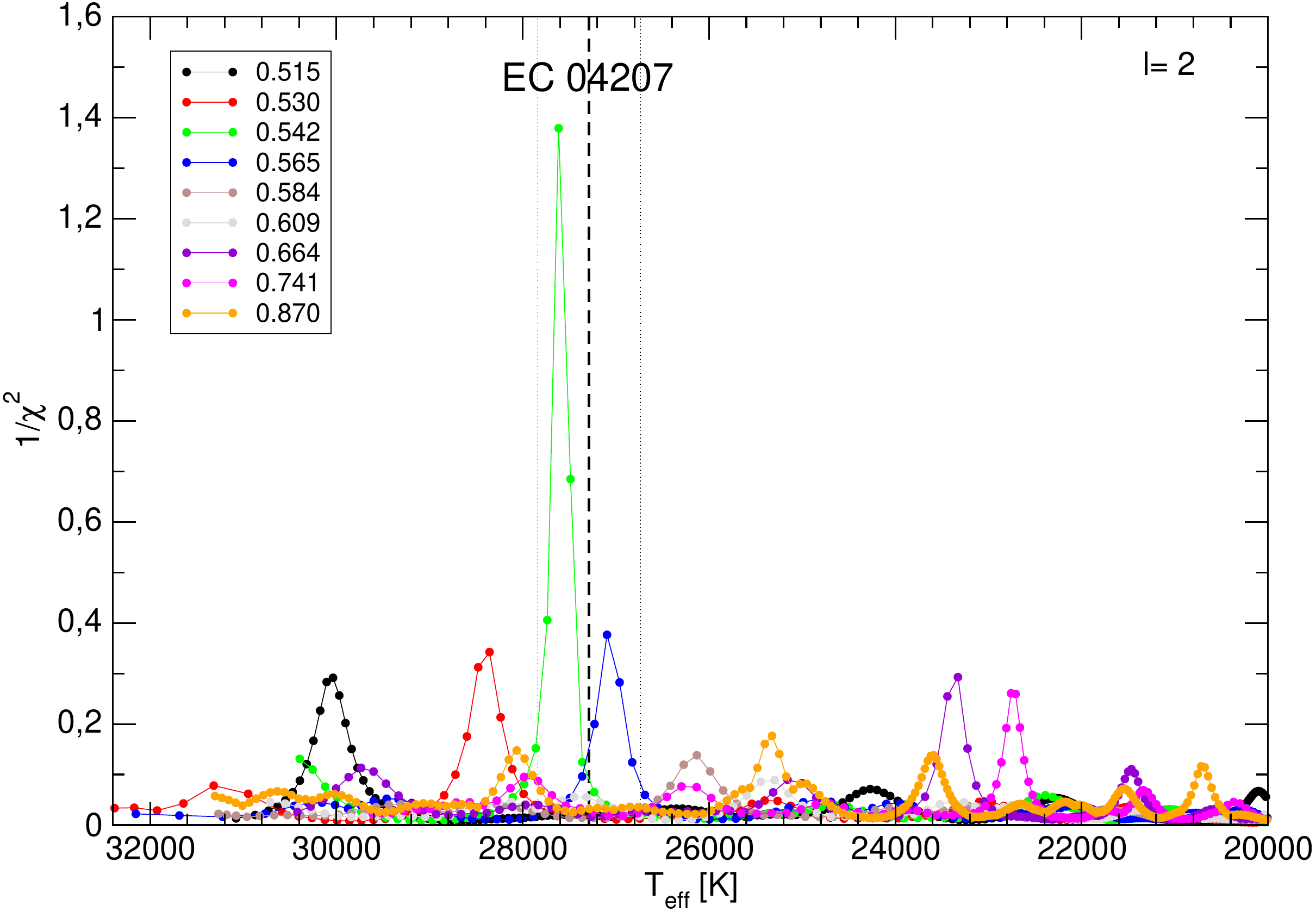}
\caption{The inverse of the quality function of the period fit with $\ell= 2$ in terms of 
the effective temperature, shown with different colors for the different 
stellar masses, corresponding to the four periods of EC~04207 in Table \ref{table:EC04207-extended}. 
The vertical black dashed line corresponds to the spectroscopic $T_{\rm eff}$ of 
EC~04207 and the vertical dotted lines its uncertainties \citep[$T_{\rm eff}= 27\,288\pm 545$ K;][]{2007A&A...470.1079V}. A prominent maxima corresponds to the best-fit model selected 
for EC~04207, with $M_{\star}= 0.542 M_{\odot}$ and $T_{\rm eff}= 27\,614$ K.}
\label{chi2-EC04207} 
\end{figure} 

\begin{table*}
\centering
\caption{Observed and theoretical periods of the asteroseismological solution
for EC~04207 [$M_{\star}= 0.542 M_{\odot}$, $T_{\rm eff}= 27\,614$ K]. Periods are in
seconds  and rates of period change  (theoretical) are  in units of $10^{-13}$ s/s. 
$\delta \Pi_i= \Pi^{\rm O}_i-\Pi_k$ represents  the period differences, 
$\ell$ the harmonic degree, $k$ the radial order, $m$ the azimuthal index.  
The last column gives information about the pulsational stability/instability
nature  of  the modes.}
\begin{tabular}{lc|cccccc}
\hline
\noalign{\smallskip}
$\Pi_i^{\rm O}$ & $\ell^{\rm O}$ & $\Pi_k$ & $\ell$ & $k$ & $\delta \Pi_k$ & $\dot{\Pi}_k$ & 
Unstable\\ 
(s) & & (s) & & & (s) & ($10^{-13}$ s/s) &  \\
\noalign{\smallskip}
\hline
\noalign{\smallskip}       
336.397 & 2 &  335.356       & 2 & 12   & 1.041 &  1.510     & yes \\
423.655 & 2 &  422.653       & 2 & 16   & 1.002 &  1.594     & yes \\
447.194 & 2 &  446.476       & 2 & 17   & 0.718 &  1.754     & yes \\
599.100 & 2 &  599.646       & 2 & 24   & $-0.546$ &  1.606  & yes \\
\noalign{\smallskip}
\hline
\end{tabular}
\label{table:EC04207-asteroseismic-model}
\end{table*}

In Fig.~\ref{fig:tests-EC04207} we show the statistical tests applied to the set 
of four periods in Table \ref{table:EC04207-extended}. The tests 
strongly suggest a possible constant period spacings of $\Delta \Pi \sim 21.9$ s. There is 
another possible period spacings at 17 s,  29 s, 37 s, and 43 s, that only are seen only in the 
K-S test, but not in the I-V test. To interpret these possible spacings of periods and assume an harmonic degree value for them, it is useful to first examine how they compare to the theoretical period-spacing values of stellar models. In other words, we can assign the values  
$\ell = 1$ or $\ell = 2$, or simply rule out any indication of period spacing by plotting them together with the average of the theoretical period spacings in terms of effective 
temperature for different stellar masses. This is shown in Fig.~\ref{fig:psp-teff-EC04207}, 
where we display the average of the computed period spacing  
$\overline{\Delta \Pi_{k}}$ (computed considering a period interval of $200-600$ s) 
  with curves of different colors to distinguish the different stellar 
  masses, corresponding to $\ell= 1$ (upper panel) and $\ell= 2$ (lower panel) modes. 
We note that the strongest peak in the statistical tests, that corresponds to $\Delta \Pi= 21.85$ s (averaging the values that arise from the K-S and I-V tests), must be associated to 
$\ell= 2$ modes, indicating a stellar mass of $M_{\star} \sim 0.565 M_{\star}$. The 
corresponding period spacing with $\ell= 1$ should be close to what the 
asymptotic theory predicts, that is $21.85\ {\rm s}\times \sqrt{3}= 37.85$ s. 
Indeed, the peak at $\Delta \Pi= 36.91$ s in the K-S test is likely that 
dipole period spacing, indicating, in this case, a stellar mass of $M_{\star} \sim 0.609 M_{\odot}$ for EC~04207. Alternatively, the presence of the possible period spacings at 
$\Delta \Pi= 17.45$ s ($\ell= 2$) and $\Delta \Pi= 29.05$ s ($\ell= 1$) would indicate
a stellar mass of $M_{\star} \sim 0.82 M_{\odot}$ and $M_{\star} \sim 0.85 M_{\odot}$, 
respectively. Finally, the possible period spacing $\Delta \Pi= 43.62$ s, that should be 
associated to $\ell= 1$, would indicate a stellar mass too low ($M_{\star} \sim 0.50 M_{\odot}$) and in contradiction with the indications of its spectroscopic parameters. Although all these possible period spacings are estimated on the basis of very few periods (four in total) ---and therefore we cannot conclude anything definite regarding the stellar mass based on the period spacing--- the strong signal of a possible period spacing of 21.85 s in both statistical tests would be indicating a stellar mass of $M_{\star} \sim 0.565 M_{\star}$. It remains to explain why the corresponding dipole period spacing period, of $\sim 37$ s, appears not so noticeably and only in the K-S test. We have  recomputed the  K-S test  for the  cases in  which we ignore the period of $\sim 424$ s and the period of $\sim 447$ s (one at a time). We still get the same strong indication of a  $\sim 22$ s period spacing in both  cases. As for the
possible dipole  period spacing of $\sim 37$  s, the associated minimum appears enhanced  when we ignore the $\sim 424$ s period, and  remains unchanged when we 
ignore the $\sim 447$ s period. We conclude that the $\sim 22$ s period spacing is robust, and that the $\sim 37$ s potential period spacing  is sensitive to the presence or not of the $\sim 424$ s period.

We have performed a period-to-period fit employing the four observed periods of EC~04207 from Table \ref{table:EC04207-extended}. We have fixed the harmonic degree $\ell= 2$ 
for the all the periods, since they do fit the pattern of quadrupole period spacing derived above.
The results of our period fit are shown in Fig. \ref{chi2-EC04207}. The best fit corresponds to a 
model with a mass of $M_{\star}= 0.542 M_{\odot}$ and an effective 
temperature $T_{\rm eff}= 27\,614$ K, compatible with the effective temperature of  EC~04207. 
There is also other possible solution for $M_{\star}= 0.565 M_{\odot}$ and $T_{\rm eff}= 27\,095$ K,
although the quality of the period match is poorer. We have repeated the analysis leaving free  
the harmonic degree of the four periods of EC~04207 (allowing them to be $\ell= 1$ or $\ell= 2$), that is, disregarding the constraints imposed by the period spacing derived above. In this case (not shown), we also find the same solution of $0.542 M_{\odot}$ and $T_{\rm eff}= 27\,614$ K as before. 
 We adopt the $0.542 M_{\odot}$ model as the asteroseismological model for EC~04207. The average of the $\ell= 2$ period spacing for this model is $\overline{\Delta \Pi}_{\ell= 2}= 22.09$ s, close to the period spacing derived from the statistical tests ($\Delta \Pi_{\ell= 2}= 21.85$ s). A DBV pulsator with only $\ell= 2$ modes seems to be  very unusual. In the  case of DAVs, there is the  ultra-massive 
ZZ Ceti star BPM~37094  that appears  to  be  pulsating  mostly  with $\ell=  2$  modes \citep{2004ApJ...605L.133M, 2005ApJ...622..572B, 2019A&A...632A.119C}. As far as we are aware, there is no other DBV star with all its periods associated to
$\ell= 2$ modes,  apart from EC~04207  according to our analysis.  It must be
taken  into account,  however, that  to put  this result  on a  firmer
basis, it would  be necessary to detect more periods  and, on the other  
hand, to carry out  other independent asteroseismological analyses of 
EC~04207.

\begin{figure} 
\includegraphics[clip,width=1.0\columnwidth]{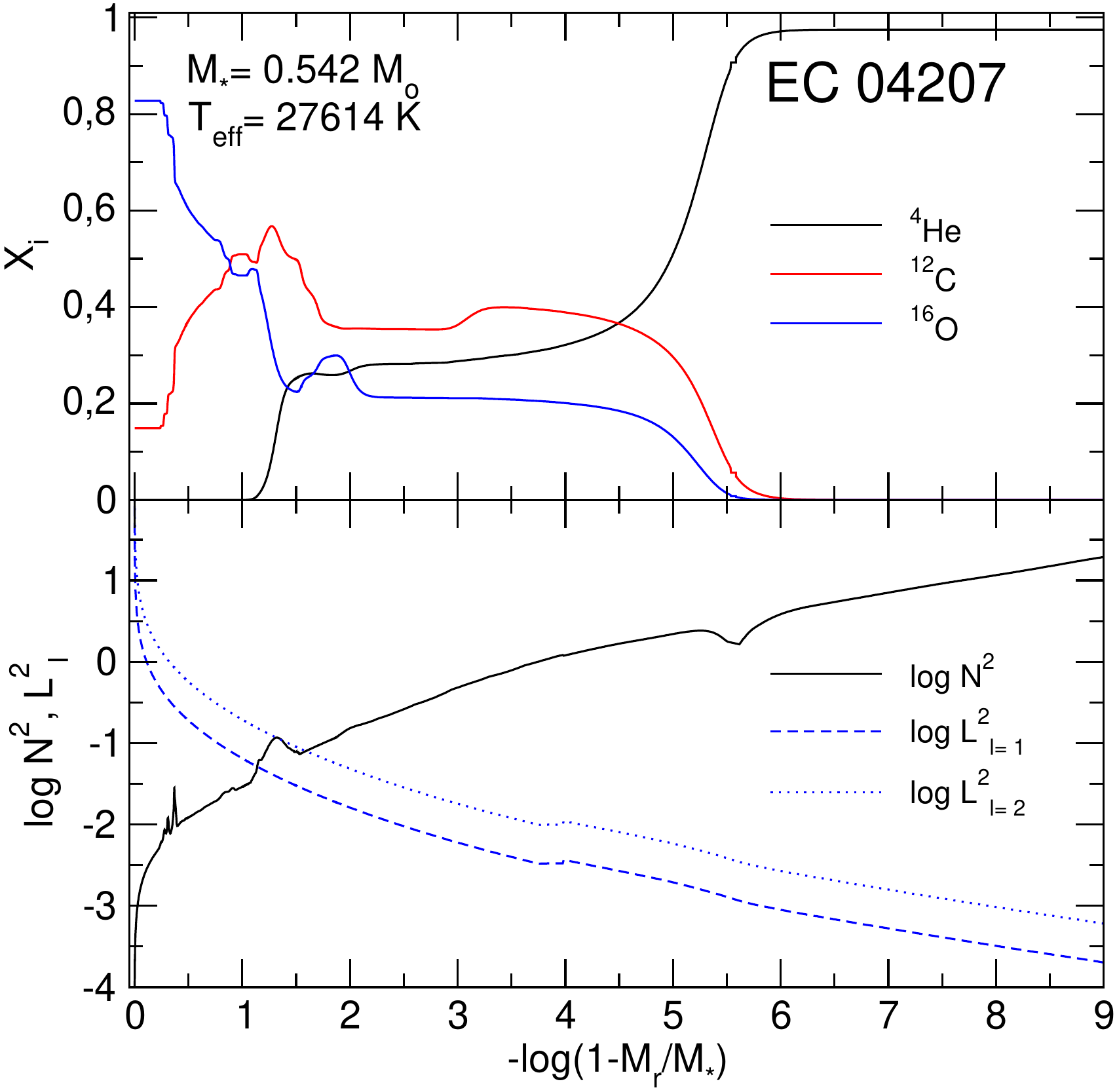}
\caption{Chemical  profiles  (upper panel)  and  the  squared 
Brunt-V\"aïs\"al\"a  and  Lamb  frequencies  for $\ell= 1$ and $\ell= 2$ (lower panel)  corresponding  to  our  asteroseismological  DB  WD  model  of EC~04207 with  a  stellar  mass $M_{\star}= 0.542 M_{\odot}$ and an effective temperature $T_{\rm eff}= 27\,614 $ K.}
\label{EC04207-modelo} 
\end{figure}

In Table \ref{table:EC04207-asteroseismic-model} we show a comparison of the observed and theoretical periods and mode identifications, along with the theoretical rate of period change and stability nature of each mode. We get, for this model, $\overline{\delta \Pi_i}= 0.83$ s, $\sigma= 0.85 $ s, and BIC= 0.16. We show the characteristics of this model in Table  \ref{astero-EC04207}. The chemical composition profiles and the Brunt-V\"ais\"al\"a and Lamb frequencies of this model are displayed Fig. \ref{EC04207-modelo}).   We find that all the periods of EC~04207 are unstable according to our nonadiabatic computations, in line with the fact that we observe these oscillation periods in the pulsation spectrum of this star (see last column of Table \ref{table:EC04207-asteroseismic-model}).

Based on the asteroseismological model for EC~04207, we can assess a seismological distance, 
as we did for PG~1351 and EC~20058. We find a magnitude absolute in the {\it Gaia} photometry of 
$M_G= 10.126$ mag using a DB WD model atmosphere. Using the observed 
{\it Gaia} magnitude, $m_{G}= 15.229$ mag, we obtained  
$d_{\rm s}= 104.88^{+1.54}_{-1.94}$ pc. The seismological distance is $\sim 12 \%$ higher than the 
astrometric distance measured by {\it Gaia} EDR3, of $91.48^{+0.22}_{-0.23}$ pc. This discrepancy is an indication of the existence of uncertainties in the modeling of the interior of this DB WD star, 
and also of the scarcity of periods exhibited by this target, which make that the asteroseismological 
methods, in this case, lead to a seismological model that is not entirely well constrained.

We close this section by  noting that,  if instead of  using  the  $T_{\rm eff}= 27\,288 
\pm  545$  K of  \citep{2007A&A...470.1079V} we used the effective temperature derived  by \citep{2014A&A...568A.118K}, that is $T_{\rm eff}= 25\,970 \pm 500$ K, the stellar  mass 
suggested by  the derived quadrupole period spacing  of $\sim 19$ s  would be slightly higher,
of $\sim 0.585 M_{\odot}$ (as compared  with $\sim 0.565 M_{\odot}$). On the other  hand,
adopting the lower $T_{\rm eff}$ assessed by \citep{2014A&A...568A.118K} we could not find a 
seismological model for EC~04207.

\begin{table}
\centering
\caption{The main characteristics of the DBV star EC~04207 according to the adopted asteroseismological model.} 
\begin{tabular}{l|cc}
\hline
\hline 
Quantity & Spectroscopy &  Asteroseismology \\
         & Astrometry   &         \\ 
\hline
$T_{\rm eff}$ [K]                           & $27\,288 \pm 545$  & $27\,614\pm 100$       \\
$M_{\star}$ [$M_{\odot}$]                   & $0.515\pm0.023$    & $0.542^{+0.017}_{-0.008}$ \\ 
$\log g$ [cm/s$^2$]                         & $7.808\pm0.058$    & $7.872^{+0.027}_{-0.021}$ \\ 
$\log (L_{\star}/L_{\odot})$                & $\hdots$           & $-0.980\pm 0.008$     \\  
$\log(R_{\star}/R_{\odot})$                 & $\hdots$           & $-1.850\pm 0.007$     \\  
$(X_{\rm C}, X_{\rm O})_{\rm c}$            & $\hdots$           & $0.15, 0.83$        \\
$M_{\rm He}/M_{\star}$                      & $\hdots$           & $1.32\times10^{-2}$ \\
$d$  [pc]                                   & $91.48^{+0.22 (a)}_{-0.23}$ & $104.88^{+1.54}_{-1.94}$ \\ 
$\pi$ [mas]                                 & $10.93\pm0.03^{(a)}$       & $9.53^{+0.18}_{-0.13}$ \\ 
\hline
\hline
\end{tabular}
\label{astero-EC04207}
{\footnotesize  References: (a) Gaia DR3.}
\end{table}

\subsection{WDJ152738.4-450207.4}
\label{WDJ1527}

\begin{figure} 
\includegraphics[clip,width= 1.0\columnwidth]{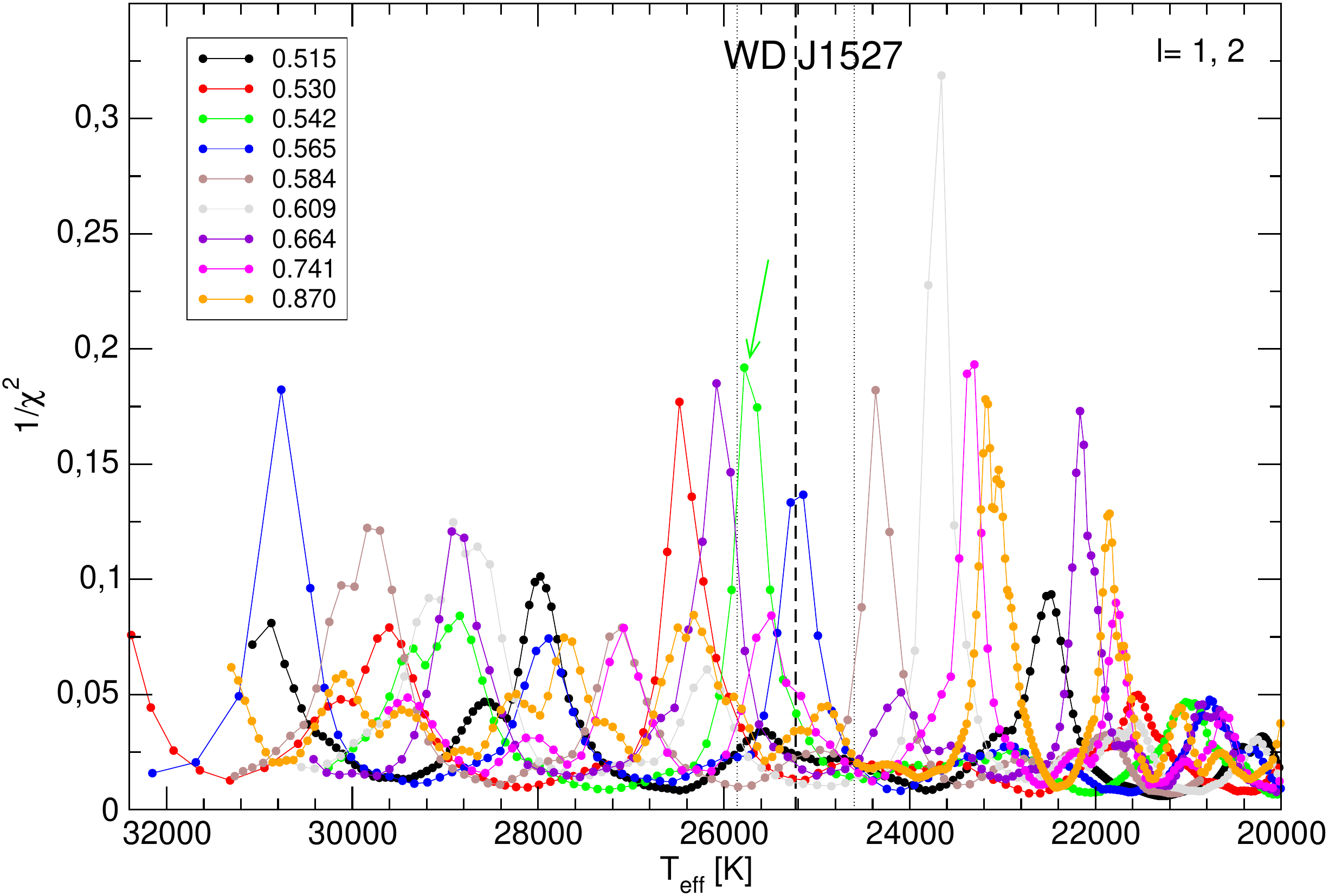}
\caption{The inverse of the quality function of the period fit in terms of 
the effective temperature, shown with different colors for the different 
stellar masses, corresponding to three of the five periods of WD~J1527 in Table~\ref{table:WDJ1527}. The vertical black dashed line corresponds to the 
spectroscopic $T_{\rm eff}$ of WD~J1527 and the vertical dotted lines its uncertainties ($T_{\rm eff}= 25\,228 \pm 630$ K). A local maxima (marked with a green arrow), 
corresponds to a possible representative model for WD~J1527 
compatible with spectroscopy (see the text).}
\label{chi2-WDJ1527} 
\end{figure}

\begin{table*}
\centering
\caption{Observed and theoretical periods of the representative model
of WD~J1527 [$M_{\star}= 0.542 M_{\odot}$, 
  $T_{\rm eff}= 25\,780$ K]. Periods are in
  seconds  and rates of period change  (theoretical) are  in units of
  $10^{-13}$ s/s. $\delta \Pi_i= \Pi^{\rm O}_i-\Pi_k$ represents  the
  period differences, $\ell$ the harmonic degree, $k$ the radial
  order, $m$ the azimuthal index.  The last column provides information
  about the pulsational stability/instability   nature  of  the
  modes.}
\begin{tabular}{lc|cccccc}
\hline
\noalign{\smallskip}
$\Pi_i^{\rm O}$ & $\ell^{\rm O}$ & $\Pi_k$ & $\ell$ & $k$ & $\delta \Pi_k$ & $\dot{\Pi}_k$ & 
Unstable\\ 
(s) & & (s) & &  & (s) & ($10^{-13}$ s/s) &  \\
\noalign{\smallskip}
\hline
\noalign{\smallskip}       
233.26   & ? & 235.619 & 2 &  7 & $-2.189$ & 0.927 & yes \\
352.93   & ? & 351.457 & 2 & 12 &  1.023 & 1.340 & yes \\
436.98   & ? & 440.544 & 2 & 16 & $-3.484$ & 1.562 & yes \\
649.397 & ? & 646.283 & 2 & 25 &  3.114 & 2.282 & yes \\ 
702.895 & 1 & 701.448 & 1 & 15 &  1.244 & 2.344 & yes \\
745.877 & ? & 744.651 & 1 & 16 &  1.226 & 2.751 & yes \\
\noalign{\smallskip}
\hline
\end{tabular}
\label{table:WDJ1527-representative-model}
\end{table*}

This is a new DBV star discovered with the {\sl TESS} data. The star exhibits
5 periods between $\sim 649$ s and $\sim 746$ s (Table \ref{table:WDJ1527}). 
Three periods ($701.754$ s, $702.895$ s, and $704.241$ s) appear to be the components
of a rotational triplet, suggesting a rotation period of $P_{\rm rot}\sim 2.3$~d.
This rotation-period value is within what is usual to find in WDs \citep[from 1 h to 4.2 d; see][]{2015ASPC..493...65K,2017ApJS..232...23H}. The rotation period 
of $\sim 2.3$~d is assessed by considering the \emph{average} of the frequency splitting
observed during the sector 38 of \emph{TESS}, since clearly the frequency splittings 
are time dependent (for some unknown reason). In summary, we have  only three independent pulsation periods (assumed to be $m= 0$ modes) of WD~J1527 available from {\sl TESS} for our asteroseismological analysis. If these 
periods were associated with low radial-order modes, they would 
be useful to establish strong asteroseismological constraints on the 
internal structure of WD~J1527 \citep[see, for example, the cases of the ZZ Ceti stars G226$-$29,  G117$-$B15A, and R~548;][]{1992ApJ...399L..91F,1995ApJ...447..874K,1998ApJS..116..307B}.  
However, these three periods correspond 
to intermediate or high radial orders, lowering their asteroseismic potential. As we mentioned in Sect. \ref{obs-WDJ1527}, we acquired
additional ground based observations in the hopes of increasing the number
of detected periods. Observations with the 1.6~m OPD Telescope allowed us to find additional periods of 233.26~s, 352.93~s, 436.98~s, and 
701.89 s. They complement the periods detected by {\sl TESS} in such a way as to cover the short-period regime of the pulsation spectrum of the star. The addition of the short periods of 233.26~s, 352.93~s, 436.98~s enables an asteroseismological analysis. Since the period
of 701.4~s detected from the ground corresponds to 
the rotationally triplet centered at 702.895~s, we will not take it into account in our period fit. First, we have searched for a uniform period spacing in the pulsational 
spectrum of WD~J1527 by employing the statistical tests, but 
we did not find any conclusive result. Then, we performed a period-to-period fit, assuming that the period 702.895 s corresponds to a mode with $\ell = 1$, $m= 0$  (the central component of a rotational triplet), 
and leaving free the value of $\ell$ for the other five periods 
(233.43~s, 352.48~s, 437.06~s, 649.397~s and 745.877~s). Our results for the inverse of the quality function versus the effective temperature for different stellar masses are shown in Fig. \ref{chi2-WDJ1527}. There are several models that reproduce well the three periods of WD~J1527, particularly a model with a stellar mass $M_{\star}= 0.609 M_{\odot}$ 
at $T_{\rm eff}= 23\,658$ K. However, its effective temperature is not in good agreement with the spectroscopic effective temperature. Restricting ourselves to models that provide good agreement with the observed periods (local maxima of the inverse of $\chi^2$) but at the same time satisfy the effective-temperature constraint, the best-fit model has $M_{\star}= 0.542 M_{\odot}$ and $T_{\rm eff}= 25\,780$ K (green arrow in Fig. \ref{chi2-WDJ1527}).
We note that the mass of this model is substantially smaller than the 
spectroscopic mass of  $M_{\star}= 0.675 \pm 0.022 M_{\sun}$. 
In Table \ref{table:WDJ1527-representative-model} we compare the observed pulsation periods of WD~J1527 with the theoretical periods of this representative model and their $\ell$ and $k$ values. The asymptotic period spacing of this model is $\Delta \Pi^{\rm a}= 42.107$ s. As we did for the other 
stars in this paper, we derive a seismological distance for WD~J1527 using the $\log g$ and 
$T_{\rm eff}$ of  the representative model. We find $d_{\rm s}= 111.98$ pc, that 
is about $20 \%$ larger than the {\it Gaia} astrometric distance, of 94.01 pc.
Clearly, more observations of the WD~J1527 variability from space and/or from the ground will be necessary to be able to analyze this DBV star in depth.

\subsection{L~7$-$44}
\label{L7-44}

\begin{figure} 
\includegraphics[clip,width= 1.0\columnwidth]{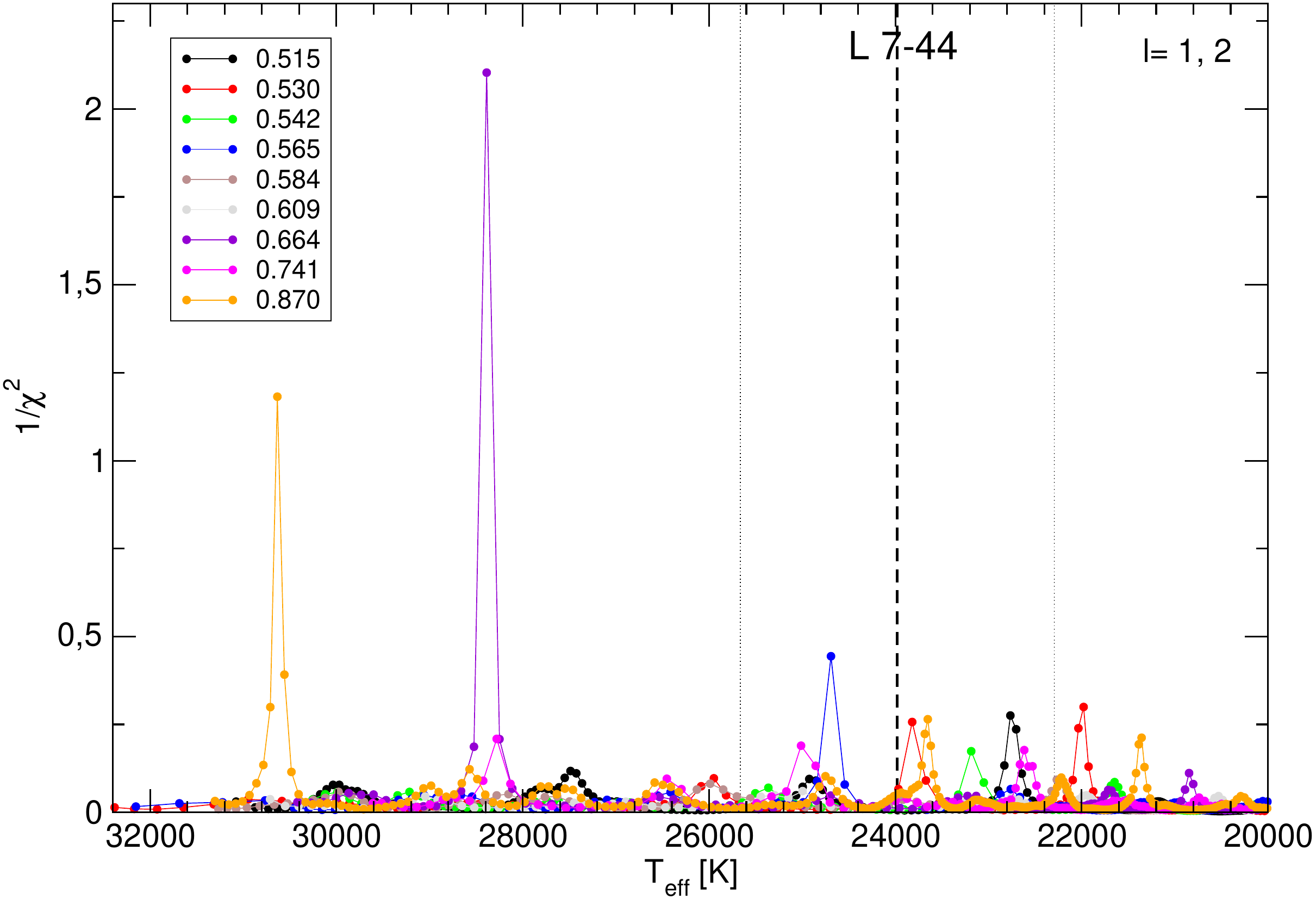}
\caption{The inverse of the quality function of the period fit in terms of 
the effective temperature, shown with different colors for the different 
stellar masses, corresponding to three of the five periods of L~7$-$44 in 
Table \ref{table:L7-44}. The vertical black dashed line corresponds to the 
spectroscopic $T_{\rm eff}$ of L~7$-$44 and the vertical dotted lines its uncertainties \citep[$T_{\rm eff}= 23\,980 \pm 1686$ K;][]{2018ApJ...857...56R}.}
\label{chi2-L7-44} 
\end{figure}

\begin{table*}
\centering
\caption{Observed and theoretical periods of a representative model
of L~7$-$44 [$M_{\star}= 0.565 M_{\odot}$, 
  $T_{\rm eff}= 24\,690$ K]. Periods are in
  seconds  and rates of period change  (theoretical) are  in units of
  $10^{-13}$ s/s. $\delta \Pi_i= \Pi^{\rm O}_i-\Pi_k$ represents  the
  period differences, $\ell$ the harmonic degree, $k$ the radial
  order, $m$ the azimuthal index.  The last column gives information
  about the pulsational stability/instability   nature  of  the
  modes.}
\begin{tabular}{lc|cccccc}
\hline
\noalign{\smallskip}
$\Pi_i^{\rm O}$ & $\ell^{\rm O}$ & $\Pi_k$ & $\ell$ & $k$ & $\delta \Pi_k$ & $\dot{\Pi}_k$ & 
Unstable\\ 
(s) & & (s) & &  & (s) & ($10^{-13}$ s/s) &  \\
\noalign{\smallskip}
\hline
\noalign{\smallskip}       
466.894  & ? & 466.720  & 2 & 17 &  0.174   & 1.159 & yes \\ 
914.955  & 1 & 914.864  & 1 & 20 &  0.091   & 2.121 & yes \\
936.750  & ? & 939.282  & 2 & 37 & $-2.532$ & 2.358 & yes \\ 
1019.138 & ? & 1017.539 & 2 & 40 &  1.599   & 2.902 & yes \\    
\noalign{\smallskip}
\hline
\end{tabular}
\label{table:L7-44-representative-model}
\end{table*}

This is another new DBV star discovered with the {\sl TESS} observations. Similarly to WD~J1527, the {\sl TESS} light curve of L~7$-$44 shows six independent periods, although three of them may belong to a rotational triplet, corresponding to a rotation period $P_{\rm rot}\sim 3.5$ d. This rotation period is within the expected range of WD rotation periods \citep{2019A&ARv..27....7C}. The periods detected fall in the range 467~s $-$ 1019~s. Since all but one of the periods are quite long, and consequently they are in the asymptotic regime of pulsations of DBV stars, this set of periods is not potentially useful for a deep asteroseismological analysis. At variance with the case of WD~J1527, for this star we do not have additional periods detected with observations from the ground available. With so few periods, for L~7$-$44 we limited ourselves to perform an exploratory asteroseismological analysis. 

We first tried to find evidence of uniform period spacings, but we find no conclusive results and we ruled out the possibility of deriving a stellar mass through this approach.
We also tried a period-to-period fit for this star. We considered the period of 914.955 s as corresponding to a mode with $\ell= 1$ since it is likely the central component ($m= 0$) of a rotational triplet, being the other $m\neq 0$ components the periods 913.655 s and 916.436 s. Thus, we have four periods available for the period fit: 466.894~s, 914.955~s, 936.750~s, and 1019.138~s. We show in Fig. \ref{chi2-L7-44} the inverse of the quality function versus the effective temperature for different stellar masses, along with the location of L~7$-$44 according to its effective temperature. There are two models that 
reproduce very well the periods of L~7$-$44, but they have very high effective temperatures. Among these two good fits, we have to discard that of $M_{\star}= 0.870 M_{\odot}$ at $T_{\rm eff}= 30\,634$ K, because it corresponds to a model that is too massive and hot, in clear contradiction with the spectroscopic mass and the effective temperature of the star, $M_{\star}= 0.630  M_{\odot}$ at $T_{\rm eff}= 23\,980\pm1686$ K. The other peak corresponds to a model with the stellar mass $M_{\star}= 0.664 M_{\odot}$ at $T_{\rm eff}= 28\,387$ K, which provides the best agreement with the observed periods.
In the range of effective temperatures allowed by spectroscopy, we find several small peaks associated to period fits of low quality. Among them, we consider that corresponding 
to a model with $M_{\star}= 0.565 M_{\odot}$ and $T_{\rm eff}= 24\,690$ K, since it represents the best-fit model in the allowed interval of $T_{\rm eff}$. In Table \ref{table:L7-44-representative-model} we show a comparison between the detected periods of L~7$-$44  and the theoretical periods of the 
model with $M_{\star}= 0.565 M_{\odot}$ at $T_{\rm eff}= 24\,690$ K. The asymptotic period spacing of this model is $\Delta \Pi^{\rm a}= 42.07$ s. We derive a seismological distance for L~7$-$44 
using the $\log g$ and $T_{\rm eff}$ of  the representative model, and find $d_{\rm s}= 75.26$ pc, that 
is $9 \%$ larger than the {\it Gaia} astrometric distance, of 69.24 pc. As in the case of WD~J1527, we emphasize that more short periods would be necessary before we can place significant
constraints on the structure of L~7$-$44 with asteroseismology. 


\begin{table*}
\renewcommand{\arraystretch}{1.5}
\centering
  \caption{The name of the studied stars (column 1), the independent periods detected 
  with TESS (column 2), and the periods measured from ground-based observations (column 3).
  Periods in boldface are the ones used in the period-to-period fits (Section \ref{asteroseismology}).}
  \begin{tabular}{lcc}
\hline
Name        &  Periods from {\sl TESS} & Periods from ground-based observations\\
            &    [s]                   & [s]       \\
\hline
PG~1351+489 & {\bf 489.260}    & {\bf 335.26}, 489.33, {\bf 584.68}, {\bf 639.63}  \\
EC~20058    & {\bf 256.852}, {\bf 280.983} & {\bf 195.0}, {\bf 204.6}, 207.6, 256.9,\\ 
            &                  & 274.7, 281.0, 286.6, {\bf 333.5},\\ 
            &                  & {\bf 350.6}, {\bf 525.4}, {\bf 539.8} \\     
EC~04207    & {\bf 336.397}, {\bf 423.655}, {\bf 447.194} & 336.4, 423.5, 447.2, {\bf 599.1} \\
WD~J1527    & {\bf 649.397}, 701.754, {\bf 702.895}, 704.241, {\bf 745.877} &
              {\bf 233.26}, {\bf 352.93}, {\bf 436.98}, 701.89  \\
L~4$-44$    &  {\bf 466.894}, {\bf 914.955}, {\bf 936.750}, {\bf 1019.138} & $\cdots$\\
\hline
\label{summary-period-astero}
\end{tabular}
\end{table*}

\section{Summary and conclusions}
\label{conclusions}

In this paper, we presented an analysis of five pulsating DBV stars observed with the {\sl TESS} mission. Three of the five analyzed targets are the already known DBV stars PG\,1351, EC\,20058 and EC\,04207. The other two objects 
are new DBVs discovered with the {\sl TESS} data:  WD~J1527 and L~7$-$44.
With the discovery of these two new DBVs, the number of
known stars of this class increases to 49. 
We examined  the variability  of  these  five DBV stars  by  analyzing  their short and ultra-short-cadence single-sector observations obtained with {\sl TESS} and  found 18 significant independent oscillation frequencies.

We detected and measured just a single frequency and a combination frequency (the first harmonic of the independent frequency) for 
PG~1351 (see Table \ref{table:PG1351}). In the case of EC~20058, we found two independent frequencies and no combination frequencies (Table \ref{table:EC20058}). These two stars have very simple FTs, 
unlike the rest of the objects. Indeed, EC~04207, 
WD~J1527 and  L~7$-$44 have relatively rich FTs. Specifically, 
for EC~04207 we detected four independent frequencies along with 
three combination frequencies (Table \ref{table:EC04207}), for  
WD~J1527 we found five independent frequencies and one combination frequency
(Table \ref{table:WDJ1527}), and finally, for L~7$-$44 we measured six independent frequencies and no combination frequencies (Table \ref{table:L7-44}). In the
case of WD~J1527, we measured four additional independent frequencies from ground-based 
observations (Table \ref{table:WDJ1527-OPD}). We also examined the running Fourier transform (sFT) of each target to investigate the temporal evolution of the pulsation modes. As in other compact pulsating stars, including sdBV, GW Vir, and DAV stars, we find variable amplitudes in the sFTs of  WD~J1527 (Fig. \ref{fig:sFTJ15}) and  L~7$-$44 (Fig. \ref{fig:FT451}). For two targets, EC~04207 and WD~J1527, we estimated the possible rotation periods from $g$-mode 
frequency splittings. In the former case, the rotation period of EC~04207 would be either 
1.14 or 2.28 d depending on the missing azimuthal order. We show in Fig. 
\ref{fig:FTEC04207_doublet} the peaks corresponding to the frequencies at 
$2359.142 ~\mu$Hz and $2361.678 ~\mu$Hz that can be considered as two components of a rotationally split dipole mode. In the latter case, the rotation period of WD~J1527 is constrained to 2.3 days. 
Fig. \ref{fig:sFTJ15} shows a detected rotational triplet with a splitting of $\sim 2.51~\mu$Hz for WD~J1527. These findings are in line with what has been discovered for other types of 
pulsating WDs, such as GW Vir variables, the rotation periods of which ranges from 5 hours to a few days \citep{2019A&ARv..27....7C,2021A&A...645A.117C,2022MNRAS.513.2285U} and DAV pulsating stars, for which it ranges from 1 hour to a 4.2 days \citep{2015ASPC..493...65K,2017ApJS..232...23H}.
According to the atmospheric parameters derived for WD~J1527 from spectroscopic data, we found that the star is well within  the  DBV  instability  strip with an effective temperature 25\,228\,K and surface gravity 8.124\,dex.

We also performed a detailed asteroseismological analysis of these stars on the basis of the fully evolutionary models of DB WDs computed by \cite{2009ApJ...704.1605A}.  In Table \ref{summary-period-astero}
we show the periods employed for the asteroseismological period fits  of
each target, including their origins (ground-based and {\sl TESS} data). We summarize below our asteroseismological findings for each star:

\begin{table*}
\renewcommand{\arraystretch}{1.5}
\centering
  \caption{The name of the studied stars (column 1), the effective temperature and gravity of their asteroseismological or 
  representative models (columns 2 and 3), the apparent and absolute {\it Gaia} magnitudes (columns 4 and 5), and the seismological and astrometric distances (columns 6 and 7). Column 8 gives the discrepancy in the distances.}
  \begin{tabular}{lccccccc}
\hline
Name       &  $T_{\rm eff}$ & $\log g$  & $m_G$ & $M_G$     & $d_{\rm s}$ & $d_{\it Gaia}$ & Distance \\
           &    [K]         & [cgs]     & [mag]   &   [mag]  & [pc]        & [pc]  & discrepancy \\
\hline
PG~1351+489     & 25\,775 & 8.103 &  16.678 & 10.564 & $167.05^{+2.31}_{-2.26}$ & $175.73^{+1.53}_{-1.58}$ &  4.94 \% \\
EC~20058$-$5234 & 25\,467 & 8.062 &  15.756 & 10.518 & $111.60^{+1.53}_{-1.50}$ & $117.95^{+0.68}_{-0.75}$ &  5.38 \% \\
EC~04207$-$4748 & 27\,614 & 7.872 &  15.229 & 10.126 & $104.88^{+1.54}_{-1.94}$ &  $91.48^{+0.22}_{-0.23}$ &  12.8 \% \\ 
WD~J1527$-$4502 & 25\,780 & 7.877 &  15.482 & 10.236 & $111.98$                 &  $94.01^{+0.34}_{-0.37}$ &  16.0 \% \\
L~7$-$44        & 24\,690 & 7.921 &  14.737 & 10.354 &  $75.26$                 &  $69.24^{+0.14}_{-0.12}$ &   8.0 \% \\
\hline
\label{distances}
\end{tabular}
\end{table*}

\begin{itemize}
\item PG~1351: {\sl TESS} detected only one period for this star, which agrees with 
one of the four periods detected from ground-based observations. We searched for
a constant period spacing and found hints of a period spacing of $\sim 31$ s 
for $\ell= 1$ modes and $\sim 19$ s for for $\ell= 2$ modes. These possible period 
spacings suggest a high mass for PG~1351, in the range $0.74-0.87 M_{\odot}$. 
The period-to-period fit carried out on this star 
indicates an asteroseismological model with a stellar mass of $M_{\star}= 0.664\pm 0.013M_{\odot}$,
which is smaller than the mass indicated by the period spacings, although larger 
than the spectroscopic mass ($M_{\star}= 0.558\pm 0.027M_{\odot}$). 

\item EC~20058: For this star, {\sl TESS} detected only two periods, which are 
in perfect agreement with two of the 11 periods found from ground-based observations. 
The search for constant period spacings suggests the existence of 
a $\ell= 1$ period spacing of $41.4$  s and a $\ell= 2$ period spacing of $23.8$ s, which
indicate a stellar mass of $0.53 M_{\odot}$ and $0.55 M_{\odot}$, respectively. According to this analysis, the 11 periods of EC~20058 are a mix of $\ell= 1$ and $\ell= 2$ periods,
without rotational splitting multiplets present. An alternative analysis, 
proposed by \cite{2008MNRAS.387..137S}, consists of a period spectrum of both dipole and quadrupole periods, but with the presence of rotational triplets. A period-to-period fit analysis considering the above two possibilities results in the adoption of an asteroseismological model with a stellar mass of $0.664 M_{\odot}$, substantially larger than the mass range indicated by the period spacings ($0.53 M_{\odot}-0.55M_{\odot}$) and in 
agreement with the spectroscopic mass, of $M_{\star}= 0.614\pm 0.030M_{\odot}$. 

\item EC~04207: For this star, {\sl TESS} detected three periods, which are in 
agreement with three of the four periods detected from the ground. {\sl TESS} failed 
to detect a period measured from ground-based telescopes ($\sim 600$ s). 
A search for a constant period spacing strongly suggest a $\ell= 2$ uniform period 
separation of $\sim 21.9$ s, compatible with a 
stellar mass of $M_{\star}= 0.565 M_{\odot}$. A period-to-period fit 
analysis for this star allowed us to find an asteroseismological model with a stellar mass
of  $0.542 M_{\odot}$, somewhat higher than the spectroscopic mass ($0.515\pm 0.023M_{\odot}$),
but lower than the mass derived from the period spacing. 

\item WD~J1527: This is one of the new DBV stars discovered with {\sl TESS}. 
The star exhibits five periods between $\sim 649$ s and $\sim 746$ s, but three
periods (701.754 s, 702.895 s, and 704.241 s) appear to be the
components of a rotational triplet. We have also considered three additional periods
(233.43 s, 352.48 s, 437.0 s) detected with ground-based observations. We have 
not found a robust period spacing, nor a conclusive asteroseismological model.
We have to wait for further observations of this star ---either from space or from the 
ground--- in order to analyze it in depth.

\item L~7$-$44: This is the other new DBV star discovered with {\sl TESS}. Similar
to what happened in the case of WD~J1527, L~7$-$44 exhibits six periods, three of which 
seems to be part of a rotational triplet. Thus, we have only four independent periods. 
Clearly, this is insufficient for asteroseismological modelling, so we are again limited to searching for a representative model that replicates as closely as possible the observed periods. For a deeper study, we will have to wait for new observations of the star. 

\end{itemize}

We summarize in Table \ref{distances} the asteroseismological distances for the five stars analyzed in this work, and their comparison with their astrometric {\it Gaia} distances. We find a relatively good agreement ($\sim 5-8\%$) between both sets of distance estimates for some stars 
(PG~1351, EC~20058, and EC~04207), although for the cases of WD~J1527 and  L~7$-$44 the 
discrepancies are substantial
($\sim 13-16 \%$), meaning that the representative DB WD models used to estimate 
the distances are not entirely reliable.

The results presented in this work demonstrate that {\sl TESS} is having a great impact 
in the asteroseismology of WDs and pre-WDs ---as well as in other classes of pulsating stars--- as a worthy successor to the {\sl Kepler}/{\sl K2} mission. We envisage that, as this mission 
continues to provide uninterrupted high-quality photometric data of pulsating WD and pre-WDs, it will be possible to advance our understanding of the structure and evolution of these stars.

\begin{acknowledgements}
We  wish  to  acknowledge  the detailed  suggestions  and comments of an anonymous referee that strongly improved the original version of this work.  
This  paper includes data collected with the {\sl TESS} mission, obtained
from the MAST data archive at the Space Telescope Science Institute
(STScI). Funding for the {\sl TESS} mission is provided by the NASA Explorer
Program. STScI is operated by the Association of Universities for
Research in Astronomy, Inc., under NASA contract NAS 5–26555. 
M.U. acknowledges financial support from CONICYT Doctorado Nacional 
in the form of grant number No: 21190886 and ESO studentship program.
Part of this work was supported by AGENCIA through the Programa de
Modernizaci\'on Tecnol\'ogica BID 1728/OC-AR, and by the PIP
112-200801-00940 grant from CONICET. 
K.J.B.\ is supported by the National Science Foundation under 
Award AST-1903828. J.J.H. acknowledges support through TESS Guest 
Investigator Programs 80NSSC20K0592 and 80NSSC22K0737.
Financial support from the National Science Centre under 
project No.\,UMO-2017/26/E/ST9/00703 is acknowledged.
This research has made use of
NASA's Astrophysics Data System.
\end{acknowledgements}

\bibliographystyle{aa}
\bibliography{paper_bibliografia.bib}

\end{document}